\documentclass[tightenlines,eqsecnum,floats,aps,amsmath,amssymb,nofootinbib,prd,showpacs,notitlepage]{revtex4-1}
\pdfoutput=1 
\usepackage{amsmath,amssymb,amsfonts}
\usepackage{graphicx}
\usepackage{enumerate} % advanced enumerate environment
\usepackage{colordvi} % for color text
\usepackage{bm}% bold math
\usepackage{epsfig}
\usepackage{float}
\usepackage{verbatim}
\usepackage{subfig}
\newcommand{\bfx}{\mbox{\boldmath$x$}}
\newcommand{\bfk}{\mbox{\boldmath$k$}}

\begin{document}
\title{Theoretical Accuracy in Cosmological Growth Estimation}

\vfill
\author{Benjamin Bose$^{1}$, Kazuya Koyama$^{1}$, Wojciech A. Hellwing$^{1,2}$, Gong-Bo Zhao$^{1,3}$, Hans A. Winther$^{1}$}
\bigskip

\affiliation{$^1$Institute of Cosmology \& Gravitation, University of Portsmouth,
Portsmouth, Hampshire, PO1 3FX, UK}
\affiliation{$^2$Janusz Gil Institute of Astronomy, University of Zielona G\'ora, ul. Szafrana 2, 65-516 Zielona G\'ora, Poland}
\affiliation{$^3$National Astronomy Observatories, Chinese Academy of Science, Beijing, 100012, P.R.China}

\bigskip
\vfill
\date{today}
\begin{abstract}
We elucidate the importance of the consistent treatment of gravity-model specific non-linearities when estimating the growth of cosmological structures from redshift space distortions (RSD). Within the context of standard perturbation theory (SPT), we compare the predictions of two theoretical templates with redshift space data from COLA (COmoving Lagrangian Acceleration) simulations in the normal branch of DGP gravity (nDGP) and General Relativity (GR). Using COLA for these comparisons is validated using a suite of full N-body simulations for the same theories. The two theoretical templates correspond to the standard general relativistic perturbation equations and those same equations modelled within nDGP.  Gravitational clustering non-linear effects are accounted for by modelling the power spectrum up to one loop order and redshift space clustering anisotropy is modelled using the Taruya, Nishimichi and Saito (TNS) RSD model. Using this approach, we attempt to recover the simulation's fiducial logarithmic growth parameter $f$. By assigning the simulation data with errors representing an idealised survey with a volume of $10\mbox{Gpc}^3/h^3$, we find the GR template is unable to recover fiducial $f$ to within 1$\sigma$ at $z=1$ when we match the data up to $k_{\rm max}=0.195h$/Mpc. On the other hand, the DGP template recovers the fiducial value within $1\sigma$. Further, we conduct the same analysis for sets of mock data generated for generalised models of modified gravity using SPT, where again we analyse the GR template's ability to recover the fiducial value. We find that for models with enhanced gravitational non-linearity, the theoretical bias of the GR template becomes significant for stage IV surveys. Thus, we show that for the future large data volume galaxy surveys, the self-consistent modelling of non-GR gravity scenarios will be crucial in constraining theory parameters.
\end{abstract}
\maketitle
%%%%%%%%%%%%%%%%%%%%%%%%%%%%%%%%%%%%%%%%%%%%%%%%%%%%%%%
%%%%%%%%%%%%%%%%%%%%%%%%%%%%%%%%%%%%%%%%%%%%%%%%%%%%%%%
\section{Introduction}
%%%%%%%%%%%%%%%%%%%%%%%%%%%%%%%%%%%%%%%%%%%%%%%%%%%%%%%
%%%%%%%%%%%%%%%%%%%%%%%%%%%%%%%%%%%%%%%%%%%%%%%%%%%%%%%
The accelerated expansion of the universe has been the focus of strenuous research since its discovery almost 20 years ago \cite{Riess:1998cb,Perlmutter:1998np} with the concordance model of cosmology performing very well in explaining observational data on large scales. Most of this success is based on a homogenous and isotropic solution within Einstein's GR. This standard model ($\Lambda$ Cold Dark Matter -LCDM) assumes that GR is the valid theory of gravity at large scales. Additionally, GR does just as well locally, having passed all tests within the solar system with flying colours. Further, a recent remarkable measurement made by the LIGO consortium has also confirmed the existence of gravitational waves \cite{Abbott:2016blz}, an early GR predicted phenomena. On the other hand, theoretically the standard model of cosmology leaves a lot to be desired with an ubiquitous and vastly dominant dark energy component having to be introduced. A natural explanation for this energy has lead to the largest fine tuning problem in physics \cite{Weinberg:1988cp,Martin:2012bt} and the resulting search for a more natural solution has brought researchers to the tricky task of modifying GR \cite{Hu:2007nk,Starobinsky:1980te,Dvali:2000hr,Gomes:2013ema,Hassan:2011zd,deRham:2011by,deRham:2010ik,deRham:2010kj,Maartens:2010ar}. 
\newline
\newline
A common feature of these modifications is the introduction of an additional fifth force sourced by the modified theory's additional degree of freedom. To deal with observational constraints, these modifications make use of the so called screening mechanisms, shielding matter from the fifth force at small scales (See \cite{Koyama:2015vza,Sakstein:2015oqa,Clifton:2011jh} for reviews). This vastly diminishes the distinguishability of these models at small scales, while the unscreened fifth force outside the screened regime makes them very different from GR. The large scale structure (LSS) of the universe has thus become an ideal laboratory for testing the phenomenology of modifications to gravity, a fact widely discussed in the literature \cite{Uzan:2000mz,Lue:2004rj,Ishak:2005zs,Knox:2005rg,Koyama:2006ef,Chiba:2007rb,Amendola:2007rr,Simpson:2012ra,Terukina:2012ji,Terukina:2013eqa,Yamamoto:2010ie,Jain:2007yk,Zhao:2008bn,Zhao:2009fn,Asaba:2013xql}. 
\newline
\newline
One of the most powerful sources of cosmological information is the signal encoded in anisotropic galaxy clustering observed in the redshift space \footnote{Redshift space is a coordinate system where the radial distance to a galaxy is obtained by its observed redshifts.}, an effect commonly denoted as Redshift Space Distortions (RSD) \cite{Kaiser:1987qv}.This effect is due to the non-linear mapping between real position and redshift space, the linear part coming from the usual cosmological redshift-distance relation while the non-linear part arising from a galaxy's peculiar velocity. In most modified gravity scenarios the fifth force couples to matter, enhancing matter's peculiar velocity and producing an imprint on the redshift space clustering statistics \cite{Linder:2007nu,Guzzo:2008ac,Yamamoto:2008gr,Song:2008qt,Song:2010bk,Guzik:2009cm,Song:2010fg,Asaba:2013mxj,Hellwing:2014nma}. 
\newline
\newline
At linear scales, the observed galaxy clustering is usually translated into a measurement of the parameter combination $f\sigma_8$ where $f=d \ln(F_1)/ d\ln(a)$, with $F_1$ being the linear growth of structure and $a$ being the scale factor, while $\sigma_8$ normalises the linear power spectrum (see eg. \cite{Peebles1980}). In the last decade, the extraction of the $f\sigma_8$ parameter from the observed RSD signal in galaxy spectroscopic surveys has become a common practice, and a number of such estimates have provided us with current state-of-the-art cosmological constraints. Here, among others, we can mention the measurements using luminous red galaxy (LRG) sample of the Sloan Digital Sky Survey \cite{Oka:2013cba}, the BOSS survey galaxy samples \cite{Samushia:2013yga,Sanchez:2013uxa,Sanchez:2013tga,Gil-Marin:2015sqa,Sanchez:2016sas} and the VIPERS survey \cite{delaTorre:2013rpa}. However, an important caution needs to be made here. So far all of these state-of-the-art constraints on $f$ were obtained by employing a standard GR approach to RSD modelling, even when the aim was to place constraints on modified models of gravity. The question naturally arises: \emph{ is the flexibility of standard templates, with the inclusion of nuisance parameters such as velocity dispersions, enough to encompass MG effects or should the analysis be done in a model dependent way?} In other words, are the systematic biases induced by using a GR-based RSD approach small enough that, when compared to the statistical errors in the data, can be safely ignored? Such a question is very timely now at the advent of precision cosmology, which will be fuelled by vast amounts of data from upcoming large observational endeavours.
\newline
\newline
Whether the GR treatment gives biased results for the constraints on modified gravity model parameters or not depends on the approach used in RSD modelling and also the precision of the measurements. Clearly, if the errors in the measurements are larger than the bias introduced by the inaccurate modelling of RSD, there is no immediate call to change the RSD modelling. This condition should be tested before the analysis pipeline is applied to real data. For the TNS model \cite{Taruya:2010mx}, which has been used to measure $f \sigma_8$ using the power spectrum measurements of the BOSS survey \cite{Beutler:2016arn}, this has been done only for $f(R)$ gravity. Using this model, it was shown that the standard GR template gives a biased estimation of the model parameter assuming an ideal survey with a volume of $10 \mbox{Gpc}^3/h^3$ at $z=1$  \cite{Taruya:2013my}. On the other hand, in \cite{Barreira:2016mg} they find that with their analysis of the redshift space power spectrum for the braneworld model of gravity by Dvali, Gabadadze and Porrati (DGP) \cite{Dvali:2000hr}, negligible model bias is found. In that work the authors employ the growth rate estimation pipeline of \cite{Sanchez:2016sas}. It is clear that a general way to deal with this systematic is needed, especially in preparation for stage IV surveys such as the Dark Energy Spectroscopic Instrument (DESI) \footnote{\url{http://desi.lbl.gov/}} and the ESA/Euclid survey\footnote{\url{www.euclid-ec.org}}, which will significantly reduce observational uncertainties.
\newline
\newline
It is worth mentioning that if one is able to analyse the data in a model dependant way without cost then the answer to the above question is clear. In a previous work \cite{Bose:2016qun} we presented a code which includes model specific non-linearities when constructing the TNS redshift space power spectrum, giving us access to a wide variety of theoretical templates. The non-linearities are constructed in the context of standard Eulerian perturbation theory (SPT) (See \cite{Bernardeau:2001qr} for a comprehensive review). 
\newline
\newline
In this work we present predictions generated using the framework described in \cite{Bose:2016qun} as well as further investigate the issue of model bias, extending the work of \cite{Barreira:2016mg}, and providing a means of quickly testing the validity of constraining non-GR models using the standard GR template. In addition, our aim is to assess the level and scales at which theoretical modelling bias becomes important and starts to affect the results in a significant way. We do this first in terms of taking the analysis to increasing non-linear scales. With respect to non-linear scales, the errors on $f \sigma_8$ depend sensitively on the maximum $k$ used in the analysis. By including smaller scales the predictions of each template become more unique and so model bias becomes more pronounced. The maximum scale we can include depends on the redshift \cite{Carlson:2009it} and so there is also a redshift dependency on the systematic given our theoretical approach. Lastly, we stress-test the GR template to obtain a limit at which it becomes significantly biased due the enhanced small-scale dynamics induced by MG. We do this by increasing the modification to gravity at non-linear scales in MG mock data.  
\newline
\newline
This paper is organised as follows: In Sec.II we review the relevant theory behind the theoretical templates used to fit the RSD. In Sec.III we test the realm of validity of the framework using results from both LCDM and DGP N-body simulations as well as MG-PICOLA simulations \cite{Winther:2017jof}. We then conduct an MCMC estimation of $f$ using both the standard GR template and the full MG template for varying inclusion of scales. This is done using a suite of 20 $1\mbox{Gpc}^3/h^3$ MG-PICOLA simulations. The same analysis is then done for mock data generated using SPT in order to mimic an ideal survey. We do this for varying levels of fifth force interactions. Finally, we summarise our results and highlight future work in Sec.IV.  
%%%%%%%%%%%%%%%%%%%%%%%%%%%%%%%%%%%%%%%%%%%%%%%%%%%%%%%
%%%%%%%%%%%%%%%%%%%%%%%%%%%%%%%%%%%%%%%%%%%%%%%%%%%%%%%
\section{Theoretical Template for Growth Estimation }
\subsection{The Perturbative Treatment : SPT} 
\noindent Our starting point will be perturbations in a universe described by the Friedman-Robertson-Walker (FRW) metric in the Newtonian gauge:
%%%%%%%%%%%%%%%%%%%%%%%%%%%%%%%%%%%%%%%%%%%%%%%%%%%%%%%%%%%%%%%%%%%%%%
\begin{equation}
ds^2=-(1+2\Phi)dt^2+a^2(1-2\Psi)\delta_{ij}dx^idx^j.
\end{equation}
%%%%%%%%%%%%%%%%%%%%%%%%%%%%%%%%%%%%%%%%%%%%%%%%%%%%%%%%%%%%%%%%%%%%%%
The evolution of matter and velocity perturbations within the Hubble horizon, under the quasi-static approximation are given by the Euler and continuity equations. Specifically, we consider the evolution of the density contrast 
\begin{equation}
\delta({\bf x})=\frac{\rho_m({\bf x})-\bar{\rho}}{\bar{\rho}}
\end{equation}
where $\rho_m(\bf{x})$ and $\bar{\rho}$ are the matter densities at a given point and the background respectively. We also consider the evolution of the peculiar velocity field $v_p(\bf{x})$. At the scales of interest (\emph{ie.} well above the scales of virialised, collapsed objects), the cosmic velocity filed is to a good approximation curl-free (see eg.\cite{Pichon:1999tk,Pueblas:2008uv,Libeskind:2012ya}), thus as a potential flow it can be fully characterised by the divergence only part of the vector. In our analysis we will use the velocity divergence, here defined as:
\begin{equation}
\theta({\bf x})=-\frac{\nabla\cdot v_p({\bfx})}{aH(a) f}
\end{equation}
where $H(a)$ is the Hubble function. As usual convention, we choose to work in the Fourier space, where the many SPT expressions become simpler. Beyond linear order, mode coupling terms are introduced and in Fourier space the equations are given by (see eg.\cite{Bernardeau:2001qr})
%%%%%%%%%%%%%%%%%%%%%%%%%%%%%%%%%%%%%%%%%%%%%%%%%%%%%%%%%%%%%%%%%%%%%%
\begin{eqnarray}
&&a \frac{\partial \delta(\bfk)}{\partial a}+\theta(\bfk) =-
\int\frac{d^3\bfk_1d^3\bfk_2}{(2\pi)^3}\delta_{\rm D}(\bfk-\bfk_1-\bfk_2)
\alpha(\bfk_1,\bfk_2)\,\theta(\bfk_1)\delta(\bfk_2),
\label{eq:Perturb1}\\
&& a \frac{\partial \theta(\bfk)}{\partial a}+
\left(2+\frac{a H'}{H}\right)\theta(\bfk)
-\left(\frac{k}{a\,H}\right)^2\,\Phi(\bfk)=
-\frac{1}{2}\int\frac{d^3\bfk_1d^3\bfk_2}{(2\pi)^3}
\delta_{\rm D}(\bfk-\bfk_1-\bfk_2)
\beta(\bfk_1,\bfk_2)\,\theta(\bfk_1)\theta(\bfk_2),
\label{eq:Perturb2}
\end{eqnarray}
%%%%%%%%%%%%%%%%%%%%%%%%%%%%%%%%%%%%%%%%%%%%%%%%%%%%%%%%%%%%%%%%%%%%%%
the prime denoting a scale factor derivative w.r.t cosmic time and the mode coupling kernels, $\alpha$  and
$\beta$, are given by
%%%%%%%%%%%%%%%%%%%%%%%%%%%%%%%%%%%%%%%%%%%%%%%%%%%%%%%%%%%%%%%%%%%%%%
\begin{eqnarray}
\alpha(\bfk_1,\bfk_2)=1+\frac{\bfk_1\cdot\bfk_2}{|\bfk_1|^2},
\quad\quad
\beta(\bfk_1,\bfk_2)=
\frac{(\bfk_1\cdot\bfk_2)\left|\bfk_1+\bfk_2\right|^2}{|\bfk_1|^2|\bfk_2|^2}.
\label{alphabeta}
\end{eqnarray}
%%%%%%%%%%%%%%%%%%%%%%%%%%%%%%%%%%%%%%%%%%%%%%%%%%%%%%%%%%%%%%%%%%%%%%
At linear order, $\alpha = \beta = 0$. Gravity enters the perturbation's evolution through the Poisson term including the Newtonian Potential $\Phi$. We can parametrise a wide range of modifications to gravity by writing out the Poisson term as \cite{Koyama:2009me,Bose:2016qun}
\begin{equation}
-\left(\frac{k}{a H}\right)^2\Phi=
\frac{3 \Omega_m(a)}{2} \mu(k,a)\,\delta(\bfk) + S(\bfk),
\label{eq:poisson1}
\end{equation}
where $\Omega_m(a) = 8 \pi G \rho_m/3 H^2$. The function $S(\bfk)$ is the non-linear source term up, to the third order given by \cite{Bose:2016qun}
%%%%%%%%%%%%%%%%%%%%%%%%%%%%%%%%%%%%%%%%%%%%%%%%%%%%%%%%%%%%%%%%%%%%%%
\begin{eqnarray}
S(\bfk)&=&
\int\frac{d^3\bfk_1d^3\bfk_2}{(2\pi)^3}\,
\delta_{\rm D}(\bfk-\bfk_{12}) \gamma_2(\bfk, \bfk_1, \bfk_2;a)
\delta(\bfk_1)\,\delta(\bfk_2)
\nonumber\\
&& + 
\int\frac{d^3\bfk_1d^3\bfk_2d^3\bfk_3}{(2\pi)^6}
\delta_{\rm D}(\bfk-\bfk_{123})
\gamma_3( \bfk, \bfk_1, \bfk_2, \bfk_3;a)
\delta(\bfk_1)\,\delta(\bfk_2)\,\delta(\bfk_3)
\label{eq:Perturb3}
\end{eqnarray}
%%%%%%%%%%%%%%%%%%%%%%%%%%%%%%%%%%%%%%%%%%%%%%%%%%%%%%%%%%%%%%%%%%%%%%
The functions $\mu(k,a)$, $\gamma_2( \bfk, \bfk_1, \bfk_2; a)$  and $\gamma_3(\bfk, \bfk_1, \bfk_2, \bfk_3;a)$ encode the modification to gravity. Particularly, $\gamma_2( \bfk, \bfk_1, \bfk_2; a)$  and $\gamma_3(\bfk, \bfk_1, \bfk_2, \bfk_3;a)$ contain the non-linear part of the information about the theory of gravity, in particular about the screening mechanism. These functions will completely specify the difference between theoretical templates used to estimate the growth parameter later on. In Appendix A we specify the form of these functions for the Vainshtein screened DGP model of gravity \cite{Dvali:2000hr}. 
\newline
\newline
Classically, the approach of SPT is to solve eq.(\ref{eq:Perturb1}) and eq.(\ref{eq:Perturb2}) perturbatively, with n-th order solutions given by 
\begin{align} 
\delta_n(\boldsymbol{k}, a) &= \int d^3\boldsymbol{k}_1...d^3 \boldsymbol{k}_n \delta_D(\boldsymbol{k}-\boldsymbol{k}_{1...n}) F_n(\boldsymbol{k}_1,...,\boldsymbol{k}_n, a) \delta_0(\boldsymbol{k}_1)...\delta_0(\boldsymbol{k}_n) \label{nth1} \\ 
\theta_n(\boldsymbol{k},a) &= \int d^3\boldsymbol{k}_1...d^3 \boldsymbol{k}_n \delta_D(\boldsymbol{k}-\boldsymbol{k}_{1...n}) G_n(\boldsymbol{k}_1,...,\boldsymbol{k}_n, a) \delta_0(\boldsymbol{k}_1)...\delta_0(\boldsymbol{k}_2) \label{nth2}
\end{align}
where $\boldsymbol{k}_{1...n} = \boldsymbol{k}_1 + ...+ \boldsymbol{k}_n$ and $F_i(\bfk_1,\bfk_2...)$ and $G_i(\bfk_1,\bfk_2...)$ are the $i^{th}$ order kernels which we solve perturbatively for. Using the numerical algorithm described in \cite{Taruya:2016jdt} we can do this for the perturbations up to any order in a general way. With the perturbative solutions up to 3rd order we can construct the 1-loop power spectrum 
\begin{equation}
P^{1-{\rm loop}}_{ij}(k) = P_{0}(k) + P^{22}_{ij}(k) + P^{13}_{ij}(k)
\label{loopps}
\end{equation}
where $P_{0}(k)$ is the linear power spectrum defined as 
\begin{equation}
\langle \delta_0(\bfk) \delta_0(\bfk')\rangle =
(2\pi)^3\delta_{\rm D}(\bfk+\bfk')\,P_{0}(k).
\end{equation} 
and
\begin{align}
\langle g_i^{2}(\bfk) g_{j}^2(\bfk')\rangle &=
(2\pi)^3\delta_{\rm D}(\bfk+\bfk')\,P_{ij}^{22}(k). \label{eq:psconstraint0} \\
\langle g_i^{1}(\bfk) g_{j}^3(\bfk')
+g_i^{3}(\bfk) g_{i}^1(\bfk') \rangle &=
(2\pi)^3\delta_{\rm D}(\bfk+\bfk')\,P_{ij}^{13}(k).
\label{eq:psconstraint1}
\end{align}
where $g^i_1 = \delta_i$ and $g^i_2= \theta_i$. 
\newline
\newline
Eq.\ref{loopps} describes the first order contribution to the power spectrum using the density and velocity divergence fields as our expansion variables. The inclusion of the higher order loop terms has been shown to improve the prediction of theory \cite{Jeong:2006xd}, an improvement more pronounced at higher redshift \cite{Carlson:2009it}. The regime of applicability of SPT with a required percent-level precision was assessed by comparing with N-body simulations. Presently, the best estimates are given by \cite{Jeong:2006xd}, who found  $\Delta^2(k,a) \leq 0.4$. Here $\Delta^2(k,a) = k^3 P_0(k,a)/(2\pi^2)$ is the dimensionless power spectrum which is a function of time, and thus so are the scales of SPT applicability. This being said, it is well known that SPT suffers from convergence problems, with higher loop contributions being of comparable size to lower ones, a problem that worsens at late times \cite{Carlson:2009it}. This problem is evident from the 1-loop power spectrum expressions. Mode coupling introduces an integral over all scales producing a small scale dependency. Errors in the small scale regime thus sneak into our large scale predictions when we go beyond linear theory. This makes comparisons to data limited with the loss of very constraining small scale information. A number of analytical treatments have been proposed in the literature \cite{Taruya:2007xy,Matarrese:2007wc,Valageas:2006bi,Matsubara:2007wj,Pietroni:2008jx,Anselmi:2012cn,Blas:2015qsi} which all go a way to improving the match with N-body simulations in the mildly non-linear regime. The effective field theory of large scale structure (EFToLSS) \cite{Baumann:2010tm,Carrasco:2012cv} has also made great improvements to the range of validity of theoretical predictions albeit with the aid of N-body simulations. This makes it impractical for our general approach. 
\newline
\newline
So far we have discussed theoretical predictions for clustering statistics in real space. All depth information of the universe is confined to redshift space and so next we will discuss how to construct the redshift space power spectrum. 
\newpage

%%%%%%%%%%%%%%%%%%%%%
\subsection{A non-linear Model for RSD}
As mentioned in the introduction, RSD arises from the non-linear mapping between real and redshift space due to  contamination of the redshift distances by the contribution from the line-of-sight component of galaxy peculiar velocities. First modelled by Kaiser \cite{Kaiser:1987qv}, the anisotropy was described as a squashing effect along the line of sight because of coherent infall velocities at large linear scales.  
\begin{equation}
P_{K}^S(k,\mu) = (b+f\mu^2)^2P_{\delta \delta}(k)
\label{linkais}
\end{equation}
where $\mu$ is the cosine of the angle between the line of sight and $\bfk$ while $P_{\delta \delta}(k)$ is the linear matter power spectrum, and $b = \delta_g/\delta_m$ is the linear density bias parameter, $\delta_g$ being the galaxy density contrast  \footnote{The use of $\mu$ here should not be confused with the function $\mu(k;a)$ which will always include its arguments.}. If we move to smaller scales we find that virialised motion causes a broadening in the velocity distribution. This results in a damping effect. This highly non-linear effect was coined the Fingers of God (FoG) and is usually modelled phenomenologically\cite{Scoccimarro:2004tg,Percival:2008sh,Cole:1994wf,Peacock:1993xg,Park:1994fa,Ballinger:1996cd,Magira:1999bn}. 
\newline
\newline
We will focus our attention on the TNS model of RSD mentioned in the introduction. This model has proven to be quite successful in reproducing simulation data \cite{Nishimichi:2011jm,Taruya:2013my,Ishikawa:2013aea} and is founded on SPT. Further it has enjoyed a lot of success in the context of survey data comparisons with the WiggleZ Dark Energy Survey  \cite{Blake:2011rj}  and the Vipers survey \cite{Pezzotta:2016gbo}. For these reasons it is the natural choice for our analysis. We quote the TNS power spectrum as
 \begin{equation}
 P^S(k,\mu) = \mbox{D}_{\mbox{FoG}} (f k\mu \sigma_v) \{ P_{\delta \delta} (k) - 2  \mu^2 P_{\delta \theta}(k) + \mu^4 P_{\theta \theta} (k) + A(k,\mu) + B(k,\mu) \} 
 \label{redshiftps}
 \end{equation}
 \noindent where $P_{\delta \delta}, P_{\delta \theta}$ and $ P_{\theta \theta}$ are all at 1-loop order. The correction terms, A and B are given by 
 \begin{equation}
 A(k,\mu)=  -(k \mu) \int d^3 \boldsymbol{k'} \frac{k_z '}{k'^2} \{B_\sigma(\boldsymbol{k'},\boldsymbol{k}-\boldsymbol{k'},-\boldsymbol{k})-B_\sigma(\boldsymbol{k'},\boldsymbol{k}, -\boldsymbol{k}-\boldsymbol{k'}) \} 
 \label{Aterm}
 \end{equation}
 \begin{equation}
 B(k,\mu)= (k \mu)^2 \int d^3\boldsymbol{k'} F(\boldsymbol{k'}) F(\boldsymbol{k}-\boldsymbol{k'})
 \label{Bterm}
 \end{equation}
 where
 \begin{equation}
 F(\boldsymbol{k}) = \frac{k_z}{k^2}\left[P_{\delta \theta} (k) - \frac{k_z^2}{k^2}P_{\theta \theta} (k) \right] 
 \end{equation}
 The cross bispectrum $B_\sigma$ is given by
 \begin{equation}
 \delta_D(\boldsymbol{k}_1+ \boldsymbol{k}_2+ \boldsymbol{k}_3)B_\sigma( \boldsymbol{k}_1,\boldsymbol{k}_2,\boldsymbol{k}_3) = \langle \theta(\boldsymbol{k}_1)\{ \delta(\boldsymbol{k}_2) - \frac{k_{2z}^2}{k_2^2} \theta(\boldsymbol{k}_2)\}\{ \delta(\boldsymbol{k}_3) - \frac{k_{3z}^2}{k_3^2} \theta(\boldsymbol{k}_3)\rangle\}
 \label{lcdmbi}
 \end{equation}
 Finally $D_{FoG}(fk\mu\sigma_v)$, where $\sigma_v$ is treated as a free parameter quantifying the dispersion in velocities (expressed in RSD displacement units Mpc/$h$), is not treated perturbatively but rather phenomenologically. We choose to take an exponential form $ D_{FoG}(fk\mu\sigma_v) = \exp{(-f^2k^2 \mu^2 \sigma_v^2)}$ \cite{Peacock1992}. Again, this term provides the small scale damping of the power spectrum due to random, small scale, motion. 
\newline
\newline
The main feature of this model is the inclusion of the $A$ and $B$ correction terms which account for higher-order interactions between the density and velocity fields. This gives the model good predictive power at weakly nonlinear scales, as shown by $N$-body comparisons \cite{Taruya:2010mx,Nishimichi:2011jm,Taruya:2013my}. In GR these terms have been shown to enhance the power spectrum amplitude at the BAO scale and have a non-negligible effect on the acoustic features of the power spectrum \cite{Taruya:2010mx}.  In modified gravity theories the $B$ term is generally expected to be enhanced because of its linear growth dependance while the $A$ term involves the 2nd order perturbations so it is not obvious how it will change. In \cite{Taruya:2013quf} the authors show that these correction terms can be significantly different if the model of gravity is changed, specifically between GR and the $f(R)$ model by Hu and Sawicki \cite{Hu:2007nk}.
\newline
\newline
With the inclusion of galaxy bias, Eq.(\ref{redshiftps}) gives a prediction for an observable that can be measured from ongoing and upcoming surveys. In its presented form, one can still perform a valid comparison with matter statistics from N-body simulations. This being said, a comment on its realm of validity should also be noted: because it relies on SPT, it also suffers from a restricted realm of validity due to divergences in the PT scheme. Within the context of GR, this expression has been computed up to 2-loop order in the resummed PT scheme \cite{Taruya:2013my,Taruya:2009ir,Okamura:2011nu,Crocce:2007dt,Crocce:2012fa,Taruya:2012ut} which has a larger range of validity over the basic SPT treatment. The standard SPT treatment still gives us a good working range of scales in the quasi non-linear regime which will be the regime for our basic test of model bias.
\newline
\newline  
In the next section we present template comparisons using different data sets, highlighting when model bias becomes an issue at the level of matter statistics. 
\section{Results}
\begin{figure}[H]
 \centering
  \subfloat[]{\includegraphics[width=8.3cm, height=8.cm]{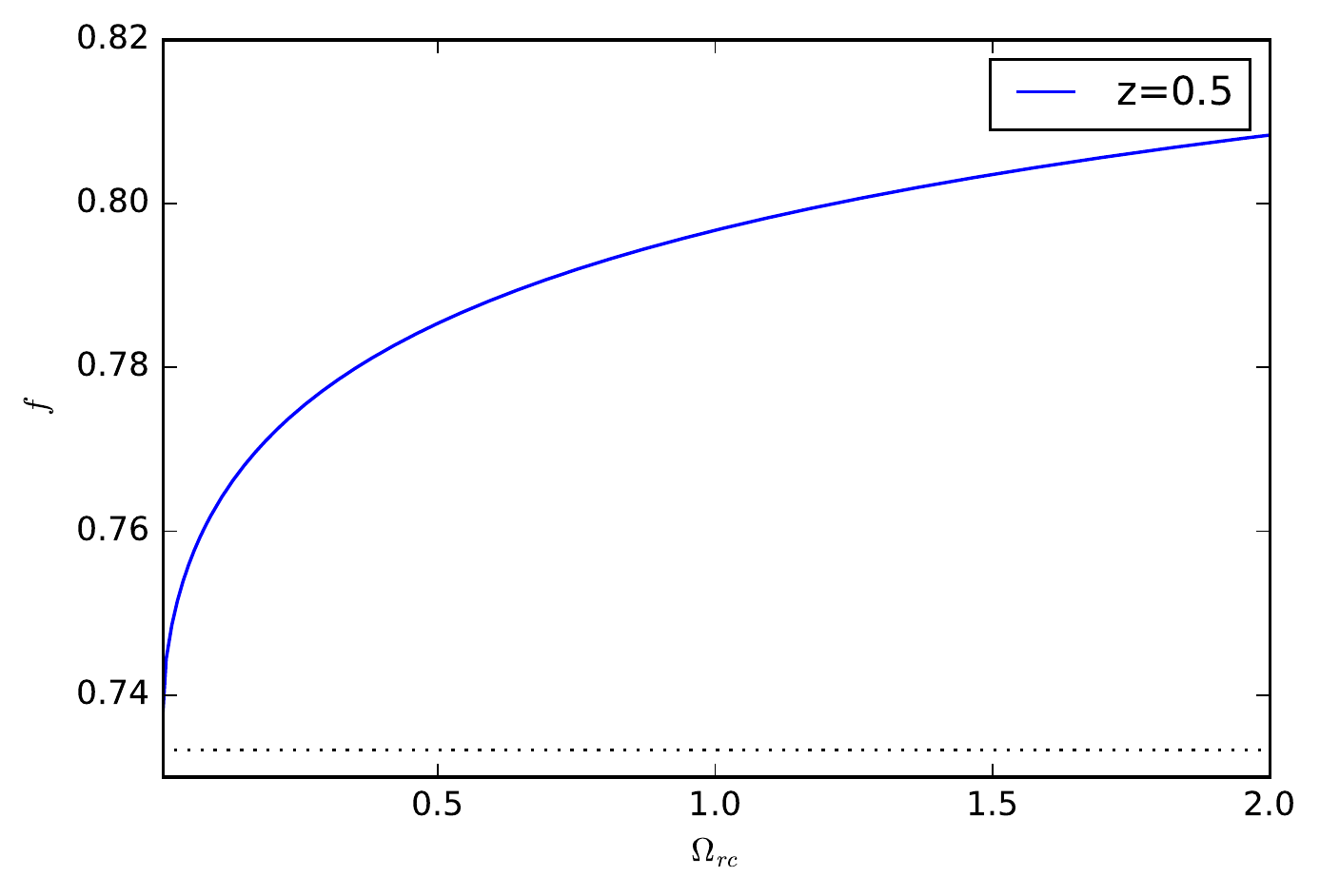}} \quad
  \subfloat[]{\includegraphics[width=8.3cm, height=8.cm]{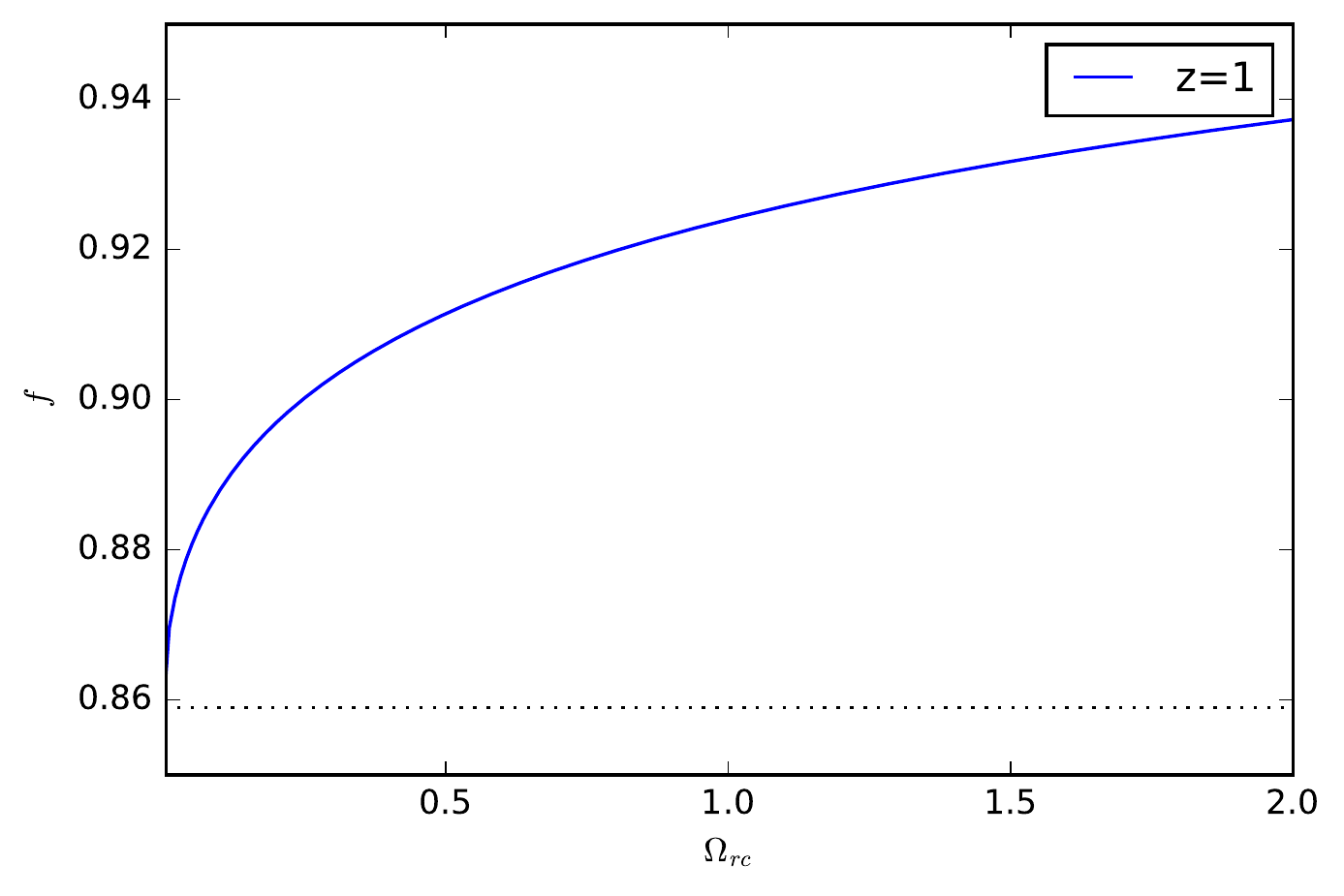}}
   \caption[CONVERGENCE ]{The logarithmic growth $f$ as a function of $\Omega_{rc}$ at $z=0.5$ (left) and $z=1$ (right). $\Omega_{rc}=0$ corresponds to GR and is marked by the dotted line. For other cosmological parameters see Sec.III. A.}
\label{forc}
\end{figure}

Our main goal here is to highlight the importance of model-specific non-linearities and biases that need to be accounted for in cosmological parameters estimation from galaxy spectroscopic surveys. We focus on the logarithmic growth rate, $f$, as the main parameter of interest. Thus, inspired by the approach presented in \cite{Oka:2013cba}, we will fix the amplitude of linear density perturbations, or equivalently $\sigma_8$, to the fiducial value. We will be dealing with data for which all cosmological parameters (\emph{ie}. $\Omega_m,\Omega_{DE},H_0$ etc) are already known \emph{a priori}, and therefore choose to keep all of them fixed during the analysis. Allowing them all to vary would just decrease the statistical significance of the estimates and fixing them does not introduce systematics. In this way we end up with only two free parameters $\{\sigma_v, f \}$ where $\sigma_v$ is the 1-dimensional velocity dispersion in the TNS model which needs to be fit to data. 
\newline
\newline
In this work we focus on GR and the normal branch of the DGP model (nDGP) with $\Omega_{rc} = 0.438$ where $\Omega_{rc} = 1/(2H_0r_c)^2$  and $r_c$ is the cross over scale (See Appendix A). The choice of the nDGP model as our guinea pig is motivated by its phenomenology, enriched by the non-linear screening Vainshtein mechanism. We assume a LCDM background in nDGP so the difference between nDGP and LCDM appears only in the structure growth. $f$ is derived from the linear versions of Eq.\ref{eq:Perturb1} and Eq.\ref{eq:Perturb2} (the right hand side as well as  $\gamma_2$ and $\gamma_3$ being set to 0) and because $\Omega_m=\Omega_m^{fiducial}$, the only free parameter is $\Omega_{rc}$ and so we will opt to parametrise $f$ by $\Omega_{rc}$. Since $\Omega_{rc}>0$, a lower bound for $f$ is also imposed. Otherwise the priors for both $\sigma_v$ and $f$ are flat. For clarity, Fig.\ref{forc} shows the relationship between $f$ and $\Omega_{rc}$ at $z=0.5$ and $z=1$. We note that $f(\Omega_{rc} = 0)$ corresponds to the logarithmic growth in LCDM. 
To get an estimate of $f$ we consider the multipoles of Eq.\ref{redshiftps}. These are modelled as 
\begin{equation}
P_\ell^{(S)}(k)=\frac{2\ell+1}{2}\int^1_{-1}d\mu P_{TNS}^{(S)}(k,\mu)\mathcal{P}_\ell(\mu)
\end{equation}
where $\mathcal{P}_\ell(\mu)$ denote the Legendre polynomials and $P_{TNS}^{(S)}(k)$ is given by eq.(\ref{redshiftps}). In order to get a robust estimation of higher order multipoles, simulations with large volumes and high mass resolution are required. Since our simulations are limited in size and resolution we will limit our modelling and analysis to the monopole and quadrupole.
\newline
\newline
In $P_{TNS}^{(S)} (k,\mu)$ we normalise the constituents to the fiducial values of the linear growth $F_1$ by using the following factor
\begin{equation}
\sigma_{{\rm norm}}(a) = \frac{F_1^{{\rm fiducial}}(a)}{F_1(a)} 
\end{equation}
with $\sigma_{{\rm norm}}^2(a)$ applied to $P^{{\rm linear}}_{ij}(k,a)$ and $\sigma_{{\rm norm}}^4(a)$ applied to $P^{13}_{ij}(k,a),P^{22}_{ij}(k,a),A(k,\mu,a)$ and $B(k,\mu,a)$. 
\newline
\newline
As stated earlier, $f$ and $\sigma_v$ are treated as free parameters and we fit the first two TNS multipoles to the data with the aim of recovering our fiducial value for $f$. To do this we perform an Markov Chain Monte Carlo (MCMC) analysis using the following likelihood function 
\begin{equation}
-2\ln{\mathcal{L}} = \sum_n \sum_{\ell,\ell'=0,2} \left(P^{(S)}_{\ell,{\rm data}}(k_n)-P^{(S)}_{\ell,{\rm model}}(k_n)\right) \mbox{Cov}^{-1}_{\ell,\ell'}(k_n)\left(P^{(S)}_{\ell',{\rm data}}(k_n)-P^{(S)}_{\ell',{\rm model}}(k_n)\right)
\label{covarianceeqn}
\end{equation}
where $\mbox{Cov}_{\ell,\ell'}$ is the covariance matrix between the different multipoles. Expressions for the covariance components can be found in Appendix C of \cite{Taruya:2010mx}. For our analysis we do not consider non-Gaussianity in the covariance (See \cite{Takahashi:2009bq} for justification of this treatment) but include the effect of shot-noise assuming the number density of an ideal future survey $\bar{n}_g= 4 \times 10^{-3} h^3 \mbox{Mpc}^{-3}$. Further, we use linear theory to estimate the covariance matrix components. This approximation has been checked to work well within $k\leq0.3h$/Mpc for the LCDM simulations used in \cite{Taruya:2010mx}. Given this, we have found it sufficient for our purposes to check that the covariance of both the density and velocity divergence fields are the same in both our LCDM and nDGP simulations within the scales of interest. Using the scaled covariance (the so-called decoherence function, see \cite{Chodorowski:2001id}) we have estimated that they agree to sub percent levels at $k\leq0.2h$/Mpc.
\newline
\newline
We consider two redshifts, $z=1$ and $z=0.5$. This is to give an idea of the trade off between enhanced non-linearity but decreased 
realm of validity of SPT at lower $z$. For both these redshifts, we assume in the MCMC analysis that the errors are those characteristic for a survey with a volume of $V_s=10 h^{-3} \mbox{Gpc}^3$ which is conservatively smaller than the upcoming Dark Energy Spectroscopic Instrument (DESI) \cite{Aghamousa:2016zmz} which is further around 3 times smaller than the Euclid spectroscopic survey \cite{Laureijs:2011gra}. Our number density, $\bar{n}_g= 4 \times 10^{-3} h^3 /\mbox{Mpc}^{3}$ and volume are comparable to that of the BOSS MGS sample \cite{Ross:2014qpa} or DESI's BGS \cite{Zhao:2013dza,Aghamousa:2016zmz}. 
\newline
\newline
Since we are only presented with $1\mbox{Gpc}^3/h^3$ box realisation of a full N-body run, using solely this data set could hamper our analysis through uncertainties connected with cosmic variance that become severe at the box-scale. Observing this we decided to eliminate that potential caveat by using an additional 20 $\times$ $1\mbox{Gpc}^3/h^3$ box realisations as our data set. We do this by employing a rather inexpensive modified gravity COLA approach (hereafter MG-PICOLA) recently  presented in \cite{Winther:2017jof} . MG-PICOLA is based on a parallel COLA implementation (PICOLA) (see \cite{Howlett:2015hfa} for details). As we have mentioned MG-PICOLA is relatively computationally inexpensive, but this advantage comes at the price of significantly limited accuracy (when compared to a same resolution  N-body run) in the non-linear regime of structure formation. We have carefully performed many tests to ascertain that MG-COLA is sufficiently accurate for the purpose of the analysis presented in this work. The reader is referred to Appendix B for the details of the MG-PICOLA tests we have performed. 
\subsection{Real and Redshift Space: Theory vs Simulations}
Here we provide comparison of our full N-body measurements and MG-PICOLA measurements to the SPT predictions. To begin, we compare the real space N-body auto and cross real space power spectra with the SPT 1-loop predictions. This provides a measure of the non-linearity captured by SPT as well as realm of validity of the PT treatment. We then compare the TNS multipoles to MG-PICOLA measurements. 
\subsubsection{Real Space Comparisons: N-body}
\begin{figure}[H]
  \captionsetup[subfigure]{labelformat=empty}
  \centering
  \subfloat[]{\includegraphics[width=8.3cm, height=8.3cm]{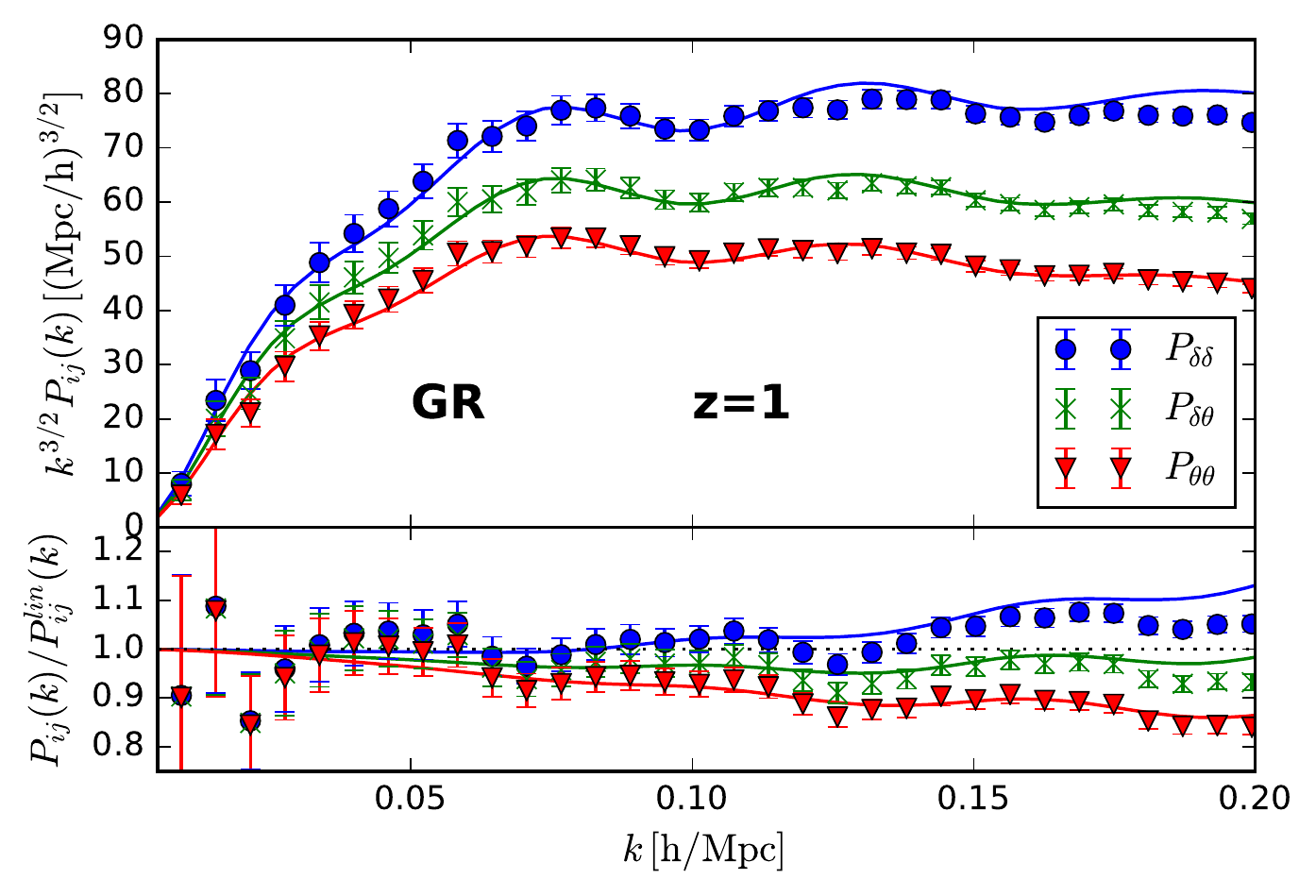}} \quad
  \subfloat[]{\includegraphics[width=8.3cm, height=8.3cm]{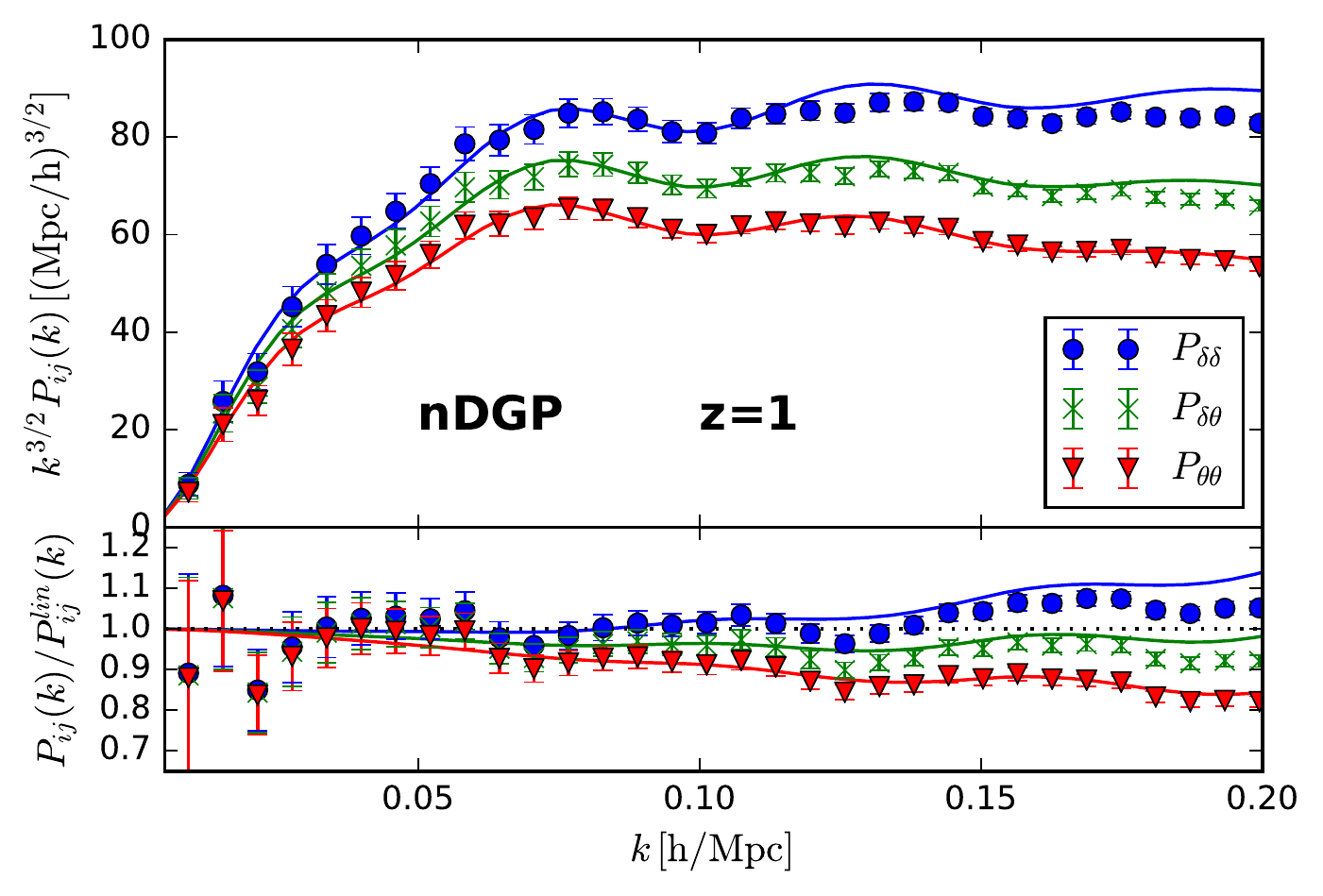}} 
  \caption[CONVERGENCE ]{SPT predictions (solid) and N-body measurements (points) of the auto and cross power spectra of density and velocity fields in real space at $z=1$ for GR (left) and $nDGP$ (right) fitted with Poisson errors assuming a $1 \mbox{Gpc}^3/h^3$ volume. The top panels show the power spectra scaled by $k^{3/2}$ and the bottom panels show the deviations from the linear predictions. }
\label{nbodypsz1}
\end{figure}

\begin{figure}[H]
  \captionsetup[subfigure]{labelformat=empty}
  \centering
  \subfloat[]{\includegraphics[width=8.3cm, height=8.3cm]{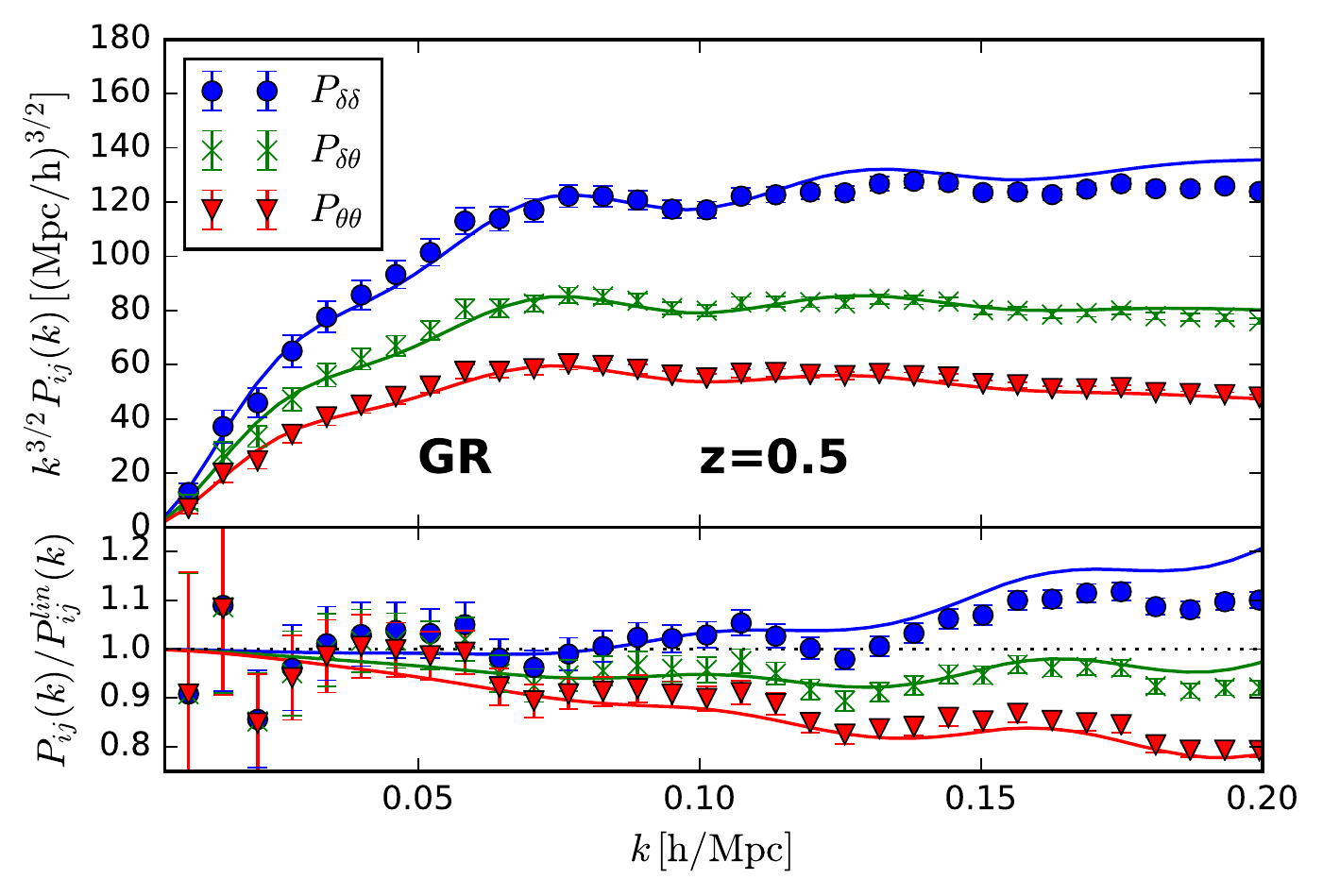}} \quad
  \subfloat[]{\includegraphics[width=8.3cm, height=8.3cm]{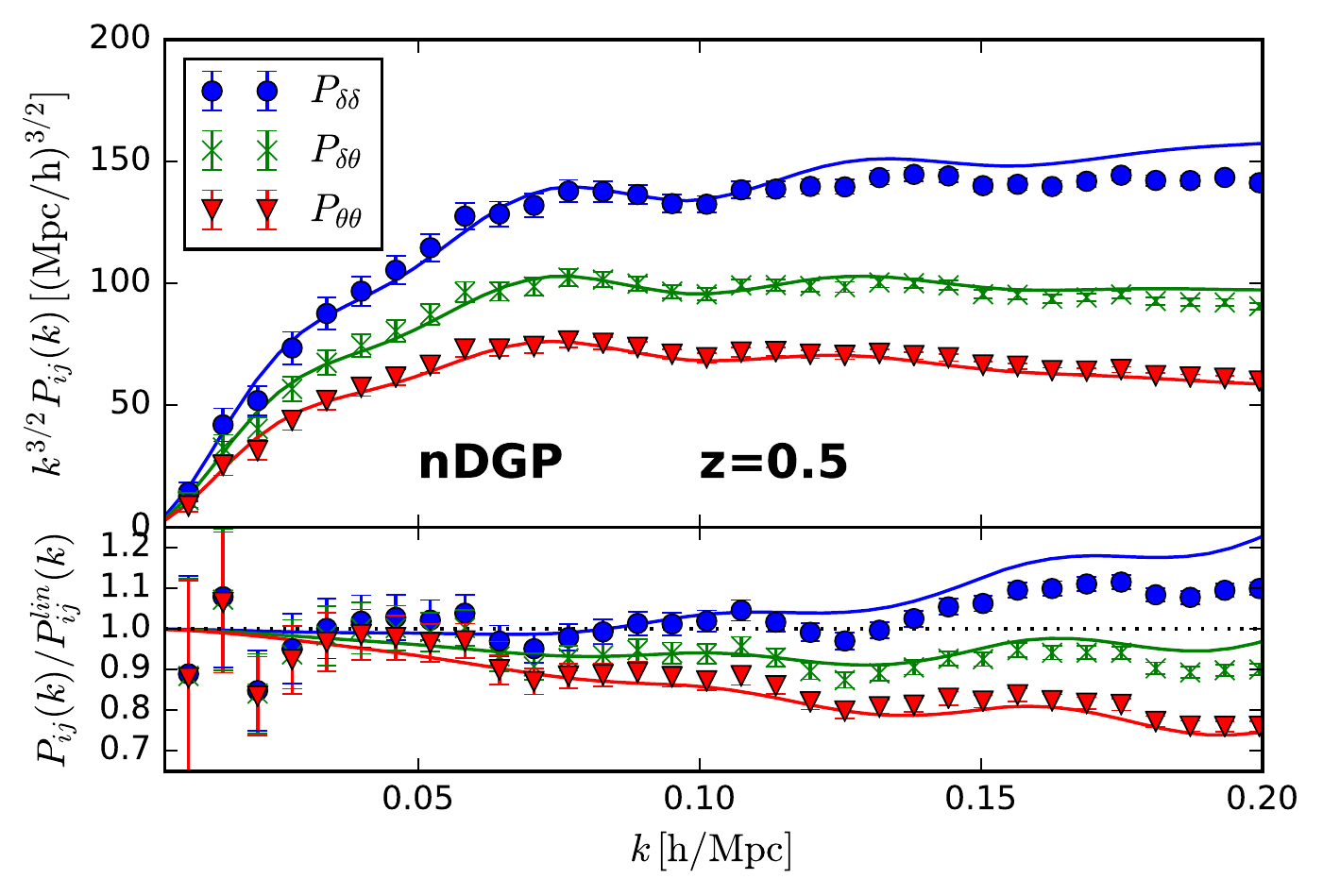}} 
  \caption[CONVERGENCE ]{SPT predictions (solid) and N-body measurements (points) of the auto and cross power spectra of density and velocity fields in real space at $z=0.5$ for GR (left) and $nDGP$ (right) fitted with Poisson errors assuming a $1 \mbox{Gpc}^3/h^3$ volume.  The top panels show the power spectra scaled by $k^{3/2}$ and the bottom panels show the deviations from the linear predictions.}
\label{nbodypsz05}
\end{figure}

\begin{figure}[H]
  \captionsetup[subfigure]{labelformat=empty}
  \centering
  \subfloat[]{\includegraphics[width=8.3cm, height=8.3cm]{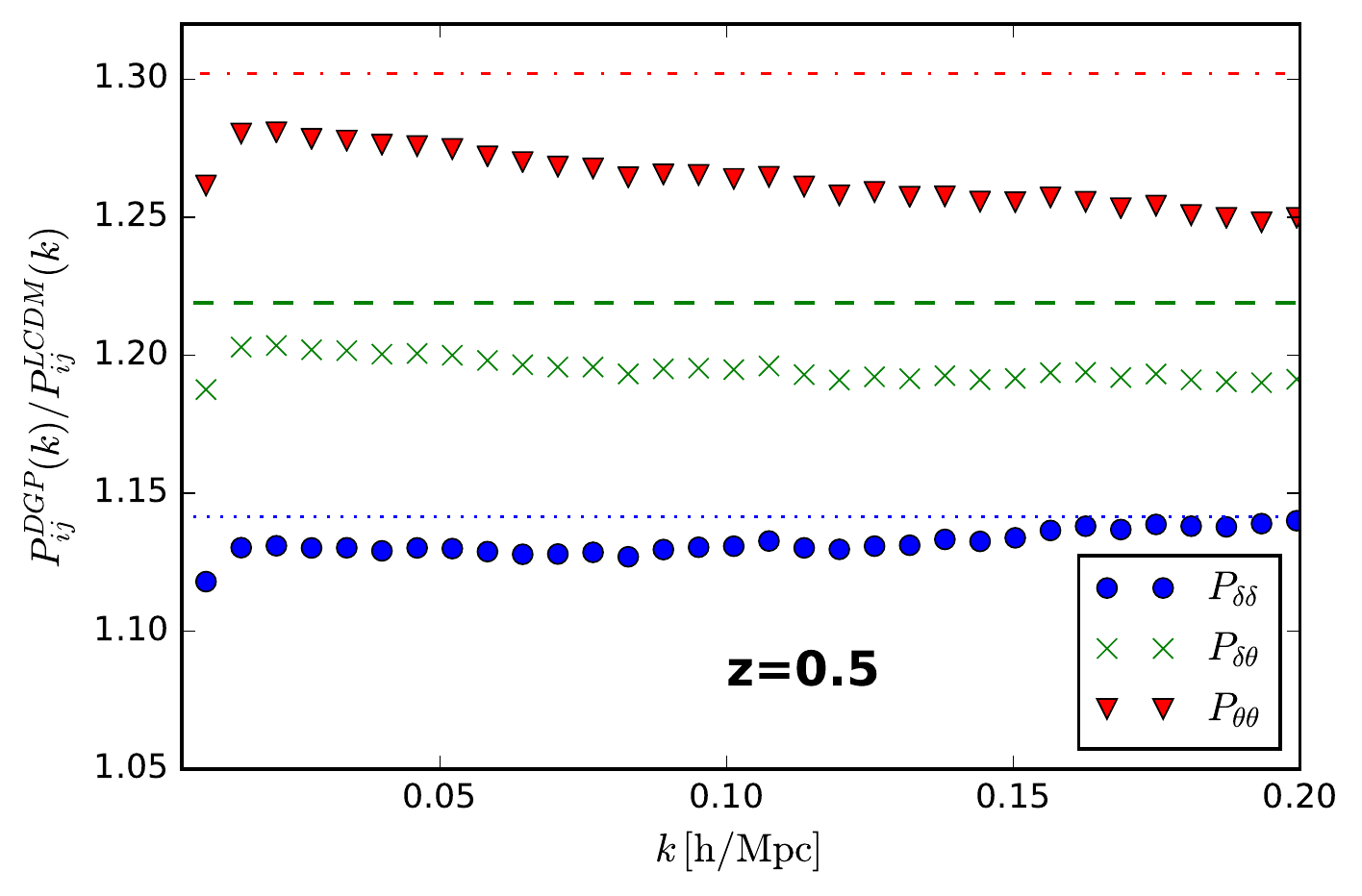}} \quad
  \subfloat[]{\includegraphics[width=8.3cm, height=8.3cm]{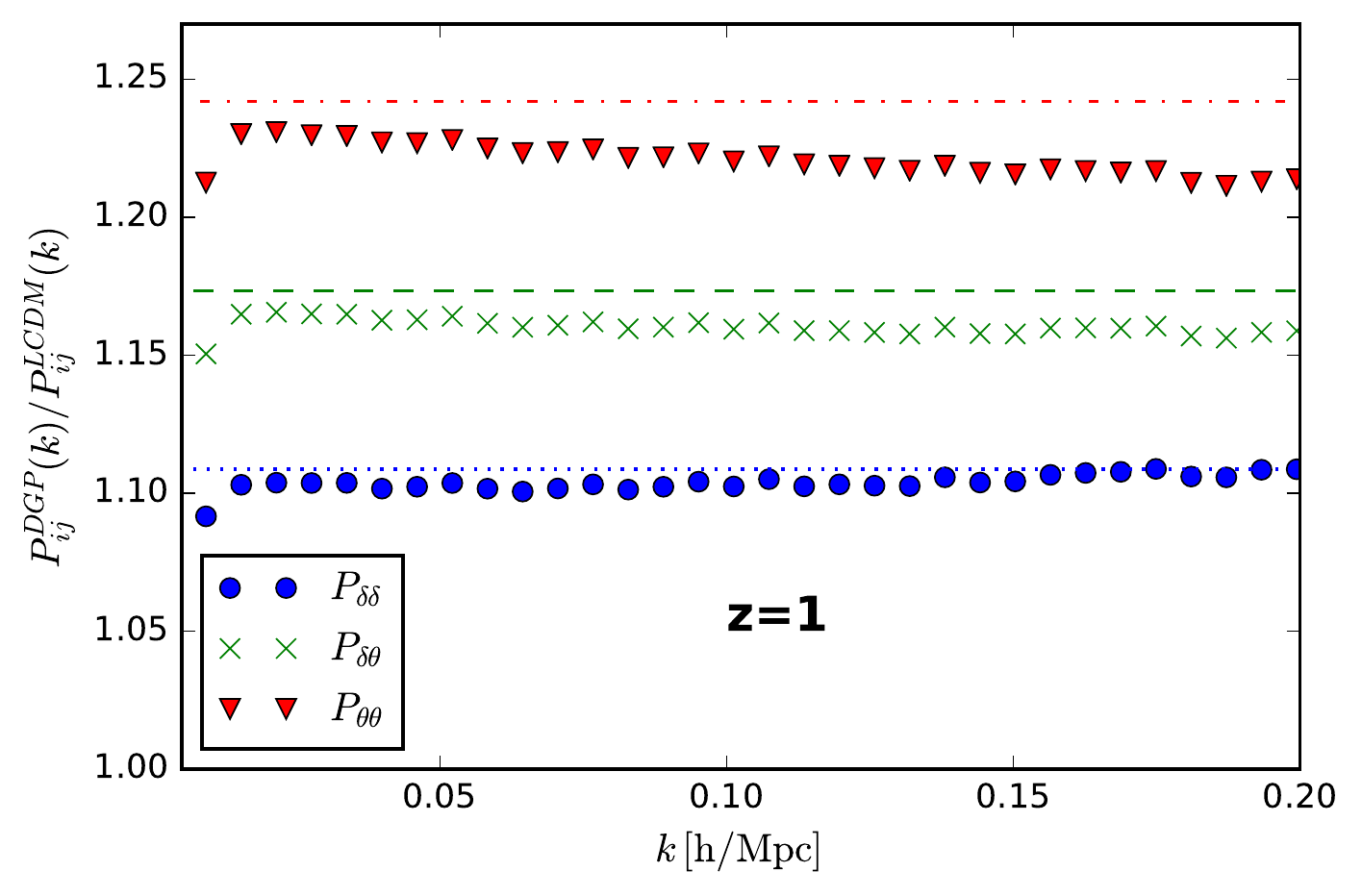}} 
  \caption[CONVERGENCE ]{The ratio of the DGP to LCDM N-body real space spectra (points) at $z=0.5$ (left) and $z=1$ (right). The linear ratios are shown as dotted, dashed and dot-dashed lines.}
\label{nbodyrat1}
\end{figure}

\begin{figure}[H]
  \captionsetup[subfigure]{labelformat=empty}
  \centering
  \subfloat[]{\includegraphics[width=8.3cm, height=8.3cm]{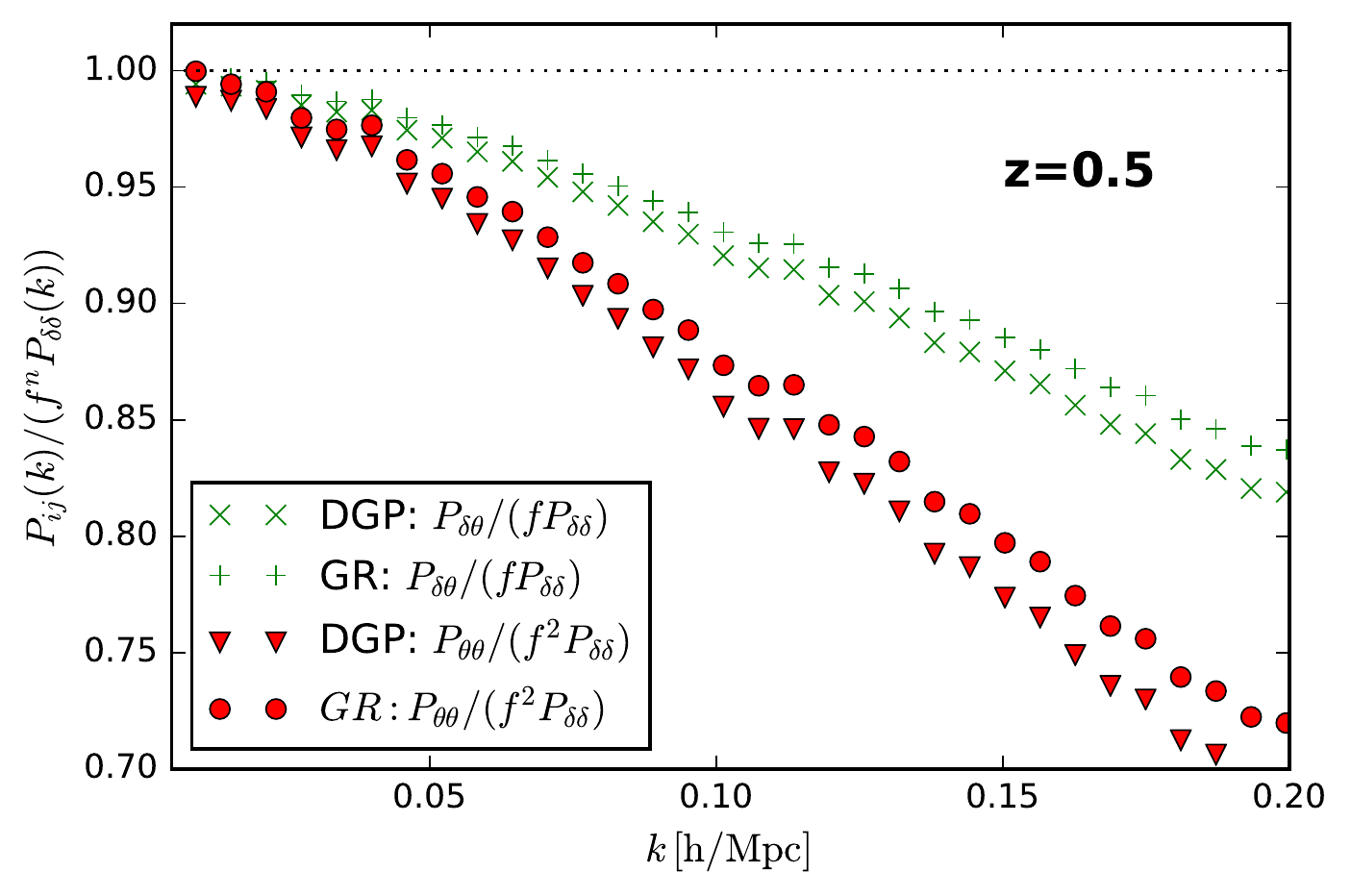}} \quad
  \subfloat[]{\includegraphics[width=8.3cm, height=8.3cm]{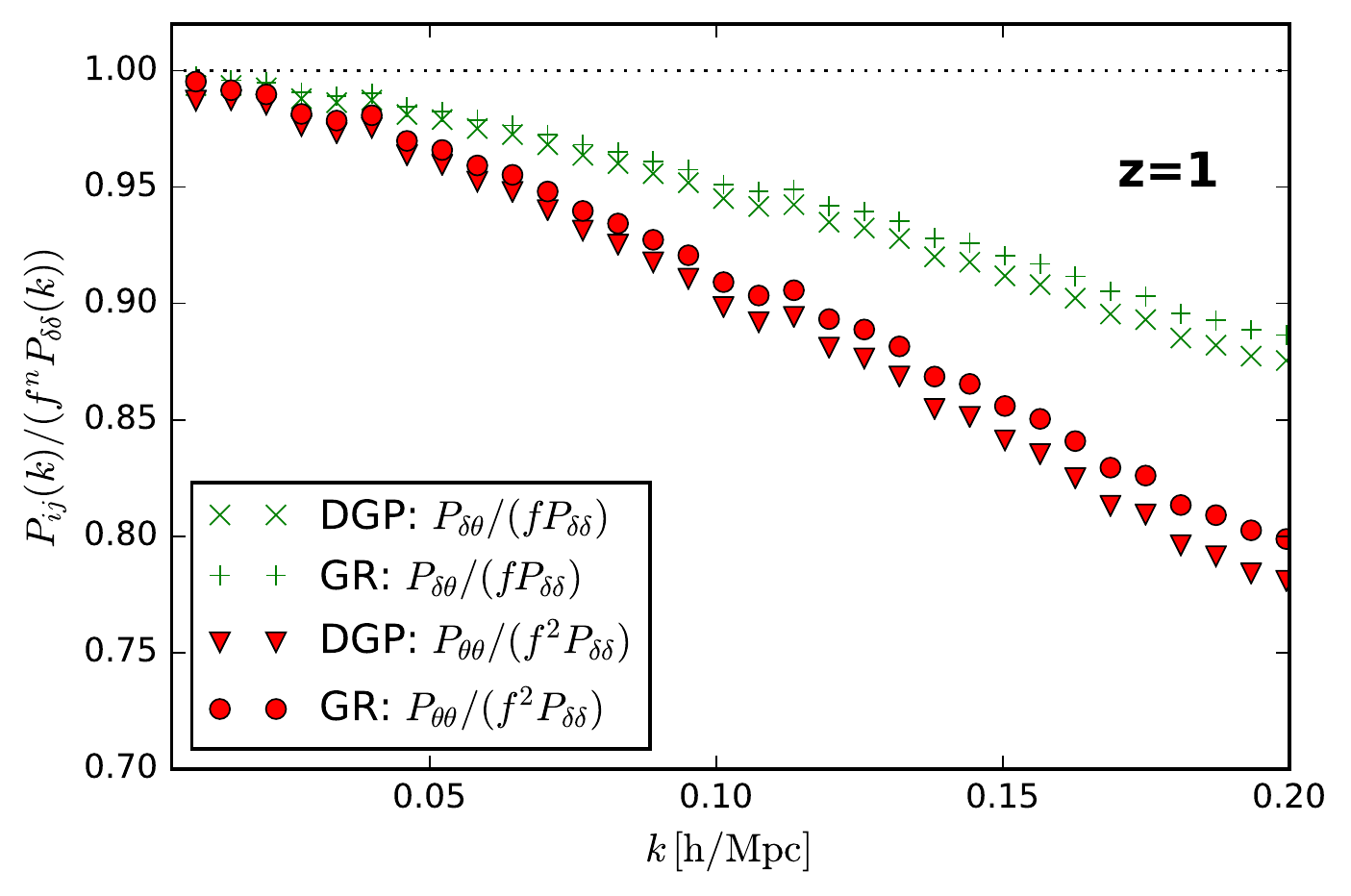}} 
  \caption[CONVERGENCE ]{The ratio of the N-body real space auto velocity (red) and cross spectra (green) to their linear predictions in LCDM (circles and pluses) and in nDGP (triangles and crosses) at $z=0.5$ (left) and $z=1$ (right).}
\label{nbodyrat2}
\end{figure}

We use cosmological N-body simulations of the nDGP model and LCDM and compare them to SPT predictions to gain an idea of the range of validity of the theoretical modelling. These simulations were run using the AMR code ECOSMOG~\citep{Li:2011vk}. The background cosmology is taken from WMAP9~\citep{Hinshaw:2012aka}: $\Omega_m = 0.281$, $h=0.697$, and $n_s=0.971$. The box length is $1024 \mbox{Mpc}/h$ with $1024^3$ dark matter particles used and a starting redshift of $49$. This design sets the resulting mass resolution at $m_p\cong7.8\times10^{10}M_{\odot}h^{-1}$ and the Nyquist fluid approximation limit of $k_{Nyq}\cong\pi h$/Mpc. The most refined AMR grid were at level 16, setting a maximal force resolution at $\epsilon=1024/2^{16}=0.015 \mbox{Mpc}/h$. The initial conditions were generated using {\tt MPGrafic}\footnote{Available at \url{http://www2.iap.fr/users/pichon/mpgrafic.html}} and both nDGP and LCDM simulations begin with the same initial seeds. The linear theory power spectrum normalisations was set to be $\sigma_8=0.844$. The nDGP simulation uses $\Omega_{rc}=1/4r_c^2H_0^2=0.438$. We choose to use a high $\sigma_8$ with these specific combination of nDGP parameters to obtain a model that would be characterised by noticeable deviations from the GR fiducial linear growth rate and at the same time would have strong enough small scale non-linearities. In this way we can expect that our nDGP bed-test should contain a signal strong enough to be detected by RSD analysis. 
\newline
\newline
We use the {\it Delaunay Tesselation Field Estimator} (DTFE) method implemented in publicly avilable DTFE code by \cite{cv2011}. The DTFE code employs {\it the Delanay Tesselation Field Estimation}, a method described in details 
in \cite{sv2000,vs2009}, which assures that the resulting smooth fields have the highest attainable resolution, 
are volume weighted and have suppressed sampling noise. The fields are then smoothed using top-hat filtering
and we proceed to obtain density $P_{\delta\delta}(k)=\langle\delta(\bfk)\delta^{*}(\bfk)\rangle$
and velocity divergence $P_{\theta\theta}(k)=\langle\theta(\bfk)\theta^*(\bfk)\rangle$ power spectra
following the method of \cite{Li:2012by,Hellwing2013}. It is well known that energetic processes connected to
the highly non-linear physics of galaxy formation affect the cosmic density field up to scales even of
tens of Megaparsecs. Thus our simplistic N-body only approach could introduce some additional scatter and
bias in the analysis. However, as recently shown by \cite{Hellwing2016} with a use of the state-of-the-art galaxy formation
simulation, the EAGLE suite \cite{Schaye:2014tpa}, the energetic baryonic processes have a negligible impact on both velocity and density fields
on the scales we consider in this study. Hence, we can be assured that our analysis will not be affected by the fact that we
ignore baryons completely in our modelling.
\newline
\newline
Fig.\ref{nbodypsz1} shows SPT does very well in modelling non-linearities at $z=1$. For the nDGP simulation we find agreement at $1(3)\%$ level up to $k=0.175 (0.18), 0.18(0.2)$ and $0.22(0.26)h$/Mpc for $P_{\delta \delta}$, $P_{\delta \theta}$ and $P_{\theta \theta}$ respectively. Given this we consider scales up to $k_{\rm max} = 0.195h$/Mpc at $z=1$. We have fit Poisson errors to this data assuming a volume of $1\mbox{Gpc}^3/h^3$ with a shot noise term of $\bar{n} = 3\times10^{-4}h^3/\mbox{Mpc}^3$ (See Eq.27 of \cite{Zhao:2013dza} for example). 
\newline
\newline
On the other hand, Fig.\ref{nbodypsz05} shows the decline in the accuracy of the SPT approach at later times.  Again for the nDGP simulation, now at $z=0.5$, we find agreement at $1(3)\%$ level up to $k=0.12 (0.18), 0.18(0.2)h$/Mpc for $P_{\delta \delta}$ and $P_{\delta \theta}$ respectively. $P_{\theta \theta}$ is found to be very noisy around the $1$-$3\%$ band within $0.1 \leq k \leq 0.2$. Given this, we consider scales up to a $k_{\rm max} = 0.147h$/Mpc at $z=0.5$. The stated ranges of applicability of our SPT modelling in nDGP are found to be very similar to the GR simulation comparisons. 
\newline
\newline
In Fig.\ref{nbodyrat1} we have plotted the ratio of the real space spectra in nDGP to LCDM simulations for $z=0.5$ and $z=1$ along with the linear predictions as dotted and dashed lines. This figure captures the effects of modified gravity on the real space spectra and growth. We note that the density spectra remain very much the same as we proceed into the quasi non-linear regime. On the other hand we find that the DGP velocity spectra becomes more suppressed as we go to smaller scales. This scale dependence is expected as larger fifth force enhancements to the velocity field can effectively reduce the correlation between particle velocities, an effect that is larger at lower redshift where there is more non-linear structure growth. In Fig.\ref{nbodyrat2} we further elucidate this point where we have plotted the ratio of velocity and cross spectra to their linear predictions for both LCDM and nDGP simulations. A first point is that clearly non-linearity becomes very important in the scales considered. Secondly, the nDGP simulation shows an enhanced non-linearity (and hence suppression of velocity correlations) over the LCDM simulation. 
\newline
\newline
These results are consistent with previous results found for other MG simulations such as in Fig.4 and Fig.7 of \cite{Winther:2015wla} and Fig.7 and Fig.8 of \cite{Winther:2017jof}). They found a pattern of a constant boost at the linear scales (reflecting enhanced growth rate) with the scale-dependance beginning to be suppressed closer to non-linear scales. At non-linear scales the 5th force in nDGP theories  starts to be screened effectively  by the Vainshtein mechanism, recovering the Newtonian value inside most of the virialised structures (see also \cite{Falck:2015rsa}). Our analysis is limited to only quasi-linear scales where the complicated scale-dependent patterns of the Vainshtein mechanism are still not well developed. At those scales the MG-physics can still be faithfully captured by the two-parameter TNS modelling.
%%%%%%%%%%%%
\subsubsection{Redshift Space Comparisons : MG-PICOLA}
\begin{figure}[H]
  \captionsetup[subfigure]{labelformat=empty}
  \centering
  \subfloat[]{\includegraphics[width=8.3cm, height=8.1cm]{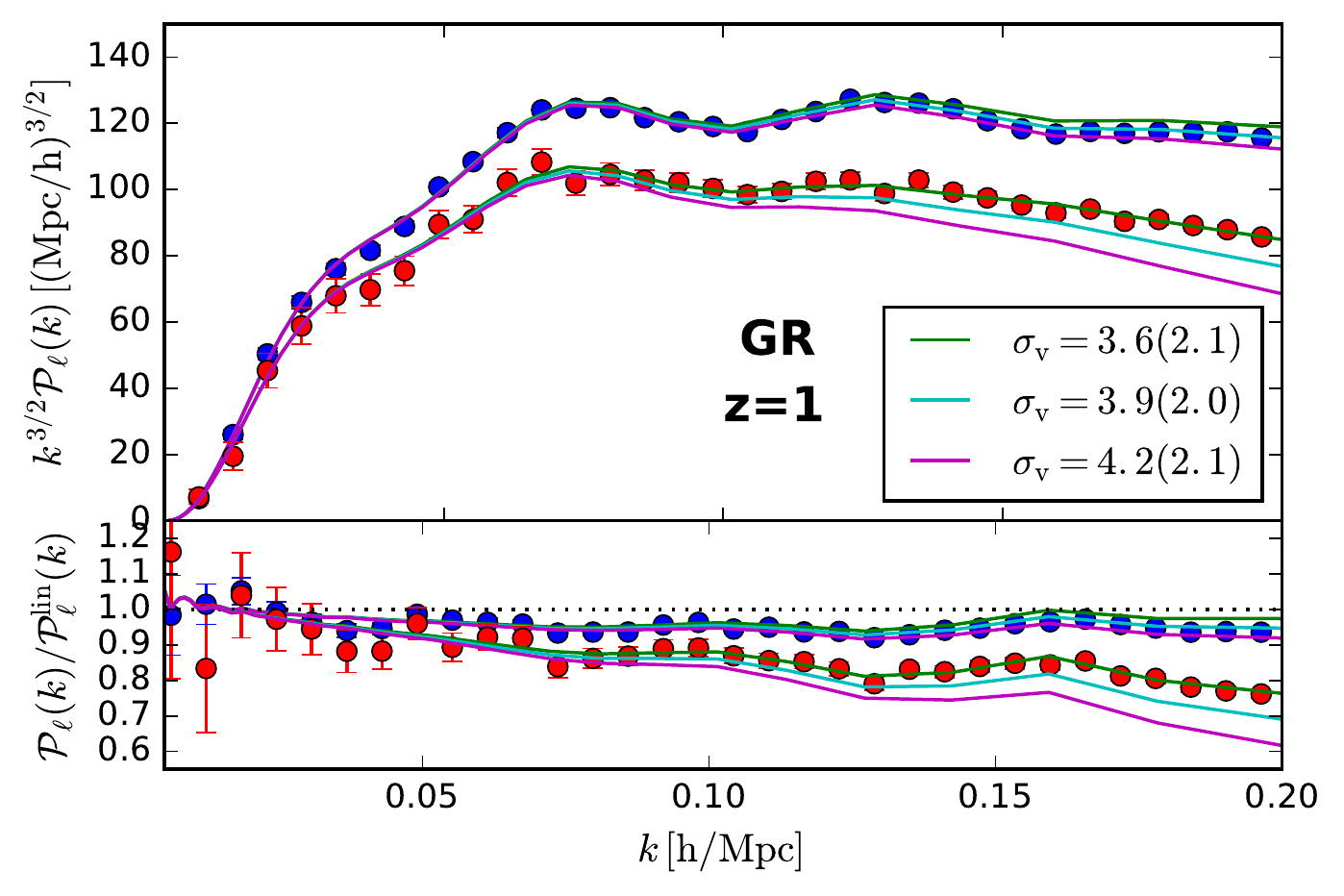}} \quad
  \subfloat[]{\includegraphics[width=8.3cm, height=8.1cm]{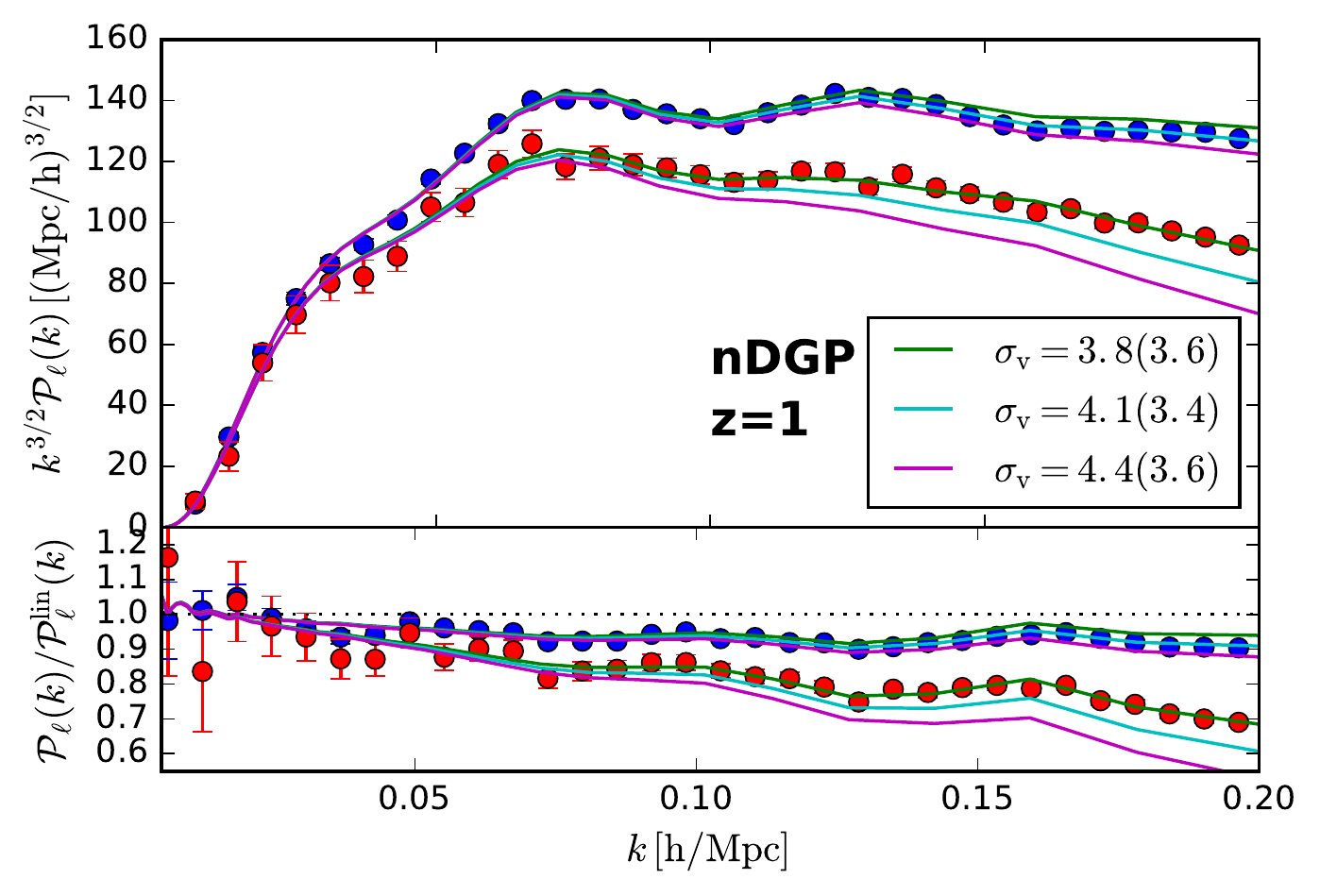}} 
  \caption[CONVERGENCE ]{TNS predictions (solid) and MG-PICOLA measurements (points) of the monopole (the upper group of points/lines) and quadrupole (the lower group of points/lines)  at $z=1$ for GR (left) and $nDGP$ (right). Three values of the TNS model parameter $\sigma_v$ are shown (in units of Mpc/$h$) with their respective reduced $\chi^2$ in brackets. The top panels show the multipoles multiplied by $k^{3/2}$ and the bottom panels show the deviations from Kaiser's linear prediction. The error bars are those from an ideal survey of $V_s=10 \mbox{Gpc}^3/h^3$. }
\label{nbodyplz1}
\end{figure}

\begin{figure}[H]
  \captionsetup[subfigure]{labelformat=empty}
  \centering
  \subfloat[]{\includegraphics[width=8.3cm, height=8.1cm]{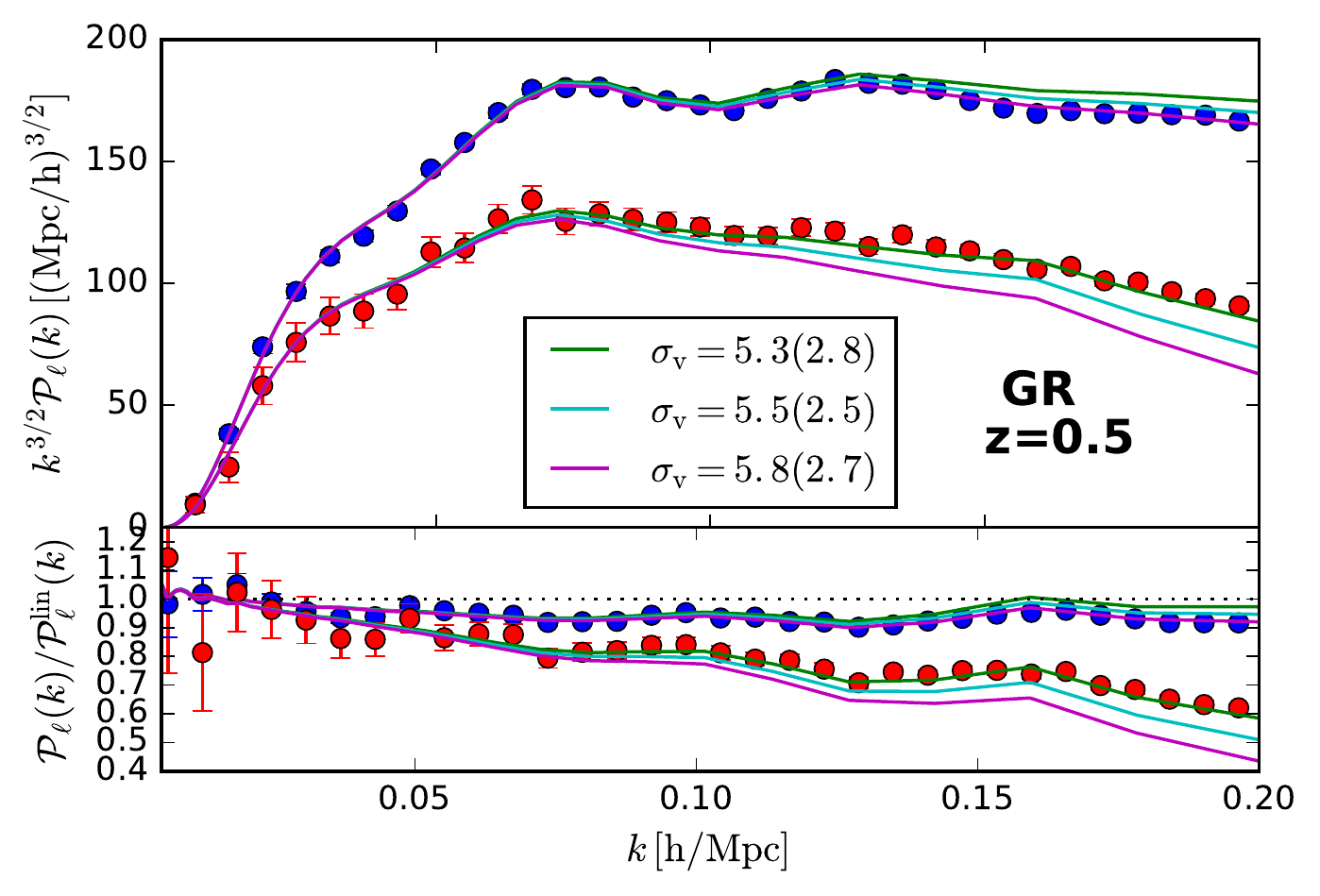}} \quad
  \subfloat[]{\includegraphics[width=8.3cm, height=8.1cm]{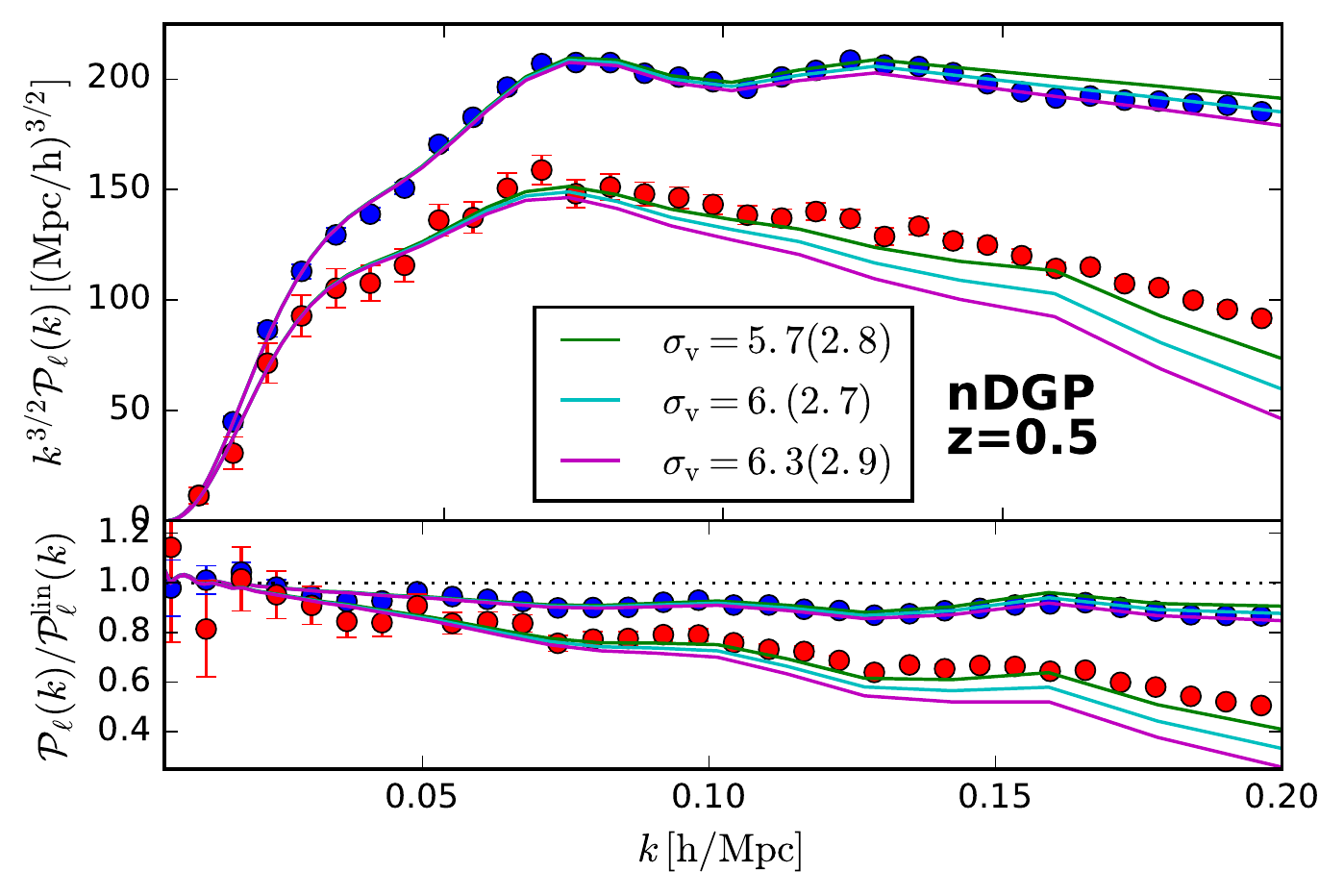}} 
  \caption[CONVERGENCE ]{TNS predictions (solid) and MG-PICOLA measurements (points) of the monopole  (the upper group of points/lines) and quadrupole (the lower group of points/lines) at $z=0.5$ for GR (left) and $nDGP$ (right). Three values of the TNS model parameter $\sigma_v$ (in units of Mpc/$h$)  are shown with their respective reduced $\chi^2$ in brackets. The top panels show the multipoles multiplied by $k^{3/2}$ and the bottom panels show the deviations from Kaiser's linear prediction. The error bars are those from an ideal survey of $V_s=10 \mbox{Gpc}^3/h^3$.}
\label{nbodyplz05}
\end{figure}

Next we take a look at the predictive power of the TNS multipoles, providing the realm of validity of the RSD modelling. The MG-PICOLA multipoles are measured using the distant-observer approximation \footnote{i.e we assume that the observer is located at a distance much greater then the boxsize ($r\gg 1024 \mbox{Mpc}/h$), and so we treat all the lines-of-sight as parallel to the chosen Cartesian axes of the simulation box. Next, we use an appropriate velocity component ($v_x, v_y$ or $v_z$) to disturb the position of a matter particle.} and averaged over three line of sight directions. We further average over 20 MG-PICOLA simulations each of $1\mbox{Gpc}^3/h^3$ volume thus ignoring the mode covariance at and above box-size scales. This should correspond to an ideal survey with a resulting volume of $20\mbox{Gpc}^3/h^3$. 
\newline
\newline
Fig.\ref{nbodyplz1}  shows the monopole and quadrupole predictions at $z=1$ for three different values of $\sigma_v$ where we have fit using 32 $k$-bins up to $k_{\rm max}=0.195h$/Mpc. The reduced $\chi^2_{\rm red} = -2\ln{\mathcal{L}}/{\rm N_{dof}}$ is shown in brackets, where $N_{dof}$ is the total degrees of freedom at that $k_{\rm max}$ which equals twice the number of bins minus the number of parameters. In this case $N_{dof}=62$($=32\times2-2$), since we have 2 parameters. Similarly, Fig.\ref{nbodyplz05} shows the same results at $z=0.5$ where we have fit using 24 $k$-bins ($N_{dof}=46$) up to $k_{\rm max}=0.147h$/Mpc as dictated by the real space power spectra comparisons. 
\newline
\newline
\noindent We find that the fit of PT is comparable between GR and nDGP simulations at both redshifts, although GR does slightly better at small scales at $z=0.5$. The quadrupole also does worse than the monopole over both models at $z=0.5$ which we can attribute to the increased dependency on the velocity auto power spectrum for which the theoretical template does worse in reproducing (See Fig.\ref{nbodypsz05}). 
\newline
\newline
As was done with the power spectra, we identify the scale at which deviations from theory are within the $1(3)\%$ region. We do this using the values of $\sigma_v$ with the lowest $\chi^2$ shown in Fig.\ref{nbodyplz1} and Fig.\ref{nbodyplz05}. This will be used to set a maximum Fourier mode for our statistical analysis. For the nDGP simulation at $z=1$ we find agreement at $1(3)\%$ level up to $k=0.24(0.25)h$/Mpc for $P_{0}$. $P_{2}$ is significantly noisier around the $1\%$ deviation region but matches PT up to $k= 0.16h$/Mpc within $3\%$. Similarly, for $z=0.5$ we find a theory-data agreement of $1(3)\%$ up to $k=0.147(0.2)h$/Mpc for $P_{0}$ while a much worse $k=0.09(0.1)h$/Mpc for $P_{2}$. Since $P_{2}$'s contribution to Eq.\ref{covarianceeqn} is significantly smaller than $P_{0}$ because of their respective errors, we decide to use $k_{\rm max}=0.147h$/Mpc at $z=0.5$ and $k_{\rm max}=0.195h$/Mpc at $z=1$ for the MCMC analysis, despite the poor quadrupole fit. In the case of GR the range of validity is found to be similar.
\newline
\newline
By having assessed and compared the range of validities of SPT for both a LCDM and nDGP cosmologies, we have gained a handle on how enhanced dynamics due to fifth force interactions degrade SPT's performance. Our comparisons indicate that the non-linearity generated by $\Omega_{rc}=0.438$ is mild enough not to significantly effect the range of validity. We caution however, this may not be the case for other models of gravity characterised by a larger degree of deviation from GR dynamics. Next we will test the capabilities of the theoretical templates when matching to the MG-PICOLA multipoles. 
\newpage
\subsection{Template Performance}
\begin{figure}[H]
  \captionsetup[subfigure]{labelformat=empty}
  \centering
  \subfloat[]{\includegraphics[width=8.3cm, height=8.3cm]{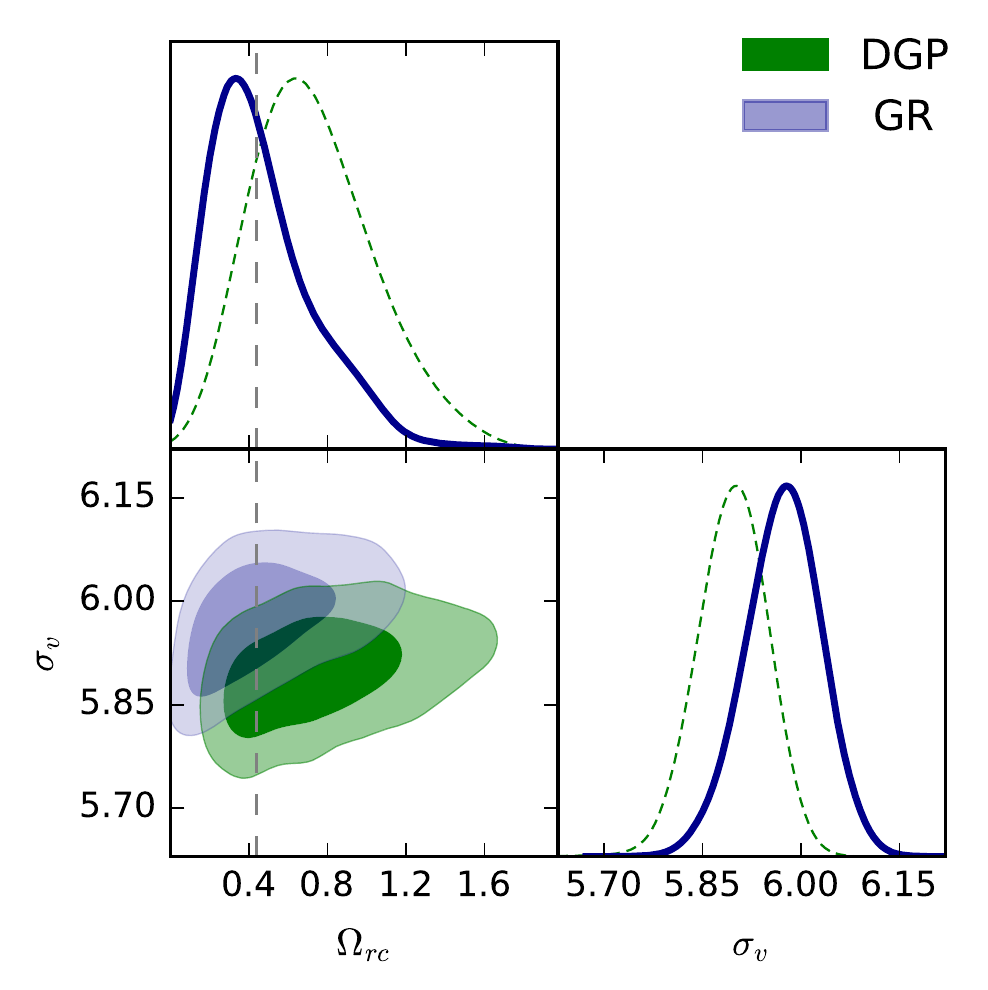}} \quad
  \subfloat[]{\includegraphics[width=8.3cm, height=8.3cm]{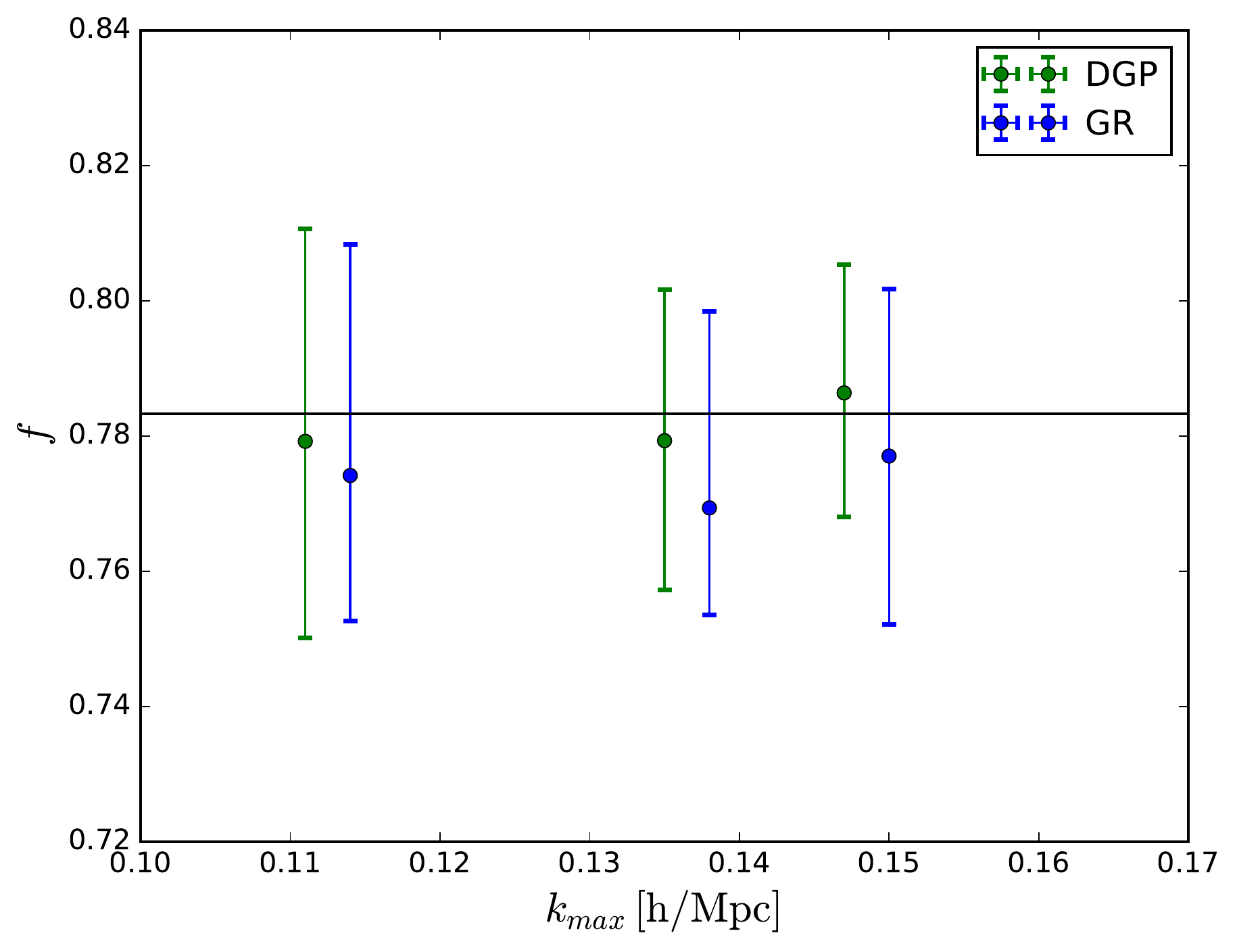}} 
  \caption[CONVERGENCE ]{Left: The $1\sigma$ and $2\sigma$ confidence contours for the DGP template and the GR template at $z=0.5$ fitting up to $k_{\rm max}=0.147h$/Mpc using 24 bins with the simulation's fiducial value for $\Omega_{rc}$ indicated by the dashed line. Right: The best fit value for $f$ as a function of $k_{\rm max}$ with the $2\sigma$ errors for the DGP and GR template. The GR template values (blue) have been slightly shifted for better visualisation. A survey of volume of $10\mbox{Gpc}^{3}/h^3$ is assumed. }
\label{nbodyz05}
\end{figure}

\begin{figure}[H]
  \captionsetup[subfigure]{labelformat=empty}
  \centering
  \subfloat[]{\includegraphics[width=8.3cm, height=8.3cm]{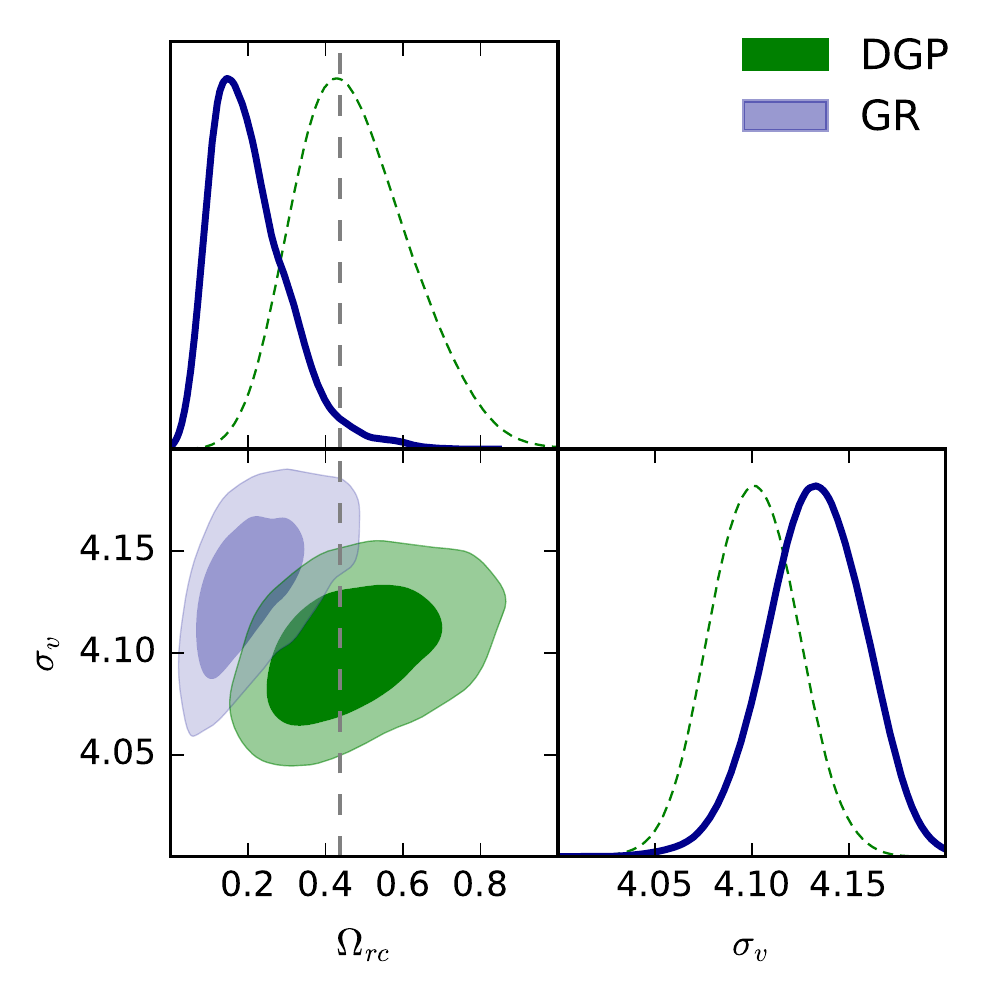}} \quad
  \subfloat[]{\includegraphics[width=8.3cm, height=8.3cm]{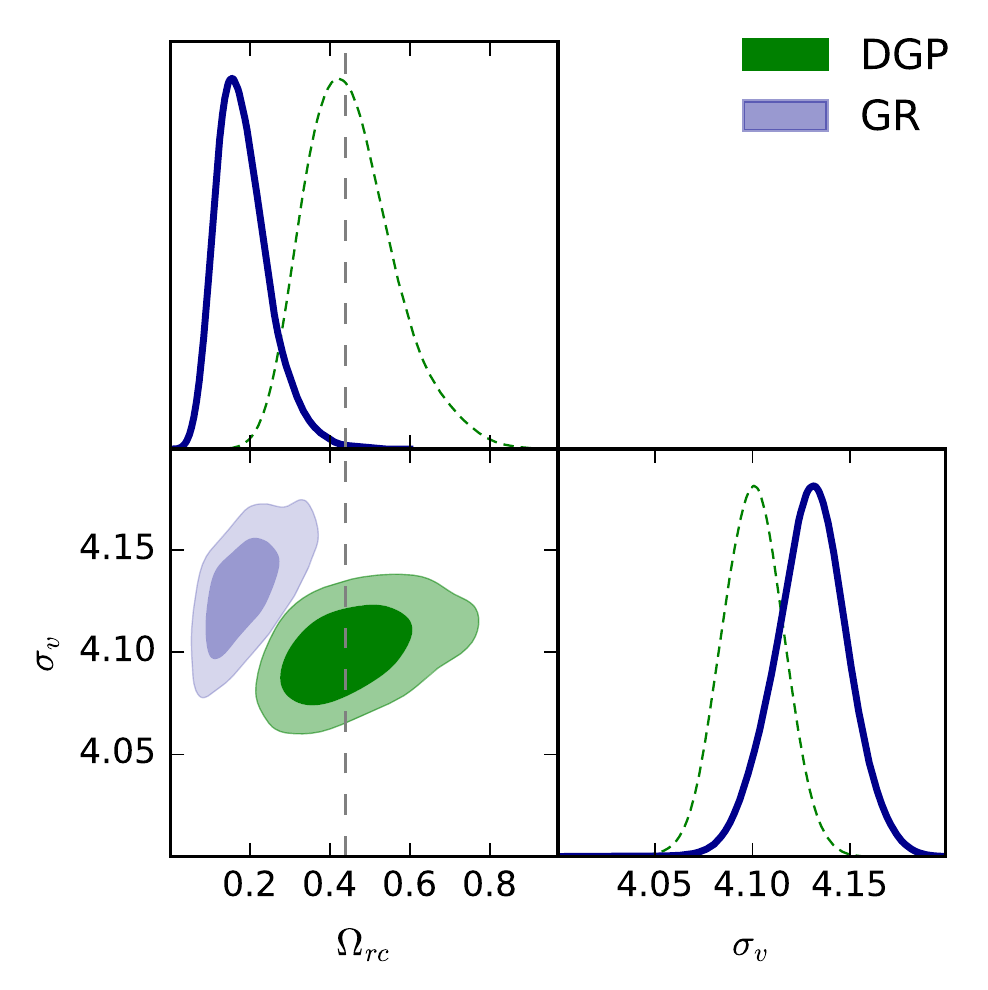}} 
  \caption[CONVERGENCE ]{Left: The $1\sigma$ and $2\sigma$ confidence contours for the DGP template and the GR template at $z=1$ fitting up to $k_{\rm max}=0.195h$/Mpc using 32 bins with the simulation's fiducial value for $\Omega_{rc}$ indicated by the dashed line. The survey volume is taken to be $10h^{-3} \mbox{Gpc}^{3}$. Right: Same as left plot but with the survey volume is taken to be $20 \mbox{Gpc}^{3}/h^3$ . }
\label{nbodyz1a}
\end{figure}

\begin{figure}[H]
  \captionsetup[subfigure]{labelformat=empty}
  \centering
 \includegraphics[width=12.3cm, height=8.39cm]{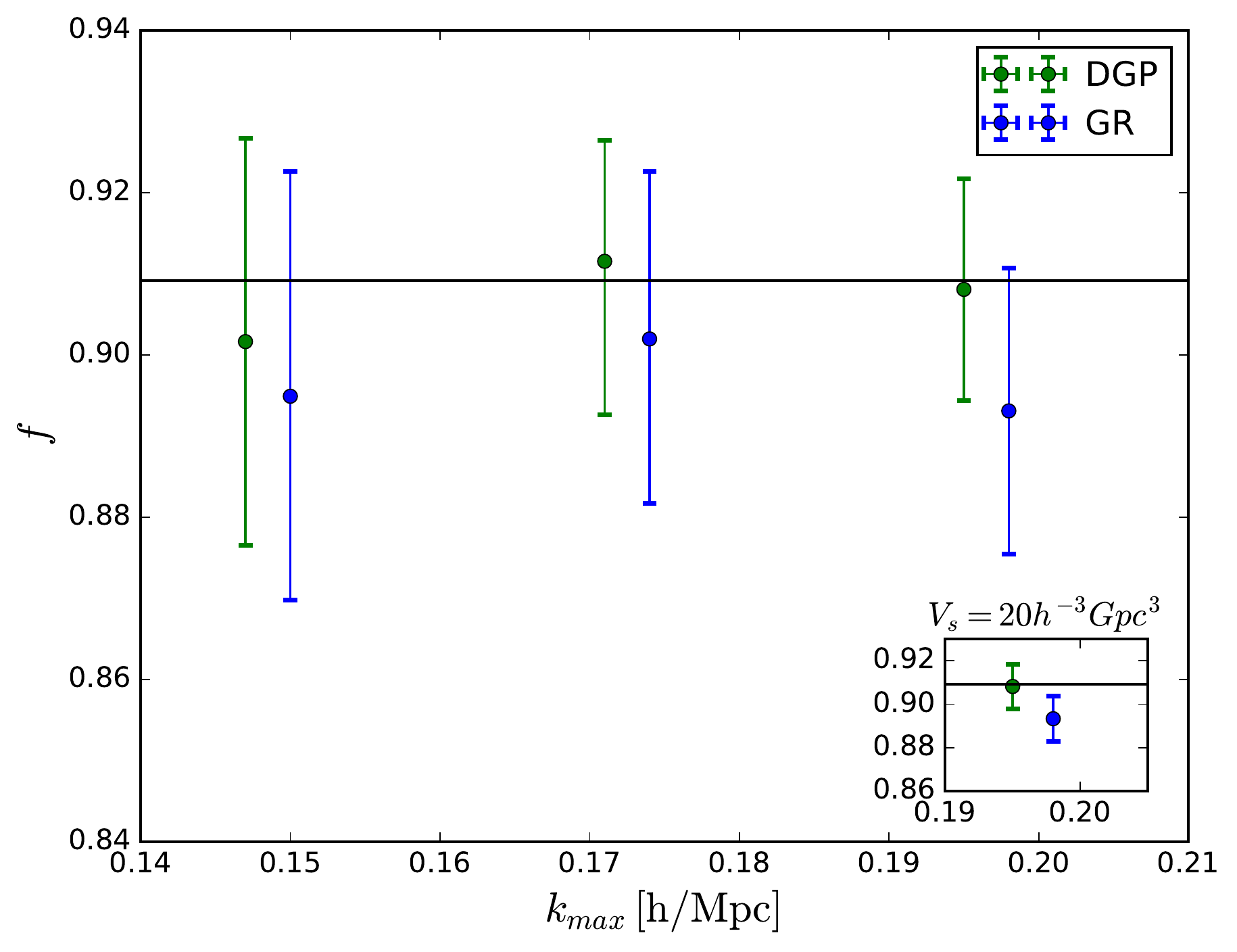} 
  \caption[CONVERGENCE ]{The best fit value for $f$ as a function of $k_{\rm max}$ with the $2\sigma$ errors for the DGP and GR templates at $z=1$. The survey volume is taken to be $10\mbox{Gpc}^{3}/h^3$  with the annotated plot containing the prediction at $k_{\rm max}=0.195h$/Mpc for an increased survey volume of $20 \mbox{Gpc}^{3}/h^3$. The GR template values (blue) have been slightly shifted for better visualisation.}
\label{nbodyz1b}
\end{figure}

\begin{figure}[H]
  \captionsetup[subfigure]{labelformat=empty}
  \centering
  \subfloat[]{\includegraphics[width=8.3cm, height=8.3cm]{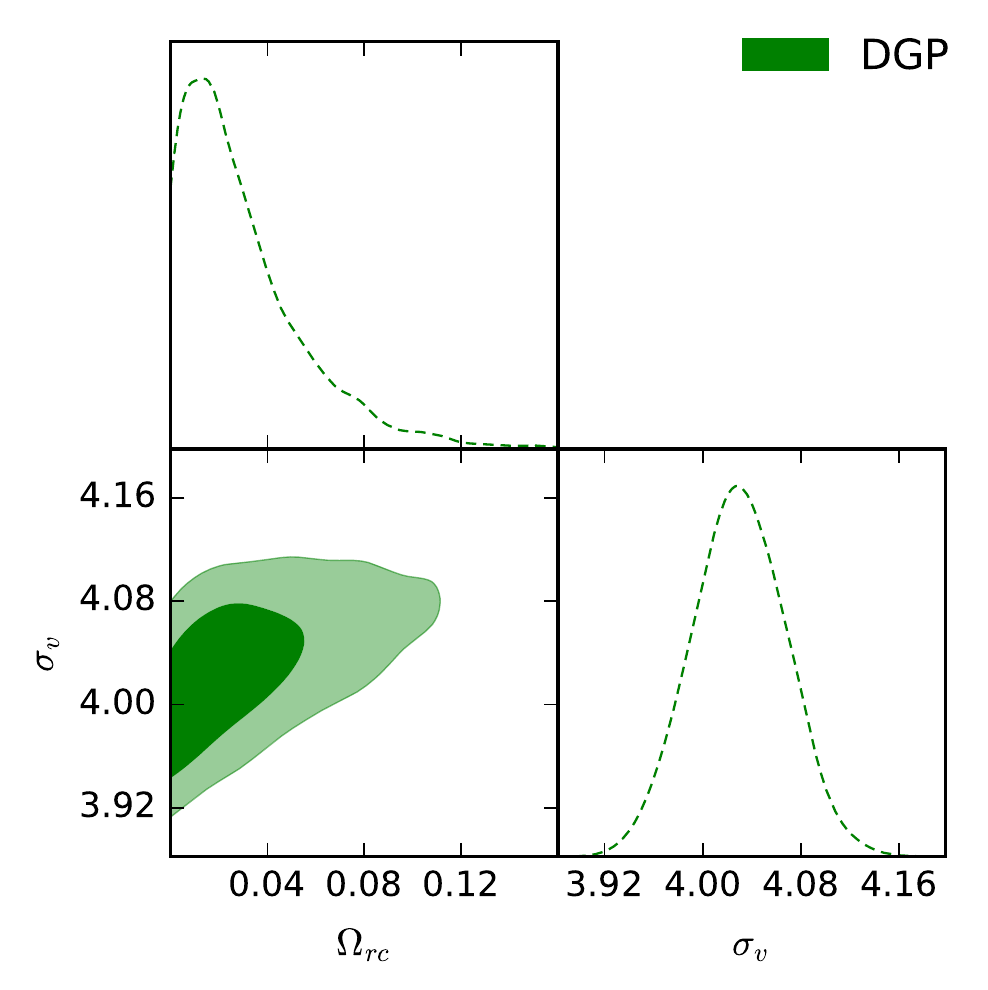}} \quad
  \subfloat[]{\includegraphics[width=8.3cm, height=8.3cm]{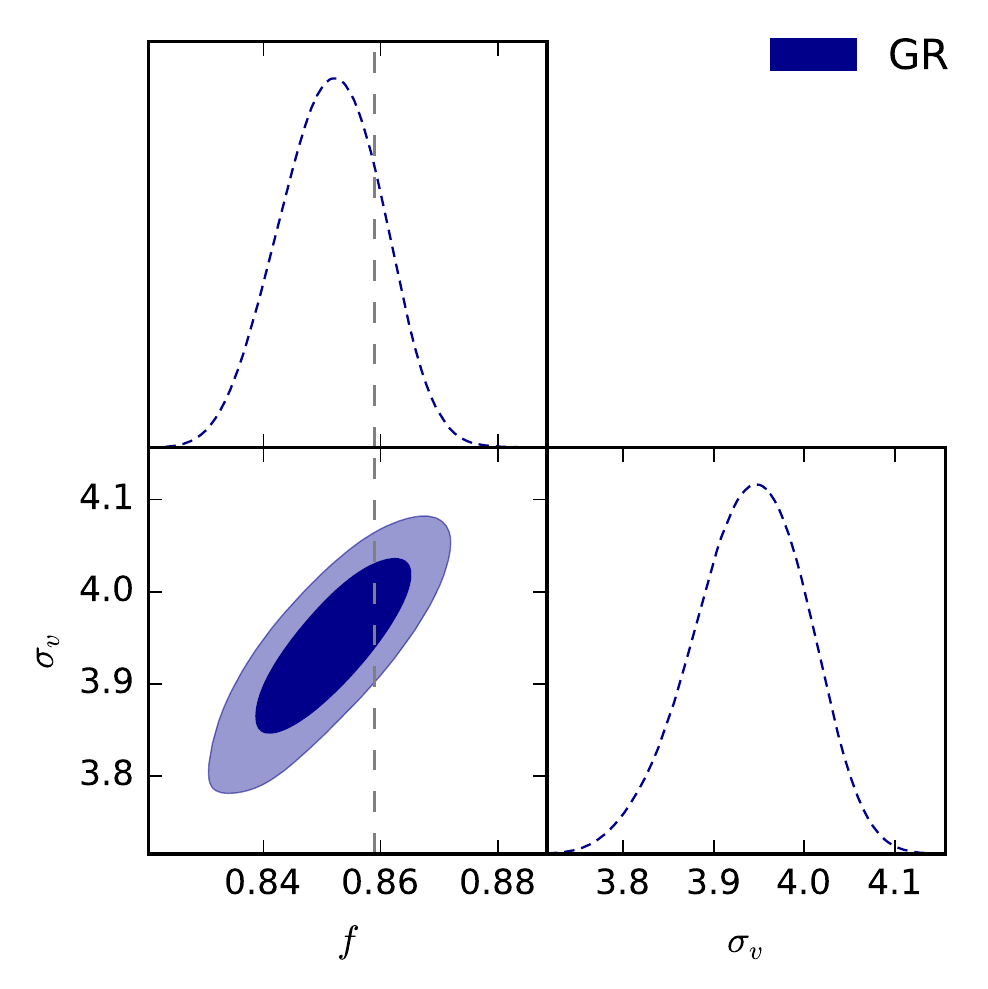}} 
  \caption[CONVERGENCE ]{Left: The $1\sigma$ and $2\sigma$ confidence contours for the DGP template at $z=1$ fitting up to $k_{\rm max}=0.171h$/Mpc using 28 bins with the simulation's fiducial value for $\Omega_{rc}=0$ (the GR simulation). Right: The $1\sigma$ and $2\sigma$ confidence contours for the GR template fitting to the same simulation as used in the left panel, using the same number of bins, but without a prior on $f$. The fiducial $f$ is marked by a dashed line. A survey of volume of $10\mbox{Gpc}^{3}/h^3$ is assumed.}
\label{nbodygrz1}
\end{figure}

\begin{figure}[H]
  \captionsetup[subfigure]{labelformat=empty}
  \centering
  \subfloat[]{\includegraphics[width=8.3cm, height=8.3cm]{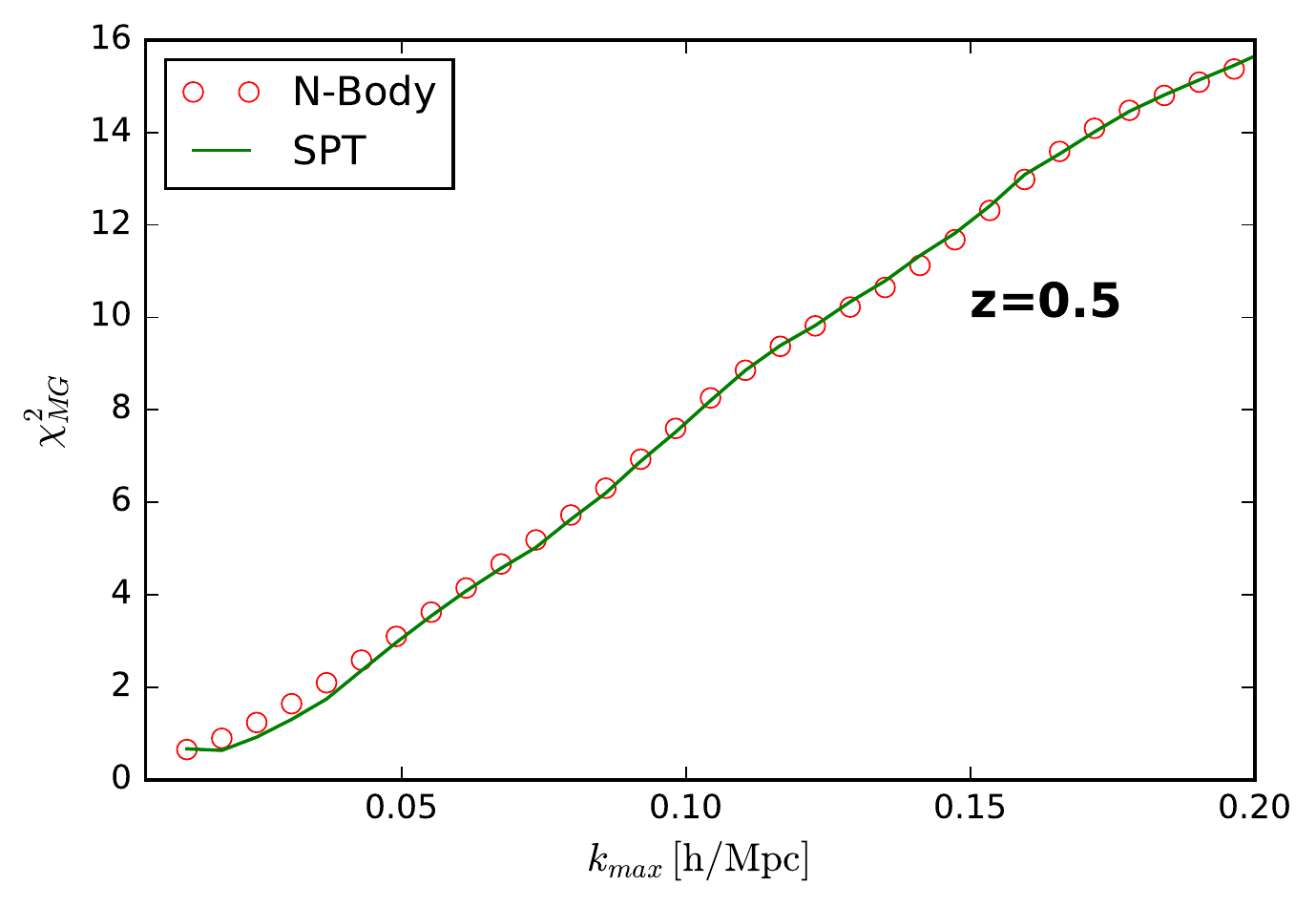}} \quad
  \subfloat[]{\includegraphics[width=8.3cm, height=8.3cm]{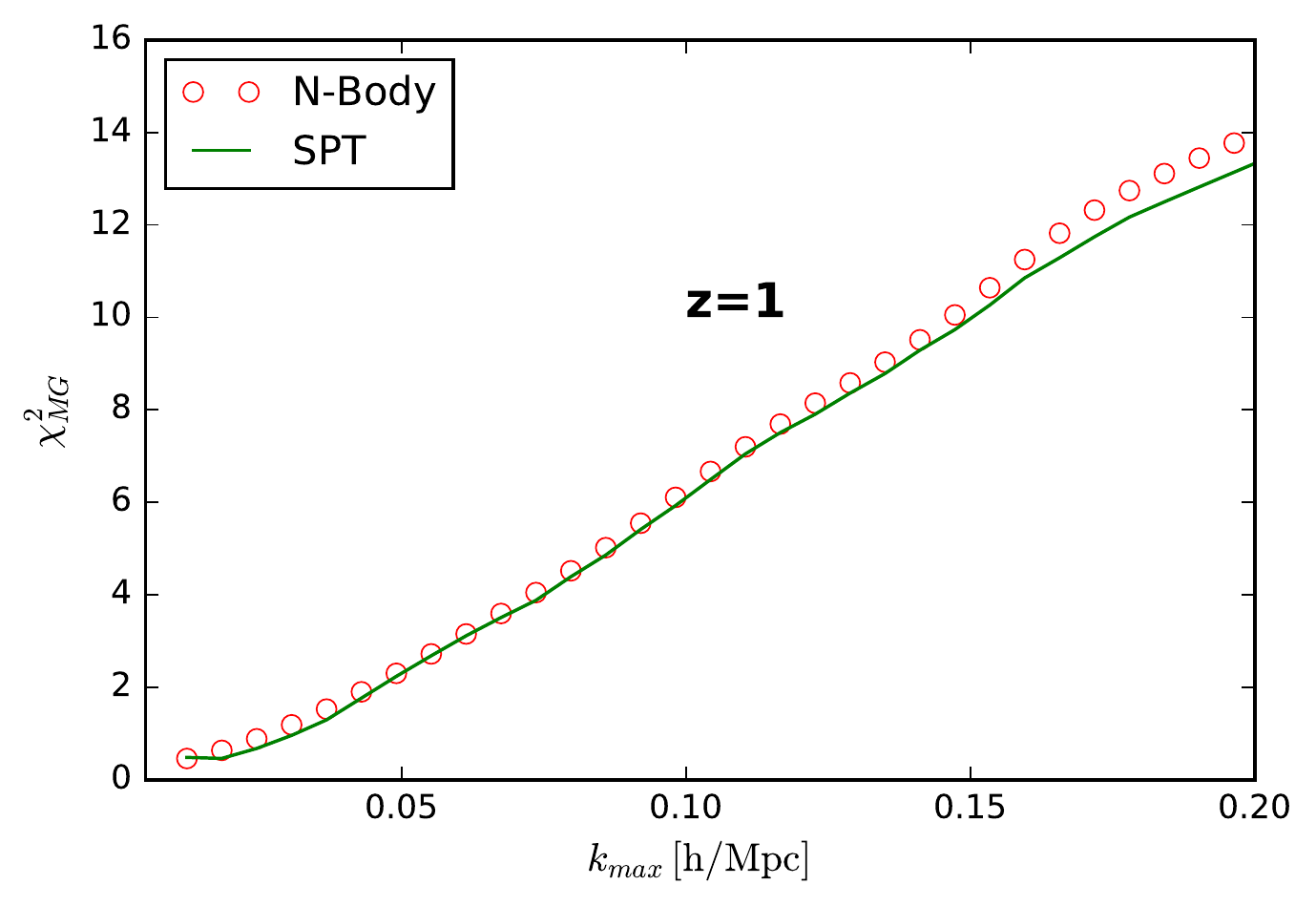}} 
  \caption[CONVERGENCE ]{The quantity computed using Eq.\ref{chisqrt} for SPT (green line) and N-body (red circles) at $z=0.5$ (left) and $z=1$ (right).}
\label{chisqrtest}
\end{figure}

We begin by explicitly defining the theoretical templates. The GR template here means that we use the TNS model for the RSD where we set $\mu=1, \gamma_2=\gamma_3=0$ but treat $f$ as a free parameter parametrised by $\Omega_{rc}$. On the other hand, for the DGP template, we use $\mu, \gamma_2, \gamma_3$ as given in Appendix A  to model the non-linearity in the DGP model properly.  It is a common practice to use a GR template to estimate $f$ (eg. \cite{Beutler:2016arn}).
\newline
\newline
Our main intention in this work is to robustly assess whether using GR-only based RSD modelling (in other words, a model ignorant of any possible deviations from the GR picture of structure formation) on a non-GR universe is an accurate enough procedure to allow for a recovery of the fiducial value of the growth rate parameter $f$ from the data. Note that the DGP template encompasses a pure-GR scenario (by setting $\Omega_{rc}=0$) and so we should observe no biasing in using the DGP template for the GR data. Further, we want to determine on what scales and what amplitudes biasing becomes significant. As we fix the linear growth of density perturbations, $F_1$, both theoretical templates are equivalent on linear scales but as we include increasingly non-linear scales in the analysis the templates deviate producing the bias. This bias may be masked with the freedom of the TNS damping parameter $\sigma_v$, although the model specific non-linearities act beyond pure damping of small scale power and so $\sigma_v$ cannot perfectly substitute for incorrect perturbation modelling. In other words, model bias should still be expected at some level. 
\newline
\newline
Our analysis follows the techniques described in \cite{Lewis:2002ah}, with the MCMC algorithm walking in $\{\sigma_v,\Omega_{rc}\}$ parameter space. In order to measure when non-linearities become an issue in terms of theoretical modelling we complete the analysis for varying values of $k_{\rm max}$ up to the upper bounds found in the previous subsection. By going to higher $k$-modes we suppress the statistical errors. Fig.\ref{nbodyz05} shows the results at $z=0.5$. On the left hand side we present the $1\sigma$(68$\%$ C.L) and $2\sigma$(95$\%$ C.L) contours for the DGP template (green) and GR template (blue) for a matching up to $k_{\rm max}=0.147h$/Mpc using 24 bins. The marginalised statistics are shown in the side panels and the fiducial value, $\Omega_{rc}=0.438$, is marked as a dashed line. We note a slight offset of the best fit values of $\Omega_{rc}$, although the fiducial value remains in the $1\sigma$ region for both templates. There is also an offset in the best fit $\sigma_v$ which can be interpreted as the GR template's use of this parameter to compensate for non-linear effects in the DGP-PICOLA simulations.  On the right of Fig.\ref{nbodyz05}, we see the marginalised best fit values for $f(\Omega_{rc})$ with their $2\sigma$ errors for varying scale inclusion ($k_{\rm max}=0.110,0.135,0.147h$/Mpc with GR template's values shifted slightly for better visualisation). Note that as we push to higher $k_{\max}$ the errors become smaller, as expected. We here remind the reader that there is a lower bound on $f$ imposed by $\Omega_{rc}\geq 0$.  As can be seen in both plots, the GR template comfortably accommodates the data within $68\%$ (for $k_{\rm max}=0.147h$/Mpc) within the templates' validity regime.  
\newline
\newline
Fig.\ref{nbodyz1a} shows the results at $z=1$ this time matching up to a $k_{\rm max}=0.195h$/Mpc. It is in this case that we find the GR template struggling to capture the full shape of the multipoles and in the left hand plot we see that the fiducial parameter lies outside the $1\sigma$ region and just within the $2\sigma$ region. On the other hand, the DGP template is centred around the fiducial value. On the right hand side of Fig.\ref{nbodyz1a} we show the results for an analysis similar to that of the left hand side but for a larger survey with a volume of $V_s=20\mbox{Gpc}^{3}/h^3$, which is the estimated volume of the Dark Energy Spectroscopic Instrument (DESI) \cite{Aghamousa:2016zmz}. In this case the GR template fails to capture the fiducial at the $2\sigma$ level implying that inadequate theoretical modelling for such a large survey could introduce a significant error in parameter inference.  
\newline
\newline
Fig.\ref{nbodyz1b} shows the improvement in constraints as we increase $k_{\rm max}$. We have included an annotation with the same analysis done for a larger survey with a volume of $V_s=20 \mbox{Gpc}^{3}/h^3$ fitting up to $k_{\rm max}=0.195h$/Mpc. We see that the DGP template does consistently well in reproducing the fiducial while the GR begins to fail at around $k_{\rm max}=0.171h$/Mpc. 
\newline
\newline
As a consistency check of our analysis, we make use of the GR simulations at $z=1$. Again, by assuming a survey volume of $V=10 \mbox{Gpc}^3/h^3$, we repeated the analysis and obtained constraints on $\Omega_{rc}$ using the DGP template, as well as a constraint on $f$ using the GR template without the induced prior on $f$ coming from our parametrisation using $\Omega_{rc}$. Fig.\ref{nbodygrz1} shows the results. Both contours recover the fiducial parameters within $1\sigma$. 
\newline
\newline
As seen from the left panel of Fig.\ref{nbodygrz1}, the nDGP model with $\Omega_{rc}=0.438$ can be excluded with high significance ($\gg 2 \sigma$) by a survey with a volume $V_s=10 \mbox{Gpc}^3/h^3$ fitting up to $k_{max}=0.171 h$/Mpc, if our universe is described by GR. To quantify this, we computed the following quantity using the N-body measurements
\begin{equation}
\chi^2_{MG}(k_{\rm max}) = \frac{1}{N_{dof}}\sum_{\ell} \sum^{k_{\rm max}}_k \mbox{Cov}^{-1}_{\ell,\ell}(k)  [P_{\ell}^{DGP} (k) - P_{\ell}^{LCDM}(k)]^2 
\label{chisqrt}
\end{equation}
where $\mbox{Cov}^{-1}_{\ell,\ell} $ is the covariance matrix between the multipoles assuming an ideal survey of $V_s=10 \mbox{Gpc}^{3}/h^3$. We also computed the same quantity using SPT. Fig.\ref{chisqrtest} shows the results up to $k_{\rm max}=0.2 h$/Mpc at $z=0.5$ and $z=1$. The results clearly shows that our ability to distinguish between LCDM and nDGP increases with $k_{max}$. Also we find that the $\chi^2$ from SPT is very similar to those obtained from simulation. This indicates that the TNS model is capable of providing the same level of significance of deviations from LCDM as the full non-linear treatment, making it a good estimator of structure formation in this range of scales. 
\newline
\newline
We have compiled the template results in Table.\ref{sumres1}, Table.\ref{sumres2} and Table.\ref{sumres3}. 
\newpage
\begin{table}[ht]
\caption{Summary of template performances at $z=0.5$  for nDGP simulations where fiducial $f = 0.783$.}
\centering
\begin{tabular}{c c c c c c c}
\hline \hline 
Template & $k_{max} [h/\mbox{Mpc}] $ & bins & $V_s [\mbox{Gpc}/h]^3$ & $f\pm2\sigma$ &$\sigma_v \pm 2\sigma [\mbox{Mpc}/h]$ \\ 
\hline
GR & 0.110 & 18 & 10 &$ 0.774\pm^{0.034}_{0.022}$& $5.92\pm_{0.53}^{0.43}$ \\
DGP & 0.110 & 18 & 10 &$ 0.779\pm^{0.031}_{0.029}$&$ 5.86 \pm_{0.44}^{0.42} $\\
GR & 0.135 & 22 & 10 & $0.769\pm^{0.029}_{0.016}$& $ 5.86\pm_{0.16}^{0.22}$\\
DGP & 0.135 & 22 & 10 &$ 0.779\pm^{0.022}_{0.022}$ & $5.86\pm_{0.19}^{0.19}$  \\
GR & 0.147 & 24 & 10 &$ 0.777\pm^{0.025}_{0.025}$&  $5.96\pm_{0.15}^{0.14} $\\
DGP & 0.147 & 24 & 10 &$ 0.786\pm^{0.019}_{0.018}$& $5.89\pm_{0.13}^{0.12}$ \\
\hline
\end{tabular}
\label{sumres1}
\end{table}
\begin{table}[ht]
\caption{Summary of template performances for nDGP simulations at $z=1$ where fiducial $f = 0.909$.}
\centering
\begin{tabular}{c c c c c c c}
\hline \hline 
Template & $k_{max} [h/\mbox{Mpc}] $ & bins & $V_s [\mbox{Gpc}/h]^3$ &$f \pm2\sigma$ &  $\sigma_v \pm 2\sigma [\mbox{Mpc}/h]$ \\ 
\hline
GR & 0.147 & 24 & 10 &$ 0.895\pm^{0.028}_{0.025}$ & $4.08\pm_{0.26}^{0.22}$  \\
DGP & 0.147 & 24 & 10 & $0.902\pm^{0.025}_{0.025} $&$ 4.05\pm^{0.19}_{0.20}  $\\
GR & 0.171 & 28 & 10 & $0.902\pm^{0.021}_{0.020}$ &$ 4.19\pm_{0.09}^{0.10} $\\
DGP & 0.171 & 28 & 10 &$0.912\pm^{0.015}_{0.019}$ &$ 4.16\pm^{0.10}_{0.10}  $ \\
GR & 0.195 & 32 & 10 &$ 0.893\pm^{0.018}_{0.018}$&  $4.13\pm_{0.07}^{0.06} $\\
DGP & 0.195 & 32 & 10 & $0.908\pm^{0.014}_{0.014}$& $4.10\pm^{0.05}_{0.06} $\\
GR & 0.195 & 32 & 20 &$0.893\pm^{0.012}_{0.013} $& $ 4.13\pm_{0.05}^{0.04} $\\
DGP & 0.195 & 32 & 20 &$ 0.908\pm^{0.010}_{0.010}$&$ 4.10\pm^{0.04}_{0.04}$ \\
\hline
\end{tabular}
\label{sumres2}
\end{table}
\begin{table}[ht]
\caption{Summary of template performances for GR simulations at $z=1$ where fiducial $f = 0.859$.}
\centering
\begin{tabular}{c c c c c c c}
\hline \hline 
Template & $k_{max} [h/\mbox{Mpc}] $ & bins & $V_s [\mbox{Gpc}/h]^3$ & $f\pm2\sigma$ & $\sigma_v \pm 2\sigma [\mbox{Mpc}/h]$ \\ 
\hline
GR & 0.171 & 28 & 10 &$ 0.851\pm^{0.020}_{0.020} $& $3.94\pm_{0.14}^{0.13} $ \\
DGP & 0.171 & 28 & 10 & $0.869\pm^{0.019}_{0.010}$ &$ 4.01\pm^{0.09}_{0.11}$  \\
\hline
\end{tabular}
\label{sumres3}
\end{table}
\subsection{An Ideal Survey: SPT Mock Data} 
\begin{figure}[H]
  \captionsetup[subfigure]{labelformat=empty}
  \centering
  \subfloat[]{\includegraphics[width=8.3cm, height=8.3cm]{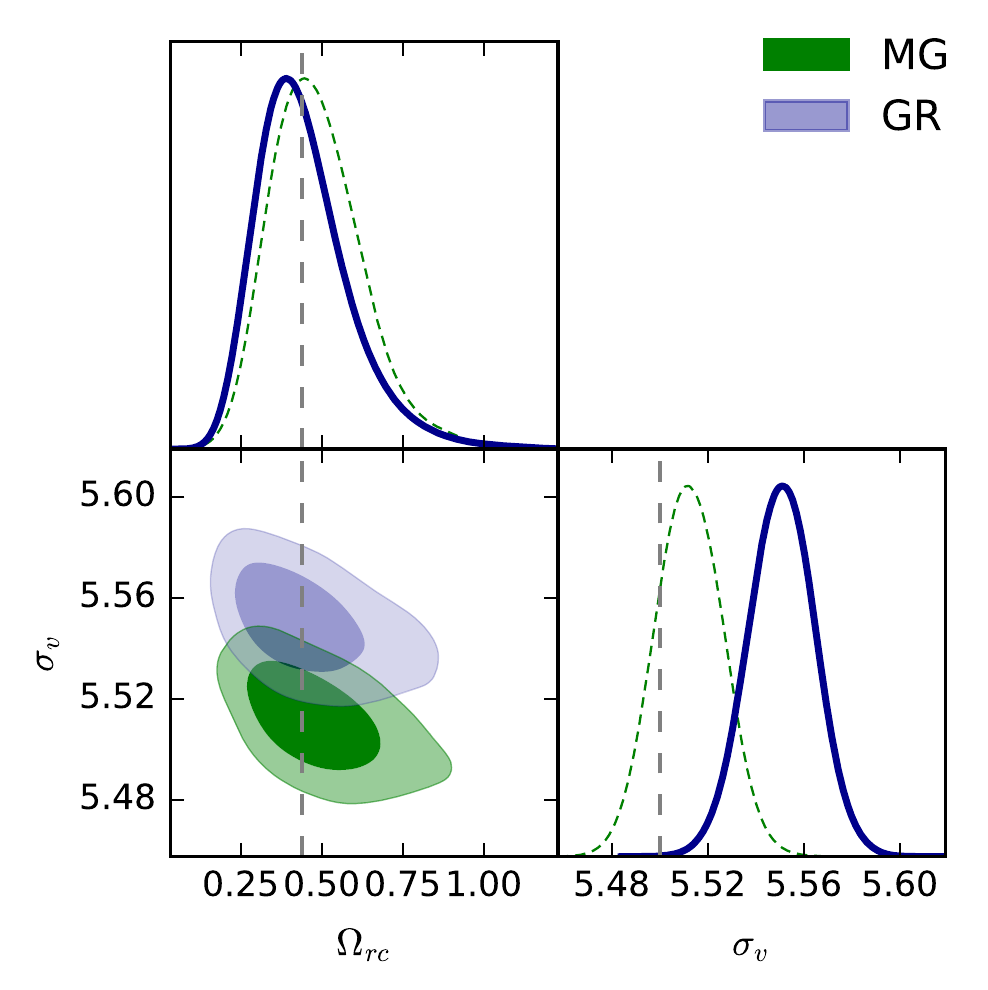}} \quad
  \subfloat[]{\includegraphics[width=8.3cm, height=8.3cm]{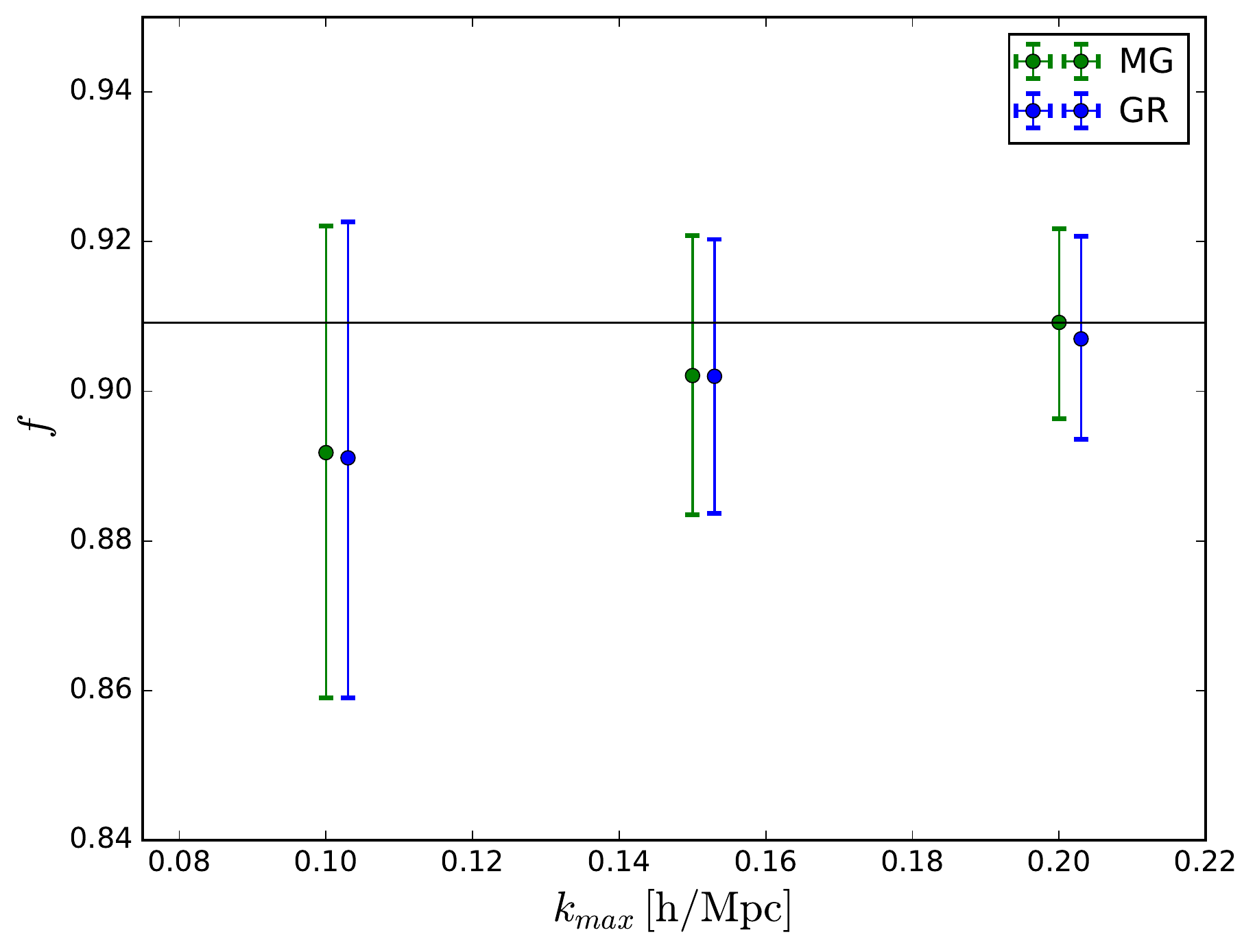}} 
  \caption[CONVERGENCE ]{Left: The $1\sigma$ and $2\sigma$ confidence contours for the $\alpha$-DGP template and the GR template at $z=1$ fitting up to $k_{\rm max}=0.2h$/Mpc using 20 bins for mock data with $\alpha =1$. The mock's fiducial values for $\Omega_{rc}$ and $\sigma_v$ indicated by the dashed line. Right: The best fit value for $f$ as a function of $k_{\rm max}$ with the $2\sigma$ errors for the $\alpha$-DGP and GR template. The GR template values (blue)  have been slightly shifted for better visualisation. A survey of volume of $10\mbox{Gpc}^{3}/h^3$ is assumed.}
\label{mocka1}
\end{figure}

\begin{figure}[H]
  \captionsetup[subfigure]{labelformat=empty}
  \centering
  \subfloat[]{\includegraphics[width=8.3cm, height=8.3cm]{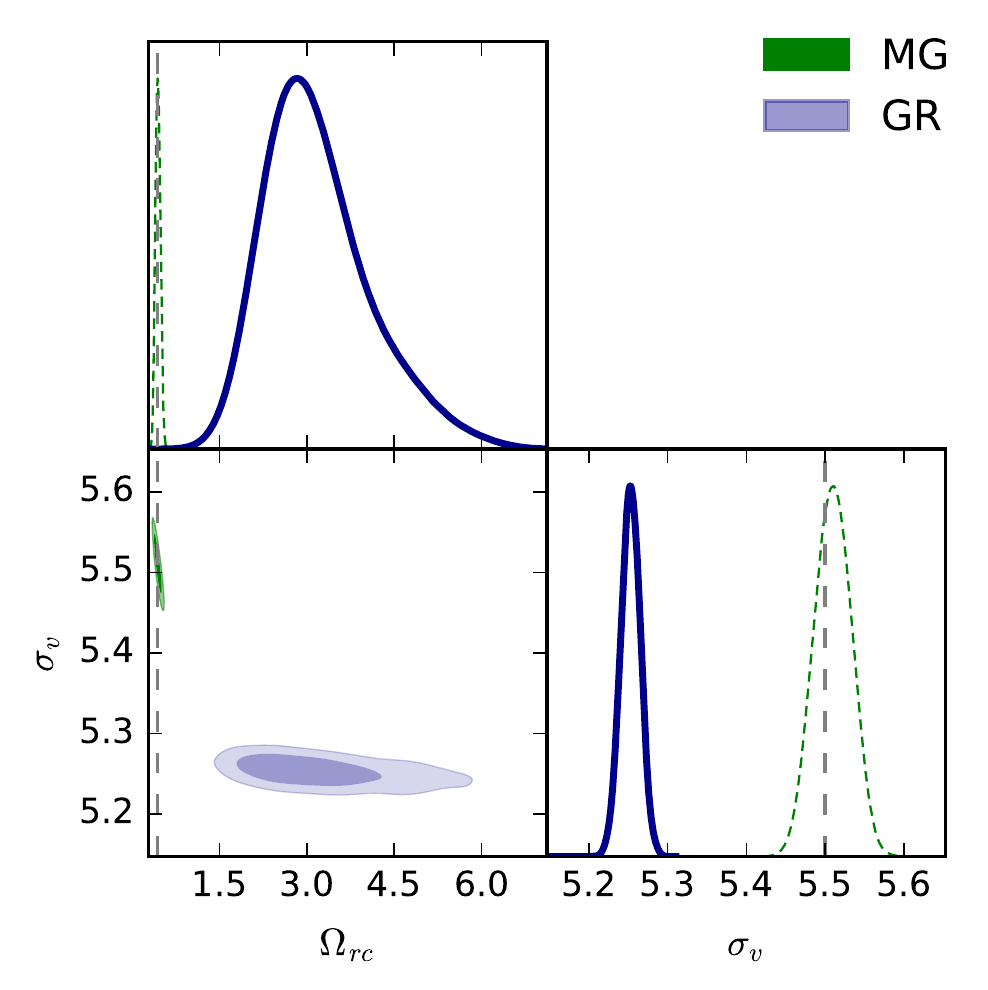}} \quad
  \subfloat[]{\includegraphics[width=8.3cm, height=8.3cm]{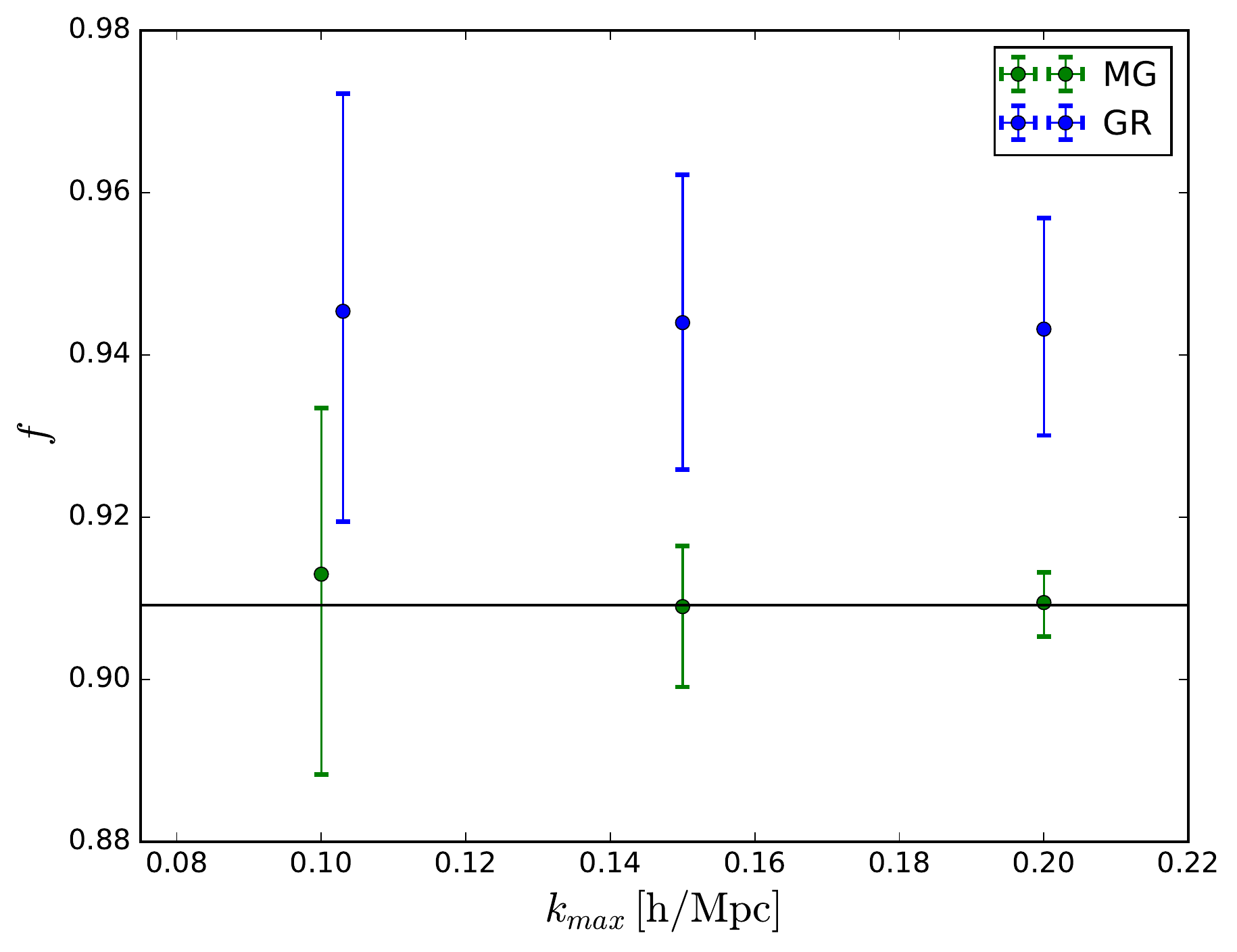}} 
  \caption[CONVERGENCE ]{Left: The $1\sigma$ and $2\sigma$ confidence contours for the $\alpha$-DGP template and the GR template at $z=1$ fitting up to $k_{\rm max}=0.2h$/Mpc using 20 bins for mock data with $\alpha =15$. The mock's fiducial values for $\Omega_{rc}$ and $\sigma_v$ indicated by the dashed line. Right: The best fit value for $f$ as a function of $k_{\rm max}$ with the $2\sigma$ errors for the $\alpha$-DGP and GR template. The GR template values (blue)  have been slightly shifted for better visualisation. A survey of volume of $10\mbox{Gpc}^{3}/h^3$ is assumed.}
\label{mocka15}
\end{figure}

\begin{figure}[H]
  \captionsetup[subfigure]{labelformat=empty}
  \centering
  \includegraphics[width=12.3cm, height=8.3cm]{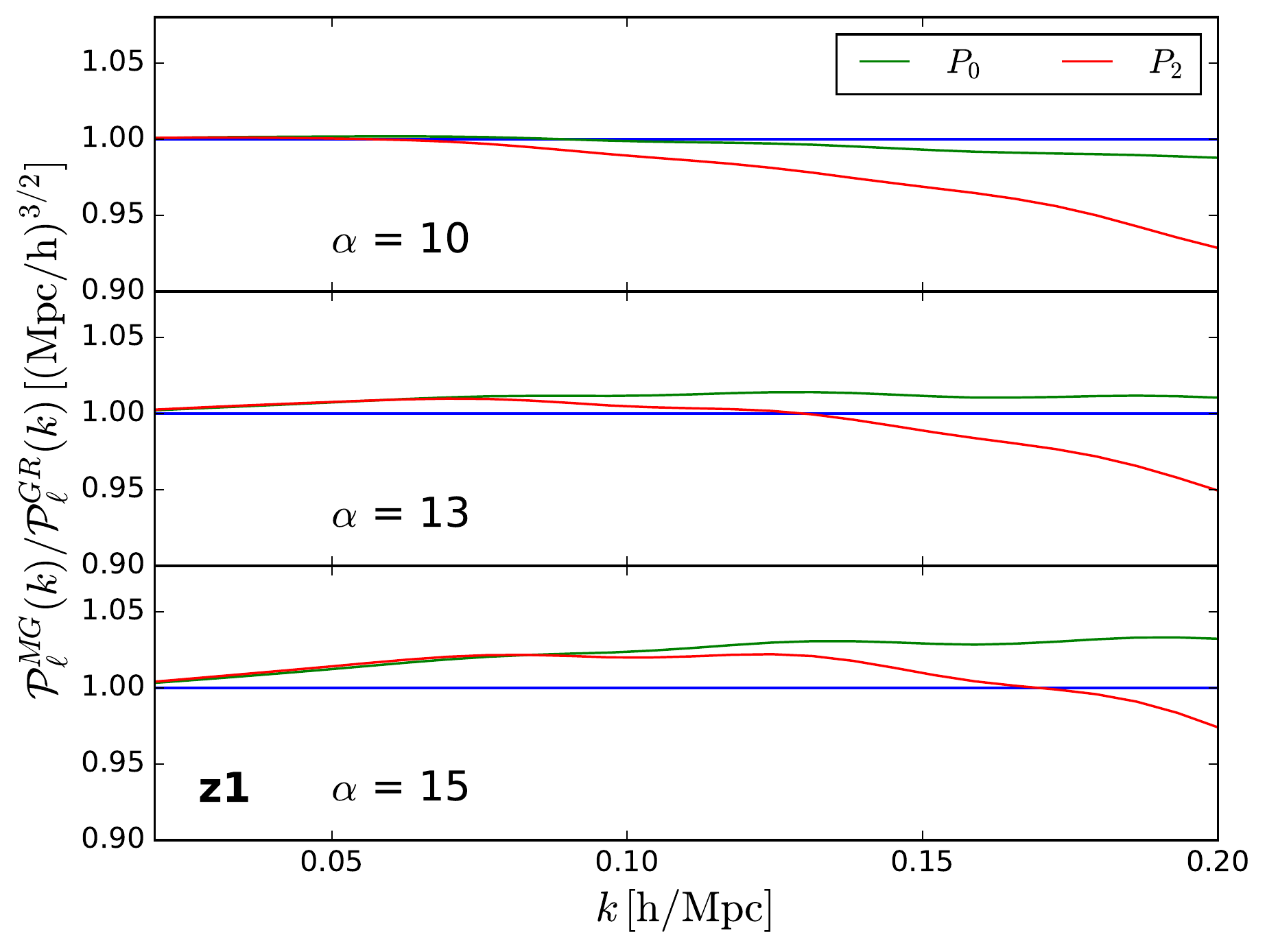}
  \caption[CONVERGENCE ]{Ratio of the GR and  $\alpha$-DGP theoretical templates' monopole (green) and quadrupole (red) predictions for varying values of $\alpha$ at $z=1$. The blue line represents no deviation in template's predictions. }
\label{alpha}
\end{figure}

Being model dependant,  we can expect that both the scales affected as well as the magnitude of theoretical bias will in general depend on the specific phenomenology of a given gravity model. Therefore, if one wants to precisely estimate the importance of such theoretical bias for a given set of real galaxy spectroscopic data, one would need to run N-body simulations for each model under consideration and then perform a similar analysis as in the previous section. This is obviously not practical and in this section we provide a means of getting a first indication of whether or not model bias is an issue for a given model, in other words, whether or not the model can be safely encompassed by the GR template within the relevant range of scales. 
\newline
\newline
We proceed as follows. First, multipole data is produced for a given model of gravity using SPT up to some valid $k_{\rm max}$. Then the covariance matrix for this data is computed as was done for the N-body data using the parameters of an ideal survey. Finally, this data is given Gaussian errors using the covariance matrix. This provides an easily produced, idealistic, simulated mock data set which can be done for any model of gravity described by the framework discussed in the first section. A statistical analysis as done in the previous section can then be performed on this data. Here we do this for the nDGP model with the same fiducial model parameter previously used, $\Omega_{rc}=0.438$ but with $\sigma_8=0.87$. All other cosmological parameters are the same as the nDGP N-body simulation. We choose a fiducial $\sigma_v=5.5 \mbox{Mpc}/h$ and use the ideal survey parameters $V_s=10\mbox{Gpc}^3/h^3$ and $\bar{n}_g= 4 \times 10^{-3} h^3 /\mbox{Mpc}^{3}$. Only $z=1$ is considered in this section and because SPT underestimates the non-linear effects we extend our statistical analysis to $k_{\rm max}=0.2h$/Mpc to include more non-linear scales. 
\newline
\newline
In addition to including more scales in the analysis one can also ask for which models does the model-bias systematic become important, or how much enhanced dynamics induced by modified gravity is needed to detect a significant deviation from the GR template. One could investigate this by creating mock data for larger values of $\Omega_{rc}$ although this becomes quite unrealistic.  In fact the value of $\Omega_{rc}=0.438$ is already ruled out by BOSS LOWZ and CMASS data to within $2\sigma$ (See \cite{Barreira:2016mg}), with the authors placing an upper bound of 0.36. What we choose to do instead is rescale the non-linear mode mixing which governs the change of the scalar field's non-linear derivative interactions - the source of screening. The rescaling is done by introducing the parameter $\alpha$. We will scale $\gamma_2$ by $\alpha$ and $\gamma_3$ by $\alpha^2$ in the Euler equation's non-linear source term $S(\bfk)$ (Eq.\ref{eq:Perturb3}). This is equivalent to enhancing the nDGP second order mode mixing term by $\alpha$ (See \cite{Koyama:2009me} for details). Then, by setting $\alpha =1$ we obtain the usual nDGP model but for values larger than unity the model changes to one with enhanced non-linearities. By tuning $\alpha$ we will be able to test the capabilities of the GR template to cover model non-linearities given an idealistic survey. Note this approach is not meant to represent a realistic or viable model but rather to be illustrative.
\newline
\newline
\newline
Fig.\ref{mocka1} show the results of the MCMC analysis for $\alpha=1$. The left hand plot shows the $1\sigma$ and $2\sigma$ contours for a matching done up to $k_{\rm max}=0.2h$/Mpc. We see that in this case both templates well recover the fiducial $\Omega_{rc}$ although the GR template fails to recover the fiducial $\sigma_v$ as expected. The right hand plot shows that the templates are comparable in their best fit value for $f$ as well as their $2\sigma$ constraints at all values of $k_{\rm max}$. We direct the reader again to Fig.\ref{nbodyz1a} which shows both best fits for $\Omega_{rc}$ when comparing to the N-body data. In contrast to results presented in Fig.\ref{mocka1}, we find the values as being substantially smaller than the fiducial value. As mentioned this may reflect a combined effect of insufficient statistics due to box size and also a use of too high a $k_{\rm max}$, where the PT-based templates begin to fail. The lower-than-fiducial preferences of both templates in Fig.\ref{nbodyz1a} would be consistent with a break down of SPT. This is because the loop corrections begin to over predict the power spectrum as SPT breaks down and so to compensate theory prefers a smaller value of $\Omega_{rc}$. Further, the breakdown of SPT for GR cosmologies at $z=1$ has been shown to be around $k_{\rm max}=0.147h$/Mpc \cite{Taruya:2010mx,Carlson:2009it}. This is curious as it would be noticed in Fig.\ref{nbodypsz1} and Fig.\ref{nbodyplz1}. Fig.\ref{nbodygrz1} suggests that if the offset is indeed due to a theoretical breakdown, it is not to a sufficient extent to push the fiducial value out of the $1\sigma$ region. In any case, we expect more light to be shed on this issue with more data. 
\newline
\newline
Finally, Fig.\ref{mocka15} illustrates that if we set $\alpha=15$ the GR template completely fails to recover the fiducial value even at large (more linear) scales, with $\Omega_{rc}=0.438$ still lying outside the $2\sigma$ errors for $k_{\rm max}=0.1h$/Mpc (right hand side plot). This is an extreme case with the non-linearities becoming far too important at linear scales. Fig.\ref{alpha} shows the deviation between the templates for $\sigma_v$ and $\Omega_{rc}$ set to fiducial values, clearly indicating when the enhancement enters the linear regime and away from $\sigma_v$'s ability to suppress it. With $\alpha=13$ we note 1-2$\%$ deviations at scales as large as $k=0.05h$/Mpc. The bias starts to become important out of $\sigma_v$'s reach at $\alpha=10$ (See Fig.\ref{nbodyplz1} for an indication of $\sigma_v$'s impact range). 
\newpage
%%%%%%%%%%%%%%%%%%%%%%%%%%%%%%%%%
\section{Summary and Conclusions}
The work presented here was motivated by the question of whether constraints of the growth rate derived with a GR PT-template give an unbiased measurement of growth for the case of a universe described by a modified gravity. This question is very relevant in the context of the next stage of cosmological surveys. We have here provided a first level analysis of the theoretical model bias systematic and have presented a method for quick assessment of the model specific significance of this systematic.
\newline
\newline
Firstly, we compared nDGP MG-PICOLA simulation data with the GR and nDGP theoretical predictions for the redshift space power spectrum. We used the TNS model of RSD which has been validated against both GR and modified gravity simulations \cite{Taruya:2010mx,Taruya:2013quf}. This was done at the level of dark matter clustering and only the first two multipoles were considered. Idealised future survey parameters were adopted in the analysis. We found out that the small-scale velocity damping term $\sigma_v$ included in the TNS model provides a flexibility through which the template can, to some extent, accommodate the enhanced small-scale clustering of the nDGP model. This was clearly indicated by higher values of $\sigma_v$ attained by the GR-template fit.
\newline
\newline
Both templates perform well in recovering the simulation's fiducial parameter at low redshift. We point out that the real space analysis done in \cite{Barreira:2016mg} concluded that no nDGP model bias is evident at redshifts up to $z=0.57$ which is consistent with our results at $z=0.5$. That being said, a full comparison of our results is difficult as there are a number of differences between their analysis and the one done here. In particular the use of different RSD models, the modelling of survey errors and their inclusion of galaxy bias which provides more fitting freedom to the GR template. We do find that at high redshift the GR prediction becomes increasingly biased and the difference between the two templates is greater. Using $V_s=20\mbox{Gpc}^3/h^3$, which will be realisable with stage IV surveys, we find systematically biased estimates of the GR template, with it failing to recover the fiducial parameter to within $2\sigma$ at $z=1$. This apparent bias might be due to specific limitations intrinsic to the SPT approach. The current analysis is left suggestive with robustness sought in additional theoretical modelling (for example including galaxy bias) or fuller treatment of non-linear scales such as using the EFT approach \cite{Baumann:2010tm,Carrasco:2012cv}. We leave for future work such analysis tailored for the detailed specifications of future generation galaxy surveys. Nonetheless, what is clear is that for the $V_s=20 \mbox{Gpc}^3/h^3$ case the $1\sigma$ regions of both templates do not intersect which implies that the template's predictions are inconsistent at that level. Again, the inclusion of galaxy bias may relieve this. 
\newline
\newline
In our second analysis we created mock data from SPT predictions by adding Gaussian noise generated using the errors derived from an idealised survey. Two data sets were created at $z=1$ using varying levels of model dependent non-linearity. This was done to simulate modified gravity models which have an enhanced non-linear source term. We find that by increasing the non-linear contributions to the higher order density and velocity perturbations, the GR template fails to recover the fiducial $\Omega_{rc}$, with the model bias being unimportant up to a non-linear contribution of around 10 times the base value. Above this the GR predictions become very biased and at 15 times the base non-linearity the GR template fails to recover the fiducial value even at scales of $k=0.1h$/Mpc. This exercise provides an indication on what scales and at what level of enhanced small-scale clustering a modified gravity model has to be consistently treated in RSD modelling in order to avoid significant theoretical biases that otherwise would diminish the desired accuracy of growth rate estimates. On this note, the creation of mock data can be done for any model of gravity within the framework discussed in \cite{Bose:2016qun} giving an avenue for assessing the importance of theoretical model bias in growth rate estimation from a given data set. The data quality of stage IV surveys indicates that this test will be important and is essential if we wish to put trusted constraints on modified gravity parameter space. 
\newline
\newline
In future analyses, the inclusion of higher multipoles would reduce the statistical errors and so further highlight model bias. On the other hand, we have only dealt with dark matter statistics and in a real survey galaxy bias needs to be included. This increases the degrees of freedom of the theoretical templates and gives the GR prediction more ability to cover the model bias systematic. The effect galaxy bias inclusion is left to a future work. We conclude that if the model exhibits sufficient non-linearity then model bias becomes an issue for modified gravity constraints using upcoming spectroscopic surveys such as DESI. This effect is more prominent at higher redshifts where we benefit from an increased spatial range of applicability of our theoretical template. Further, the improvement of theoretical predictions, such as through the EFToLSS \cite{Baumann:2010tm,Carrasco:2012cv}, will more loudly call for proper theoretical treatment of modified gravity models before attempts to obtain robust constraints are made. 
\newline
\newline
\newline
\newline
\newline
Finally, we comment on the analysis contained in this work which validate the use of MG-PICOLA \cite{Winther:2017jof} to check theoretical predictions. In Appendix B we perform a number of tests which show that the COLA approach to modified gravity consistently gives the correct non-linear predictions within  SPT's realm of validity, up to an additional damping. This can be well captured by the TNS models extra degree of freedom, and with loose constraints on $\sigma_v$ one may even push to smaller scales, with EFT approaches say, and perform similar data-theory analyses. 
\section*{Acknowledgments}
\noindent We thank the anonymous referee for useful and insightful comments and suggestions which benefited the scientific quality of this paper. BB is supported by the University of Portsmouth. KK, HAW and WAH are supported by the European Research Council through 646702 (CosTesGrav). KK is also supported by the UK Science and Technologies Facilities Council grants ST/N000668/1. WAH also acknowledges the support from the Polish National Science Center under contract $\#$UMO- 2012/07/D/ST9/02785.
%%%%%%%%%%%%%%%%%%%%%%%%%%%%%%%%%
\appendix
\section{Appendix A: DGP gravity}
\noindent In the DGP model of gravity \cite{Dvali:2000hr}  we live on a 4 dimensional brane embedded in 5 dimensional Minkowski spacetime, giving this theory a crossover scale $r_c$, which is the ratio between the 5D Newton gravitational constant and the 4D Newton  gravitational constant. $r_c$ is the only free parameter of the theory and we parametrise it as $\Omega_{rc}= 1/(4r_c^2H_0^2)$, $H_0$ being the Hubble parameter today. The modified Friedman equation in this theory is given by 
\begin{equation}
\epsilon\frac{H}{r_c} = H^2(1 - \Omega_m(a))
\end{equation}
where $\epsilon=\pm 1$. The solution for $\epsilon=1$ is known to be ghostly and so we consider what is called the normal branch with $\epsilon = -1$ (nDGP). In this branch acceleration is achieved through a dark energy constant as in GR. We also impose a background history following LCDM, done by tuning the dark energy equation of state. The non-linear interaction terms for this theory are found to be (for details on the Poisson equation form in this theory see \cite{Koyama:2009me} for example)
\begin{equation}
\mu(k;a)  = 1 + \frac{1}{3\beta(a)} 
\end{equation}
\begin{equation}
\gamma_2(k,\bfk_1,\bfk_2;a) = -\frac{H_0^2}{24 H^2 \beta(a)^3 \Omega_{rc}} \left(\frac{\Omega_{m0}}{a^3}\right)^2 (1-\mu_{1,2}^2) 
\end{equation}
\begin{equation}
\gamma_3(k,\bfk_1,\bfk_2,\bfk_3;a) = \frac{H_0^2}{144 H^2 \beta(a)^5 \Omega_{rc}^2} \left(\frac{\Omega_{m0}}{a^3}\right)^3 (1-\mu_{1,2}^2) (1-\mu_{1,23}^2)
\end{equation}
where
\begin{equation}
\beta(a)= 1+\frac{H}{\Omega_{rc}}\left(1+\frac{aH'}{3H}\right)
\end{equation}
$\mu_{i,j}$ is the cosine of the angle between $\bfk_i$ and $\bfk_j$, $\bfk_{ij}=\bfk_i+\bfk_j$ and $H'=dH/da$. 
\newpage
\section{Appendix B: N-body vs MG-PICOLA}
\begin{figure}[H]
  \captionsetup[subfigure]{labelformat=empty}
  \centering
  \subfloat[]{\includegraphics[width=8.3cm, height=8.3cm]{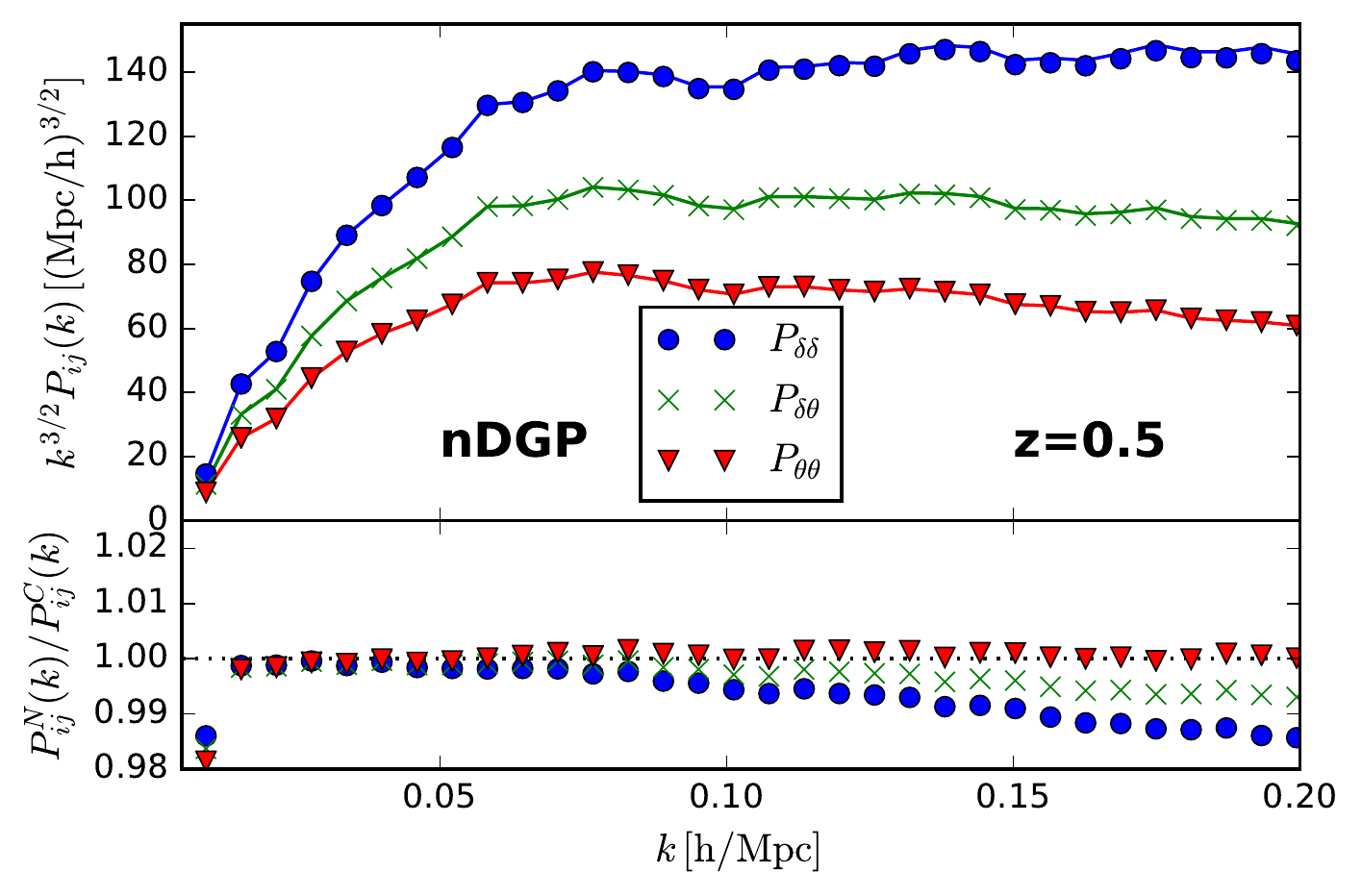}} \quad
  \subfloat[]{\includegraphics[width=8.3cm, height=8.3cm]{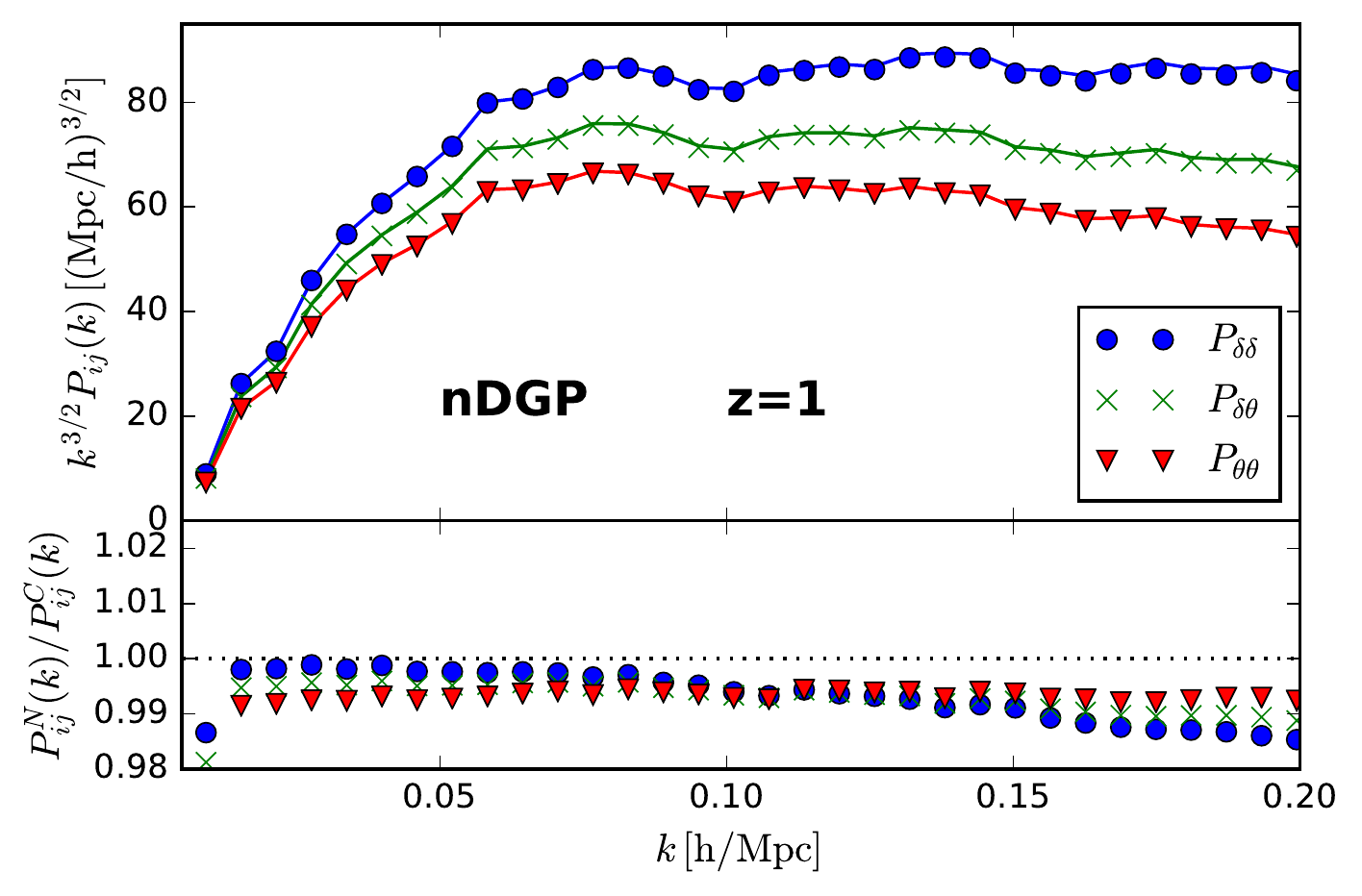}} 
  \caption[CONVERGENCE ]{COLA1 (solid) and N-body measurements (points) of the auto and cross power spectra of density and velocity fields in real space for nDGP at $z=0.5$ (left) and $z=1$ (right).  The top panels show the power spectra scaled by $k^{3/2}$ and the bottom panels show the ratio of the two measurements.}
\label{cola1vnbody1}
\end{figure}

\begin{figure}[H]
  \captionsetup[subfigure]{labelformat=empty}
  \centering
  \subfloat[]{\includegraphics[width=8.3cm, height=8.3cm]{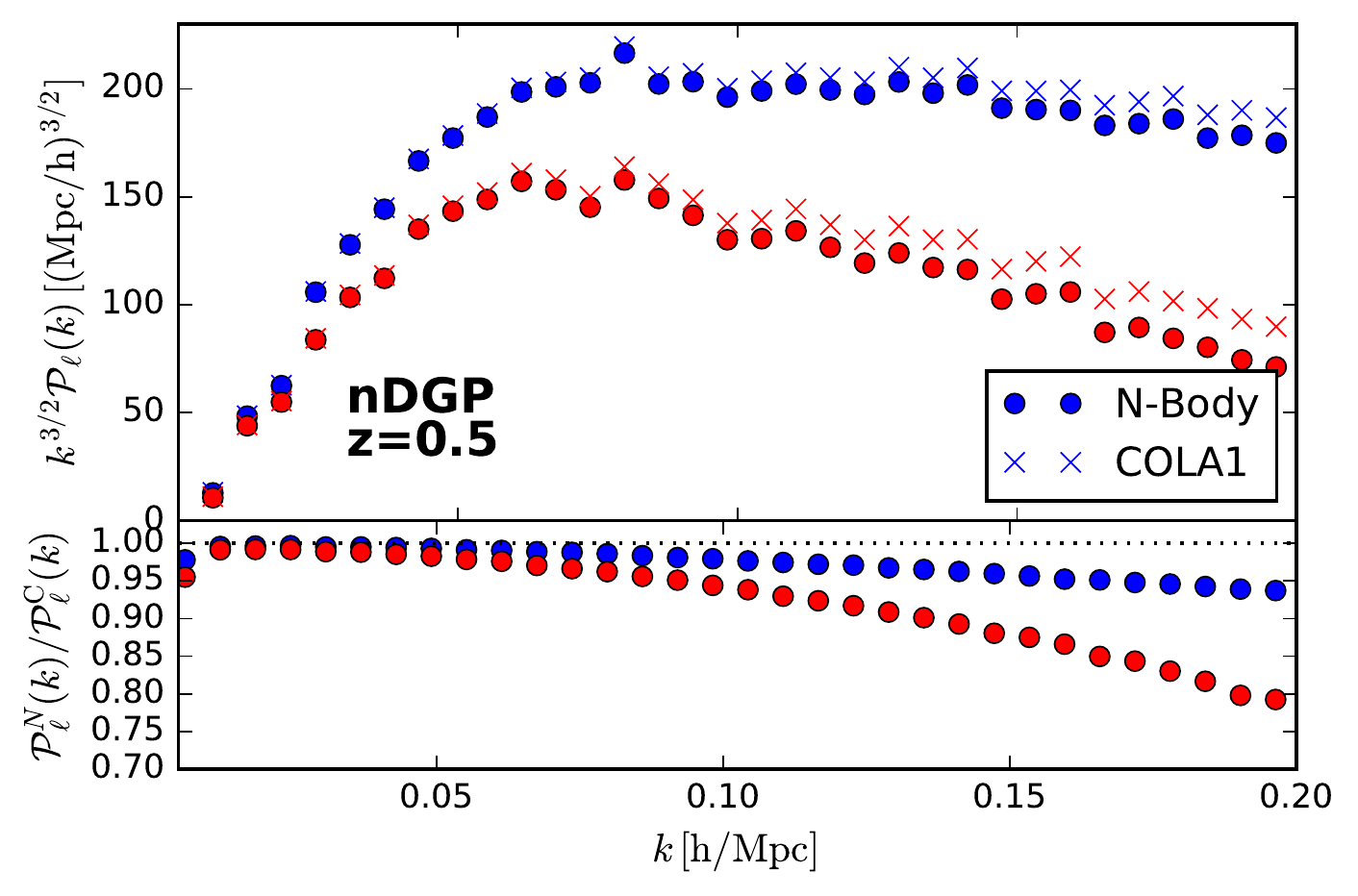}} \quad
  \subfloat[]{\includegraphics[width=8.3cm, height=8.3cm]{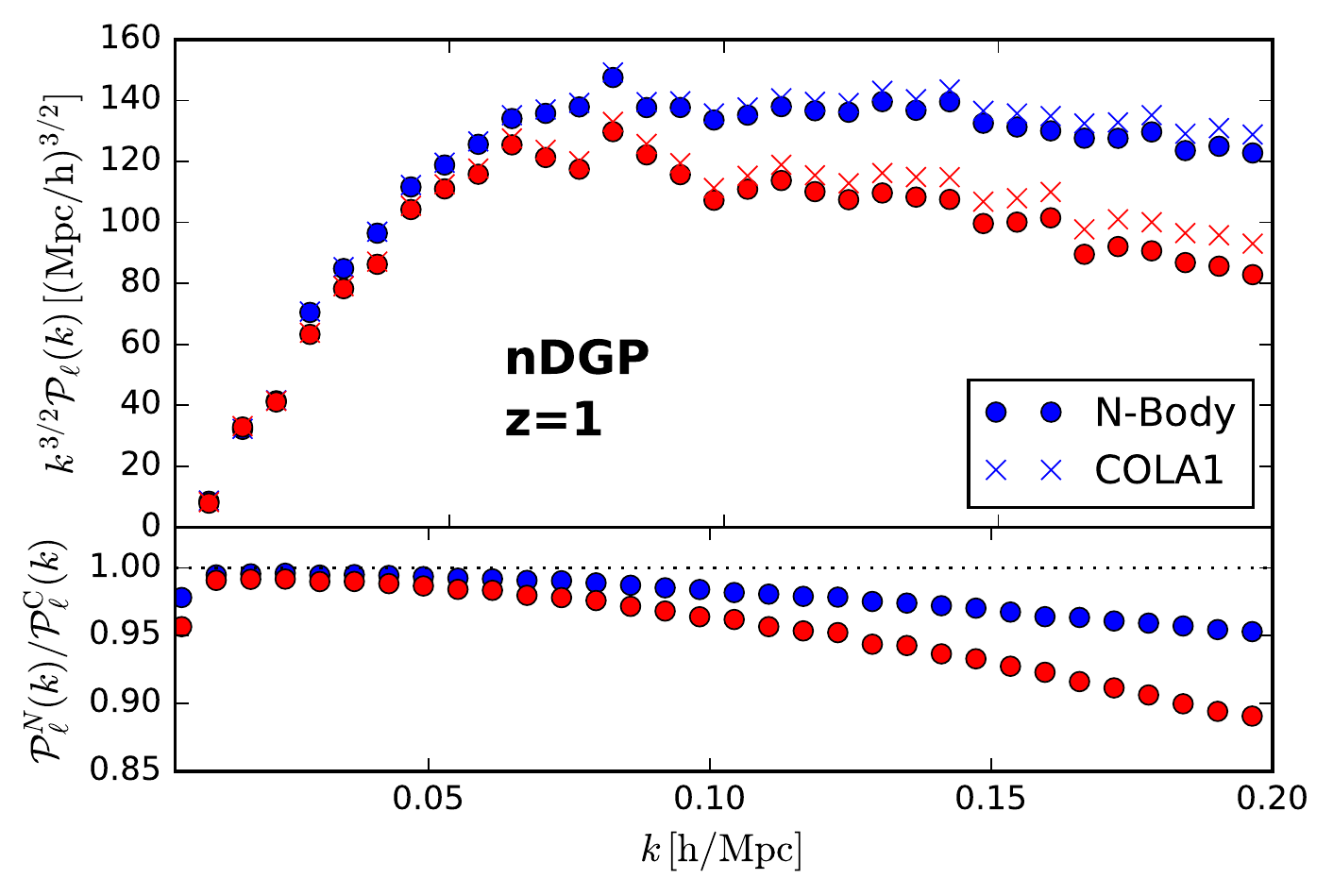}} 
  \caption[CONVERGENCE ]{COLA1 (crosses) and N-body measurements (circles) of the redshift space monopole (blue) and quadrupole (red) for nDGP at $z=0.5$ (left) and $z=1$ (right).  The top panels show the power spectra scaled by $k^{3/2}$ and the bottom panels show the ratio of the two measurements.}
\label{cola1vnbody2}
\end{figure}

\begin{figure}[H]
  \captionsetup[subfigure]{labelformat=empty}
  \centering
  \subfloat[]{\includegraphics[width=8.3cm, height=8.3cm]{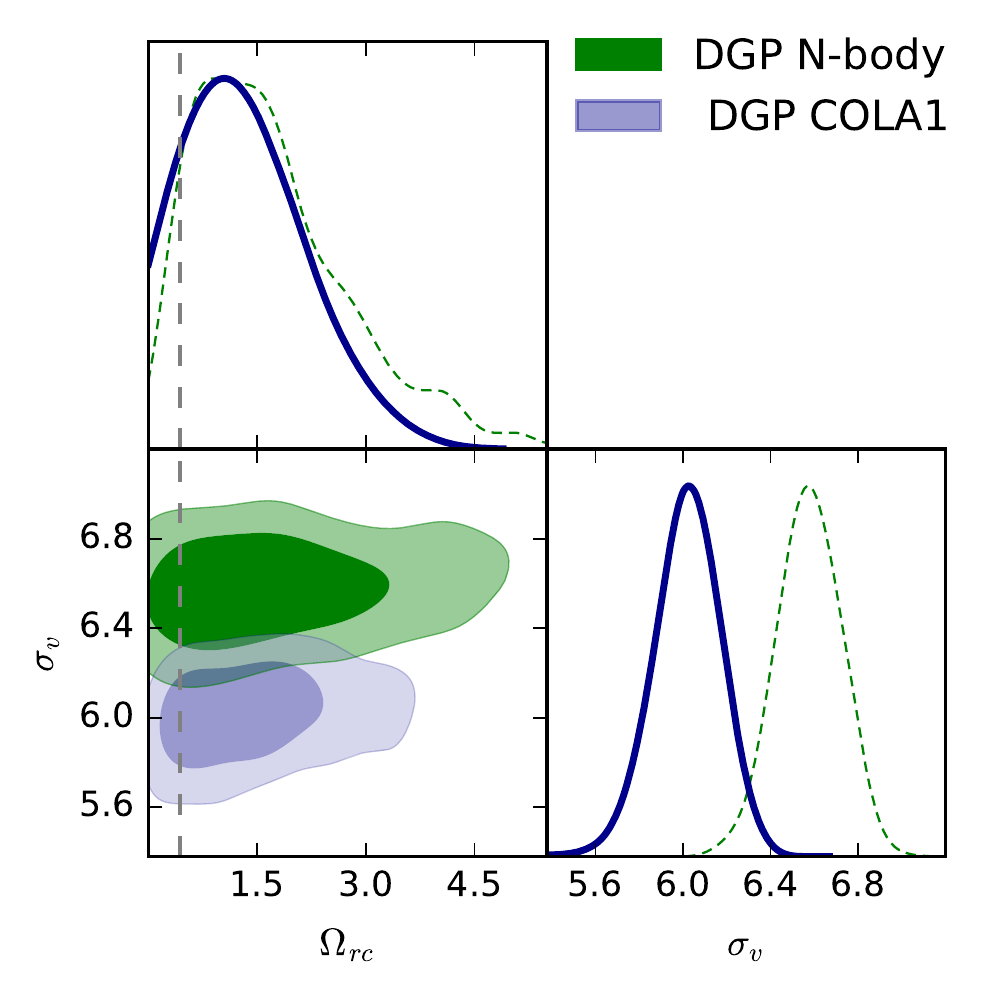}} \quad
  \subfloat[]{\includegraphics[width=8.3cm, height=8.3cm]{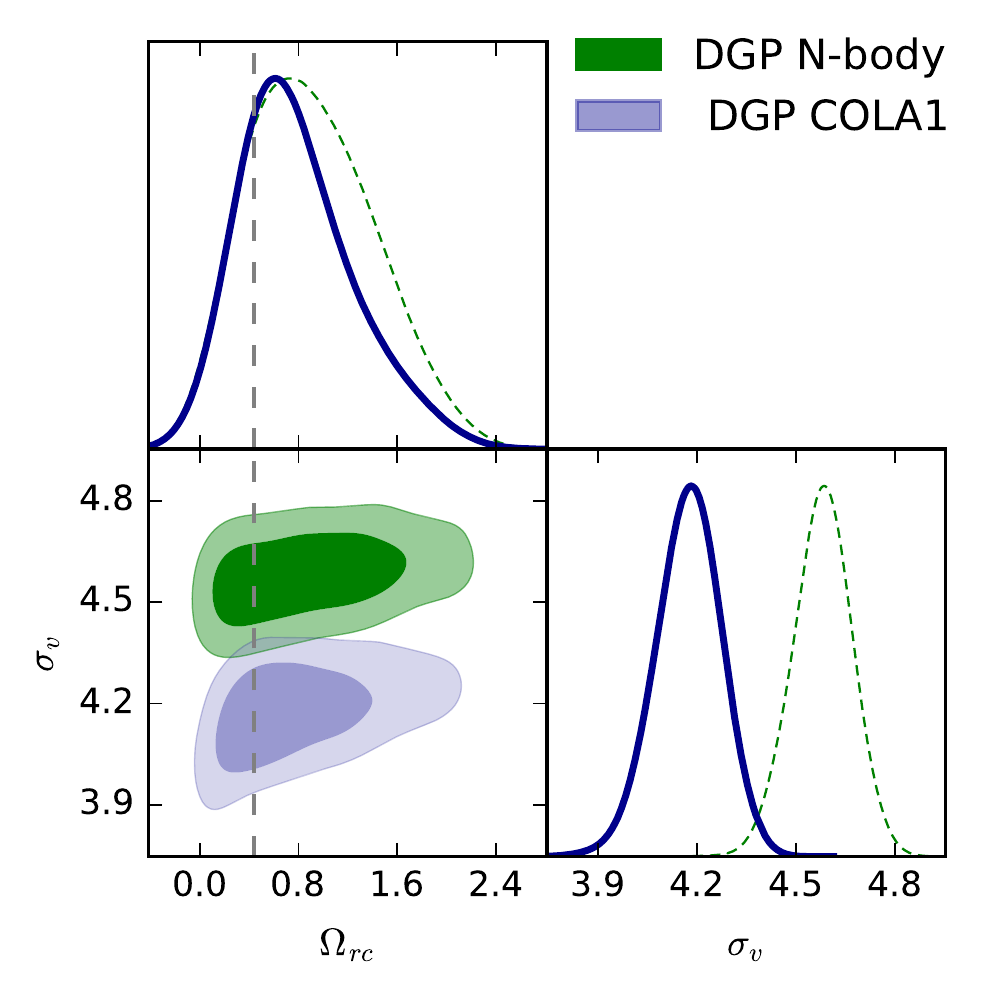}} 
  \caption[CONVERGENCE ]{ The $1\sigma$ and $2\sigma$ confidence contours for the DGP template using the N-body (green) and COLA1(blue) data sets at $z=0.5$(left) and $z=1$(right) fitting up to $k_{\rm max}=0.147h$/Mpc and $k_{\rm max}=0.171h$/Mpc respectively with the simulation's fiducial value for $\Omega_{rc}$ indicated by the dashed line. A survey of volume of $1\mbox{Gpc}^{3}/h^3$ is assumed. }
\label{cola1vnbody3}
\end{figure}

\begin{figure}[H]
  \captionsetup[subfigure]{labelformat=empty}
  \centering
  \subfloat[]{\includegraphics[width=8.3cm, height=8.3cm]{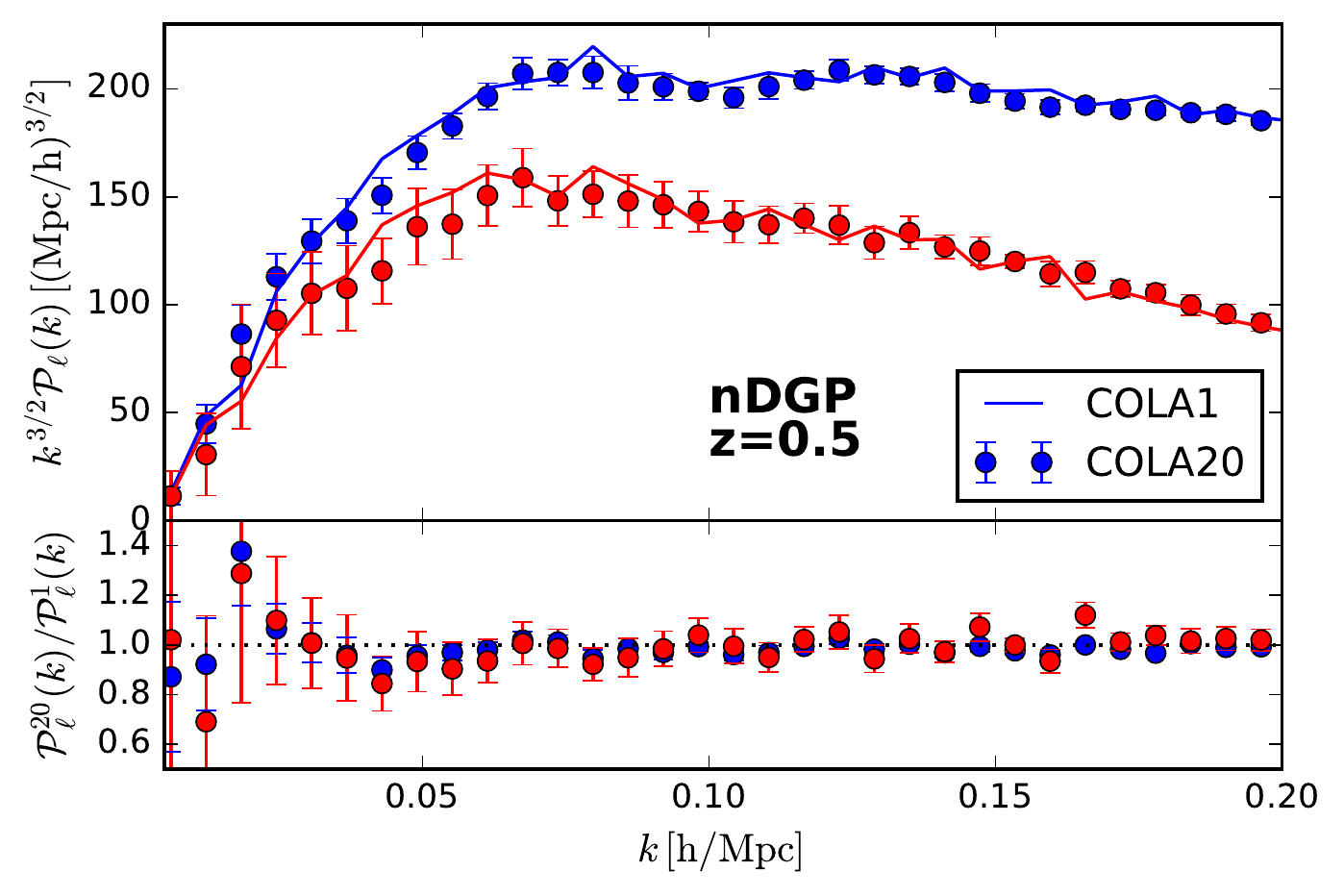}} \quad
  \subfloat[]{\includegraphics[width=8.3cm, height=8.3cm]{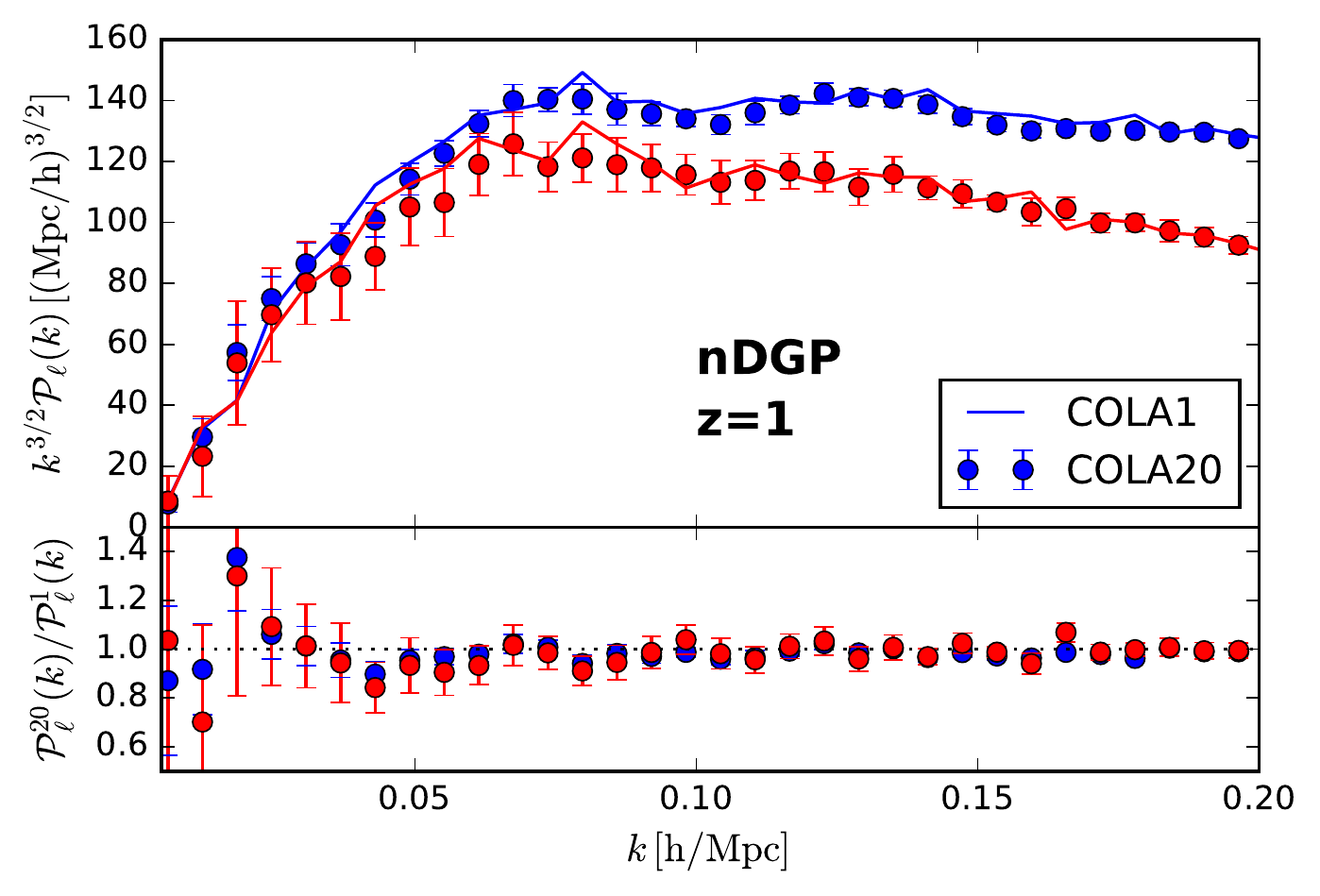}} 
  \caption[CONVERGENCE ]{ COLA1 (solid) and COLA20 (circles) of the redshift space monopole (blue) and quadrupole (red) for nDGP at $z=0.5$ (left) and $z=1$ (right).  The top panels show the power spectra scaled by $k^{3/2}$ and the bottom panels show the ratio of the two measurements. The errors bars are the variance of the 20 runs of COLA20. }
\label{cola20vcola1}
\end{figure}

In this Appendix we validate our use of MG-PICOLA in Sec.3B. We will refer to a single PICOLA simulation with the same initial conditions as N-body as COLA1 and we will refer to the averaged measurements from the 20 PICOLA runs as COLA20. We perform a number of tests listed below. 
\begin{enumerate}
\item
We compare the real space spectra from the full N-body simulation to COLA1. This serves to test the accuracy of PICOLA's evolution of structure. Fig.\ref{cola1vnbody1} shows that the COLA method reproduces the full non-linear real space spectra to within $2\%$ up to $k=0.2h$/Mpc at $z=0.5$ and $z=1$. This asserts that the full non-linear dynamics and evolution is sufficiently captured by MG-PICOLA  at the scales of interest. 
\item
We then compare the multipoles from the full N-body simulation to COLA1 and test for fiducial parameter recovery using the nDGP template for both measurements. We use a direct FFTW estimation of the multipoles from the N-body data.  Fig.\ref{cola1vnbody2} shows the multipole comparisons. The redshift space multipoles show less damping in MG-PICOLA simulations compared to the full N-body measurements due to less non-linear structures in these simulations, which give less fingers of god effects. Since the TNS model has the free parameter $\sigma_v$, that models this non-linear effect, the reduced damping can be accounted for by a smaller value of $\sigma_v$. Fig.\ref{cola1vnbody3} shows the results from an MCMC analysis using multipole data sets from COLA1 and N-body. It shows that we get a very good match in the marginalised posterior distribution of our parameter of interest $\Omega_{rc}$ and the contours are only shifted along $\sigma_v$. We use $1\mbox{Gpc}^3/h^3$ survey errors in accordance with the size of the simulations. 
\item 
Finally, we compare the redshift space multipoles from COLA1 to COLA20. This checks to see if the initial phases used in the full N-body simulation are outside the variance of the 20 runs. Fig.\ref{cola20vcola1} shows that COLA1 is within the variance of the 20 runs and nothing is unusual about the N-body's initial seeds. We also note that the initial condition for N-body was generated by {\tt MPGrafic} which uses the Zeldovich approximation while MG-PICOLA employs 2nd order Lagrangian Perturbation theory to generate initial conditions. 
\end{enumerate}
\renewcommand{\bibname}{References}
\bibliography{mybib}{}

\begin{thebibliography}{110}%
\makeatletter
\providecommand \@ifxundefined [1]{%
 \@ifx{#1\undefined}
}%
\providecommand \@ifnum [1]{%
 \ifnum #1\expandafter \@firstoftwo
 \else \expandafter \@secondoftwo
 \fi
}%
\providecommand \@ifx [1]{%
 \ifx #1\expandafter \@firstoftwo
 \else \expandafter \@secondoftwo
 \fi
}%
\providecommand \natexlab [1]{#1}%
\providecommand \enquote  [1]{``#1''}%
\providecommand \bibnamefont  [1]{#1}%
\providecommand \bibfnamefont [1]{#1}%
\providecommand \citenamefont [1]{#1}%
\providecommand \href@noop [0]{\@secondoftwo}%
\providecommand \href [0]{\begingroup \@sanitize@url \@href}%
\providecommand \@href[1]{\@@startlink{#1}\@@href}%
\providecommand \@@href[1]{\endgroup#1\@@endlink}%
\providecommand \@sanitize@url [0]{\catcode `\\12\catcode `\$12\catcode
  `\&12\catcode `\#12\catcode `\^12\catcode `\_12\catcode `\%12\relax}%
\providecommand \@@startlink[1]{}%
\providecommand \@@endlink[0]{}%
\providecommand \url  [0]{\begingroup\@sanitize@url \@url }%
\providecommand \@url [1]{\endgroup\@href {#1}{\urlprefix }}%
\providecommand \urlprefix  [0]{URL }%
\providecommand \Eprint [0]{\href }%
\providecommand \doibase [0]{http://dx.doi.org/}%
\providecommand \selectlanguage [0]{\@gobble}%
\providecommand \bibinfo  [0]{\@secondoftwo}%
\providecommand \bibfield  [0]{\@secondoftwo}%
\providecommand \translation [1]{[#1]}%
\providecommand \BibitemOpen [0]{}%
\providecommand \bibitemStop [0]{}%
\providecommand \bibitemNoStop [0]{.\EOS\space}%
\providecommand \EOS [0]{\spacefactor3000\relax}%
\providecommand \BibitemShut  [1]{\csname bibitem#1\endcsname}%
\let\auto@bib@innerbib\@empty
%</preamble>
\bibitem [{\citenamefont {Riess}\ \emph {et~al.}(1998)\citenamefont {Riess}
  \emph {et~al.}}]{Riess:1998cb}%
  \BibitemOpen
  \bibfield  {author} {\bibinfo {author} {\bibfnamefont {A.~G.}\ \bibnamefont
  {Riess}} \emph {et~al.} (\bibinfo {collaboration} {Supernova Search Team}),\
  }\href {\doibase 10.1086/300499} {\bibfield  {journal} {\bibinfo  {journal}
  {Astron. J.}\ }\textbf {\bibinfo {volume} {116}},\ \bibinfo {pages} {1009}
  (\bibinfo {year} {1998})},\ \Eprint {http://arxiv.org/abs/astro-ph/9805201}
  {arXiv:astro-ph/9805201 [astro-ph]} \BibitemShut {NoStop}%
%%CITATION = ASTRO-PH/9805201;%%
\bibitem [{\citenamefont {Perlmutter}\ \emph {et~al.}(1999)\citenamefont
  {Perlmutter} \emph {et~al.}}]{Perlmutter:1998np}%
  \BibitemOpen
  \bibfield  {author} {\bibinfo {author} {\bibfnamefont {S.}~\bibnamefont
  {Perlmutter}} \emph {et~al.} (\bibinfo {collaboration} {Supernova Cosmology
  Project}),\ }\href {\doibase 10.1086/307221} {\bibfield  {journal} {\bibinfo
  {journal} {Astrophys. J.}\ }\textbf {\bibinfo {volume} {517}},\ \bibinfo
  {pages} {565} (\bibinfo {year} {1999})},\ \Eprint
  {http://arxiv.org/abs/astro-ph/9812133} {arXiv:astro-ph/9812133 [astro-ph]}
  \BibitemShut {NoStop}%
%%CITATION = ASTRO-PH/9812133;%%
\bibitem [{\citenamefont {Abbott}\ \emph {et~al.}(2016)\citenamefont {Abbott}
  \emph {et~al.}}]{Abbott:2016blz}%
  \BibitemOpen
  \bibfield  {author} {\bibinfo {author} {\bibfnamefont {B.~P.}\ \bibnamefont
  {Abbott}} \emph {et~al.} (\bibinfo {collaboration} {Virgo, LIGO
  Scientific}),\ }\href {\doibase 10.1103/PhysRevLett.116.061102} {\bibfield
  {journal} {\bibinfo  {journal} {Phys. Rev. Lett.}\ }\textbf {\bibinfo
  {volume} {116}},\ \bibinfo {pages} {061102} (\bibinfo {year} {2016})},\
  \Eprint {http://arxiv.org/abs/1602.03837} {arXiv:1602.03837 [gr-qc]}
  \BibitemShut {NoStop}%
%%CITATION = ARXIV:1602.03837;%%
\bibitem [{\citenamefont {Weinberg}(1989)}]{Weinberg:1988cp}%
  \BibitemOpen
  \bibfield  {author} {\bibinfo {author} {\bibfnamefont {S.}~\bibnamefont
  {Weinberg}},\ }\href {\doibase 10.1103/RevModPhys.61.1} {\bibfield  {journal}
  {\bibinfo  {journal} {Rev.Mod.Phys.}\ }\textbf {\bibinfo {volume} {61}},\
  \bibinfo {pages} {1} (\bibinfo {year} {1989})}\BibitemShut {NoStop}%
%%CITATION = RMPHA,61,1;%%
\bibitem [{\citenamefont {Martin}(2012)}]{Martin:2012bt}%
  \BibitemOpen
  \bibfield  {author} {\bibinfo {author} {\bibfnamefont {J.}~\bibnamefont
  {Martin}},\ }\href {\doibase 10.1016/j.crhy.2012.04.008} {\bibfield
  {journal} {\bibinfo  {journal} {Comptes Rendus Physique}\ }\textbf {\bibinfo
  {volume} {13}},\ \bibinfo {pages} {566} (\bibinfo {year} {2012})},\ \Eprint
  {http://arxiv.org/abs/1205.3365} {arXiv:1205.3365 [astro-ph.CO]} \BibitemShut
  {NoStop}%
%%CITATION = ARXIV:1205.3365;%%
\bibitem [{\citenamefont {Hu}\ and\ \citenamefont {Sawicki}(2007)}]{Hu:2007nk}%
  \BibitemOpen
  \bibfield  {author} {\bibinfo {author} {\bibfnamefont {W.}~\bibnamefont
  {Hu}}\ and\ \bibinfo {author} {\bibfnamefont {I.}~\bibnamefont {Sawicki}},\
  }\href {\doibase 10.1103/PhysRevD.76.064004} {\bibfield  {journal} {\bibinfo
  {journal} {Phys.Rev.}\ }\textbf {\bibinfo {volume} {D76}},\ \bibinfo {pages}
  {064004} (\bibinfo {year} {2007})},\ \Eprint {http://arxiv.org/abs/0705.1158}
  {arXiv:0705.1158 [astro-ph]} \BibitemShut {NoStop}%
%%CITATION = ARXIV:0705.1158;%%
\bibitem [{\citenamefont {Starobinsky}(1980)}]{Starobinsky:1980te}%
  \BibitemOpen
  \bibfield  {author} {\bibinfo {author} {\bibfnamefont {A.~A.}\ \bibnamefont
  {Starobinsky}},\ }\href {\doibase 10.1016/0370-2693(80)90670-X} {\bibfield
  {journal} {\bibinfo  {journal} {Phys.Lett.}\ }\textbf {\bibinfo {volume}
  {B91}},\ \bibinfo {pages} {99} (\bibinfo {year} {1980})}\BibitemShut
  {NoStop}%
%%CITATION = PHLTA,B91,99;%%
\bibitem [{\citenamefont {Dvali}\ \emph {et~al.}(2000)\citenamefont {Dvali},
  \citenamefont {Gabadadze},\ and\ \citenamefont {Porrati}}]{Dvali:2000hr}%
  \BibitemOpen
  \bibfield  {author} {\bibinfo {author} {\bibfnamefont {G.}~\bibnamefont
  {Dvali}}, \bibinfo {author} {\bibfnamefont {G.}~\bibnamefont {Gabadadze}}, \
  and\ \bibinfo {author} {\bibfnamefont {M.}~\bibnamefont {Porrati}},\ }\href
  {\doibase 10.1016/S0370-2693(00)00669-9} {\bibfield  {journal} {\bibinfo
  {journal} {Phys.Lett.}\ }\textbf {\bibinfo {volume} {B485}},\ \bibinfo
  {pages} {208} (\bibinfo {year} {2000})},\ \Eprint
  {http://arxiv.org/abs/hep-th/0005016} {arXiv:hep-th/0005016 [hep-th]}
  \BibitemShut {NoStop}%
%%CITATION = HEP-TH/0005016;%%
\bibitem [{\citenamefont {Gomes}\ and\ \citenamefont
  {Amendola}(2014)}]{Gomes:2013ema}%
  \BibitemOpen
  \bibfield  {author} {\bibinfo {author} {\bibfnamefont {A.~R.}\ \bibnamefont
  {Gomes}}\ and\ \bibinfo {author} {\bibfnamefont {L.}~\bibnamefont
  {Amendola}},\ }\href {\doibase 10.1088/1475-7516/2014/03/041} {\bibfield
  {journal} {\bibinfo  {journal} {JCAP}\ }\textbf {\bibinfo {volume} {1403}},\
  \bibinfo {pages} {041} (\bibinfo {year} {2014})},\ \Eprint
  {http://arxiv.org/abs/1306.3593} {arXiv:1306.3593 [astro-ph.CO]} \BibitemShut
  {NoStop}%
%%CITATION = ARXIV:1306.3593;%%
\bibitem [{\citenamefont {Hassan}\ and\ \citenamefont
  {Rosen}(2012)}]{Hassan:2011zd}%
  \BibitemOpen
  \bibfield  {author} {\bibinfo {author} {\bibfnamefont {S.}~\bibnamefont
  {Hassan}}\ and\ \bibinfo {author} {\bibfnamefont {R.~A.}\ \bibnamefont
  {Rosen}},\ }\href {\doibase 10.1007/JHEP02(2012)126} {\bibfield  {journal}
  {\bibinfo  {journal} {JHEP}\ }\textbf {\bibinfo {volume} {1202}},\ \bibinfo
  {pages} {126} (\bibinfo {year} {2012})},\ \Eprint
  {http://arxiv.org/abs/1109.3515} {arXiv:1109.3515 [hep-th]} \BibitemShut
  {NoStop}%
%%CITATION = ARXIV:1109.3515;%%
\bibitem [{\citenamefont {de~Rham}\ and\ \citenamefont
  {Heisenberg}(2011)}]{deRham:2011by}%
  \BibitemOpen
  \bibfield  {author} {\bibinfo {author} {\bibfnamefont {C.}~\bibnamefont
  {de~Rham}}\ and\ \bibinfo {author} {\bibfnamefont {L.}~\bibnamefont
  {Heisenberg}},\ }\href {\doibase 10.1103/PhysRevD.84.043503} {\bibfield
  {journal} {\bibinfo  {journal} {Phys.Rev.}\ }\textbf {\bibinfo {volume}
  {D84}},\ \bibinfo {pages} {043503} (\bibinfo {year} {2011})},\ \Eprint
  {http://arxiv.org/abs/1106.3312} {arXiv:1106.3312 [hep-th]} \BibitemShut
  {NoStop}%
%%CITATION = ARXIV:1106.3312;%%
\bibitem [{\citenamefont {de~Rham}\ and\ \citenamefont
  {Gabadadze}(2010)}]{deRham:2010ik}%
  \BibitemOpen
  \bibfield  {author} {\bibinfo {author} {\bibfnamefont {C.}~\bibnamefont
  {de~Rham}}\ and\ \bibinfo {author} {\bibfnamefont {G.}~\bibnamefont
  {Gabadadze}},\ }\href {\doibase 10.1103/PhysRevD.82.044020} {\bibfield
  {journal} {\bibinfo  {journal} {Phys.Rev.}\ }\textbf {\bibinfo {volume}
  {D82}},\ \bibinfo {pages} {044020} (\bibinfo {year} {2010})},\ \Eprint
  {http://arxiv.org/abs/1007.0443} {arXiv:1007.0443 [hep-th]} \BibitemShut
  {NoStop}%
%%CITATION = ARXIV:1007.0443;%%
\bibitem [{\citenamefont {de~Rham}\ \emph {et~al.}(2011)\citenamefont
  {de~Rham}, \citenamefont {Gabadadze},\ and\ \citenamefont
  {Tolley}}]{deRham:2010kj}%
  \BibitemOpen
  \bibfield  {author} {\bibinfo {author} {\bibfnamefont {C.}~\bibnamefont
  {de~Rham}}, \bibinfo {author} {\bibfnamefont {G.}~\bibnamefont {Gabadadze}},
  \ and\ \bibinfo {author} {\bibfnamefont {A.~J.}\ \bibnamefont {Tolley}},\
  }\href {\doibase 10.1103/PhysRevLett.106.231101} {\bibfield  {journal}
  {\bibinfo  {journal} {Phys.Rev.Lett.}\ }\textbf {\bibinfo {volume} {106}},\
  \bibinfo {pages} {231101} (\bibinfo {year} {2011})},\ \Eprint
  {http://arxiv.org/abs/1011.1232} {arXiv:1011.1232 [hep-th]} \BibitemShut
  {NoStop}%
%%CITATION = ARXIV:1011.1232;%%
\bibitem [{\citenamefont {Maartens}\ and\ \citenamefont
  {Koyama}(2010)}]{Maartens:2010ar}%
  \BibitemOpen
  \bibfield  {author} {\bibinfo {author} {\bibfnamefont {R.}~\bibnamefont
  {Maartens}}\ and\ \bibinfo {author} {\bibfnamefont {K.}~\bibnamefont
  {Koyama}},\ }\href@noop {} {\bibfield  {journal} {\bibinfo  {journal} {Living
  Rev.Rel.}\ }\textbf {\bibinfo {volume} {13}},\ \bibinfo {pages} {5} (\bibinfo
  {year} {2010})},\ \Eprint {http://arxiv.org/abs/1004.3962} {arXiv:1004.3962
  [hep-th]} \BibitemShut {NoStop}%
%%CITATION = ARXIV:1004.3962;%%
\bibitem [{\citenamefont {Koyama}(2016)}]{Koyama:2015vza}%
  \BibitemOpen
  \bibfield  {author} {\bibinfo {author} {\bibfnamefont {K.}~\bibnamefont
  {Koyama}},\ }\href {\doibase 10.1088/0034-4885/79/4/046902} {\bibfield
  {journal} {\bibinfo  {journal} {Rept. Prog. Phys.}\ }\textbf {\bibinfo
  {volume} {79}},\ \bibinfo {pages} {046902} (\bibinfo {year} {2016})},\
  \Eprint {http://arxiv.org/abs/1504.04623} {arXiv:1504.04623 [astro-ph.CO]}
  \BibitemShut {NoStop}%
%%CITATION = ARXIV:1504.04623;%%
\bibitem [{\citenamefont {Sakstein}(2015)}]{Sakstein:2015oqa}%
  \BibitemOpen
  \bibfield  {author} {\bibinfo {author} {\bibfnamefont {J.}~\bibnamefont
  {Sakstein}},\ }\href@noop {} {\bibfield  {journal} {\bibinfo  {journal}
  {arXiv:1502.04503}\ } (\bibinfo {year} {2015})}\BibitemShut {NoStop}%
%%CITATION = ARXIV:1502.04503;%%
\bibitem [{\citenamefont {Clifton}\ \emph {et~al.}(2012)\citenamefont
  {Clifton}, \citenamefont {Ferreira}, \citenamefont {Padilla},\ and\
  \citenamefont {Skordis}}]{Clifton:2011jh}%
  \BibitemOpen
  \bibfield  {author} {\bibinfo {author} {\bibfnamefont {T.}~\bibnamefont
  {Clifton}}, \bibinfo {author} {\bibfnamefont {P.~G.}\ \bibnamefont
  {Ferreira}}, \bibinfo {author} {\bibfnamefont {A.}~\bibnamefont {Padilla}}, \
  and\ \bibinfo {author} {\bibfnamefont {C.}~\bibnamefont {Skordis}},\ }\href
  {\doibase 10.1016/j.physrep.2012.01.001} {\bibfield  {journal} {\bibinfo
  {journal} {Phys.Rept.}\ }\textbf {\bibinfo {volume} {513}},\ \bibinfo {pages}
  {1} (\bibinfo {year} {2012})},\ \Eprint {http://arxiv.org/abs/1106.2476}
  {arXiv:1106.2476 [astro-ph.CO]} \BibitemShut {NoStop}%
%%CITATION = ARXIV:1106.2476;%%
\bibitem [{\citenamefont {Uzan}\ and\ \citenamefont
  {Bernardeau}(2001)}]{Uzan:2000mz}%
  \BibitemOpen
  \bibfield  {author} {\bibinfo {author} {\bibfnamefont {J.-P.}\ \bibnamefont
  {Uzan}}\ and\ \bibinfo {author} {\bibfnamefont {F.}~\bibnamefont
  {Bernardeau}},\ }\href {\doibase 10.1103/PhysRevD.64.083004} {\bibfield
  {journal} {\bibinfo  {journal} {Phys. Rev.}\ }\textbf {\bibinfo {volume}
  {D64}},\ \bibinfo {pages} {083004} (\bibinfo {year} {2001})},\ \Eprint
  {http://arxiv.org/abs/hep-ph/0012011} {arXiv:hep-ph/0012011 [hep-ph]}
  \BibitemShut {NoStop}%
%%CITATION = HEP-PH/0012011;%%
\bibitem [{\citenamefont {Lue}\ \emph {et~al.}(2004)\citenamefont {Lue},
  \citenamefont {Scoccimarro},\ and\ \citenamefont {Starkman}}]{Lue:2004rj}%
  \BibitemOpen
  \bibfield  {author} {\bibinfo {author} {\bibfnamefont {A.}~\bibnamefont
  {Lue}}, \bibinfo {author} {\bibfnamefont {R.}~\bibnamefont {Scoccimarro}}, \
  and\ \bibinfo {author} {\bibfnamefont {G.~D.}\ \bibnamefont {Starkman}},\
  }\href {\doibase 10.1103/PhysRevD.69.124015} {\bibfield  {journal} {\bibinfo
  {journal} {Phys.Rev.}\ }\textbf {\bibinfo {volume} {D69}},\ \bibinfo {pages}
  {124015} (\bibinfo {year} {2004})},\ \Eprint
  {http://arxiv.org/abs/astro-ph/0401515} {arXiv:astro-ph/0401515 [astro-ph]}
  \BibitemShut {NoStop}%
%%CITATION = ASTRO-PH/0401515;%%
\bibitem [{\citenamefont {Ishak}\ \emph {et~al.}(2006)\citenamefont {Ishak},
  \citenamefont {Upadhye},\ and\ \citenamefont {Spergel}}]{Ishak:2005zs}%
  \BibitemOpen
  \bibfield  {author} {\bibinfo {author} {\bibfnamefont {M.}~\bibnamefont
  {Ishak}}, \bibinfo {author} {\bibfnamefont {A.}~\bibnamefont {Upadhye}}, \
  and\ \bibinfo {author} {\bibfnamefont {D.~N.}\ \bibnamefont {Spergel}},\
  }\href {\doibase 10.1103/PhysRevD.74.043513} {\bibfield  {journal} {\bibinfo
  {journal} {Phys.Rev.}\ }\textbf {\bibinfo {volume} {D74}},\ \bibinfo {pages}
  {043513} (\bibinfo {year} {2006})},\ \Eprint
  {http://arxiv.org/abs/astro-ph/0507184} {arXiv:astro-ph/0507184 [astro-ph]}
  \BibitemShut {NoStop}%
%%CITATION = ASTRO-PH/0507184;%%
\bibitem [{\citenamefont {Knox}\ \emph {et~al.}(2006)\citenamefont {Knox},
  \citenamefont {Song},\ and\ \citenamefont {Tyson}}]{Knox:2005rg}%
  \BibitemOpen
  \bibfield  {author} {\bibinfo {author} {\bibfnamefont {L.}~\bibnamefont
  {Knox}}, \bibinfo {author} {\bibfnamefont {Y.-S.}\ \bibnamefont {Song}}, \
  and\ \bibinfo {author} {\bibfnamefont {J.~A.}\ \bibnamefont {Tyson}},\ }\href
  {\doibase 10.1103/PhysRevD.74.023512} {\bibfield  {journal} {\bibinfo
  {journal} {Phys. Rev.}\ }\textbf {\bibinfo {volume} {D74}},\ \bibinfo {pages}
  {023512} (\bibinfo {year} {2006})},\ \Eprint
  {http://arxiv.org/abs/astro-ph/0503644} {arXiv:astro-ph/0503644 [astro-ph]}
  \BibitemShut {NoStop}%
%%CITATION = ASTRO-PH/0503644;%%
\bibitem [{\citenamefont {Koyama}(2006)}]{Koyama:2006ef}%
  \BibitemOpen
  \bibfield  {author} {\bibinfo {author} {\bibfnamefont {K.}~\bibnamefont
  {Koyama}},\ }\href {\doibase 10.1088/1475-7516/2006/03/017} {\bibfield
  {journal} {\bibinfo  {journal} {JCAP}\ }\textbf {\bibinfo {volume} {0603}},\
  \bibinfo {pages} {017} (\bibinfo {year} {2006})},\ \Eprint
  {http://arxiv.org/abs/astro-ph/0601220} {arXiv:astro-ph/0601220 [astro-ph]}
  \BibitemShut {NoStop}%
%%CITATION = ASTRO-PH/0601220;%%
\bibitem [{\citenamefont {Chiba}\ and\ \citenamefont
  {Takahashi}(2007)}]{Chiba:2007rb}%
  \BibitemOpen
  \bibfield  {author} {\bibinfo {author} {\bibfnamefont {T.}~\bibnamefont
  {Chiba}}\ and\ \bibinfo {author} {\bibfnamefont {R.}~\bibnamefont
  {Takahashi}},\ }\href {\doibase 10.1103/PhysRevD.75.101301} {\bibfield
  {journal} {\bibinfo  {journal} {Phys. Rev.}\ }\textbf {\bibinfo {volume}
  {D75}},\ \bibinfo {pages} {101301} (\bibinfo {year} {2007})},\ \Eprint
  {http://arxiv.org/abs/astro-ph/0703347} {arXiv:astro-ph/0703347 [astro-ph]}
  \BibitemShut {NoStop}%
%%CITATION = ASTRO-PH/0703347;%%
\bibitem [{\citenamefont {Amendola}\ \emph {et~al.}(2008)\citenamefont
  {Amendola}, \citenamefont {Kunz},\ and\ \citenamefont
  {Sapone}}]{Amendola:2007rr}%
  \BibitemOpen
  \bibfield  {author} {\bibinfo {author} {\bibfnamefont {L.}~\bibnamefont
  {Amendola}}, \bibinfo {author} {\bibfnamefont {M.}~\bibnamefont {Kunz}}, \
  and\ \bibinfo {author} {\bibfnamefont {D.}~\bibnamefont {Sapone}},\ }\href
  {\doibase 10.1088/1475-7516/2008/04/013} {\bibfield  {journal} {\bibinfo
  {journal} {JCAP}\ }\textbf {\bibinfo {volume} {0804}},\ \bibinfo {pages}
  {013} (\bibinfo {year} {2008})},\ \Eprint {http://arxiv.org/abs/0704.2421}
  {arXiv:0704.2421 [astro-ph]} \BibitemShut {NoStop}%
%%CITATION = ARXIV:0704.2421;%%
\bibitem [{\citenamefont {Simpson}\ \emph {et~al.}(2013)\citenamefont
  {Simpson}, \citenamefont {Heymans}, \citenamefont {Parkinson}, \citenamefont
  {Blake}, \citenamefont {Kilbinger} \emph {et~al.}}]{Simpson:2012ra}%
  \BibitemOpen
  \bibfield  {author} {\bibinfo {author} {\bibfnamefont {F.}~\bibnamefont
  {Simpson}}, \bibinfo {author} {\bibfnamefont {C.}~\bibnamefont {Heymans}},
  \bibinfo {author} {\bibfnamefont {D.}~\bibnamefont {Parkinson}}, \bibinfo
  {author} {\bibfnamefont {C.}~\bibnamefont {Blake}}, \bibinfo {author}
  {\bibfnamefont {M.}~\bibnamefont {Kilbinger}},  \emph {et~al.},\ }\href
  {\doibase 10.1093/mnras/sts493} {\bibfield  {journal} {\bibinfo  {journal}
  {Mon.Not.Roy.Astron.Soc.}\ }\textbf {\bibinfo {volume} {429}},\ \bibinfo
  {pages} {2249} (\bibinfo {year} {2013})},\ \Eprint
  {http://arxiv.org/abs/1212.3339} {arXiv:1212.3339 [astro-ph.CO]} \BibitemShut
  {NoStop}%
%%CITATION = ARXIV:1212.3339;%%
\bibitem [{\citenamefont {Terukina}\ and\ \citenamefont
  {Yamamoto}(2012)}]{Terukina:2012ji}%
  \BibitemOpen
  \bibfield  {author} {\bibinfo {author} {\bibfnamefont {A.}~\bibnamefont
  {Terukina}}\ and\ \bibinfo {author} {\bibfnamefont {K.}~\bibnamefont
  {Yamamoto}},\ }\href {\doibase 10.1103/PhysRevD.86.103503} {\bibfield
  {journal} {\bibinfo  {journal} {Phys.Rev.}\ }\textbf {\bibinfo {volume}
  {D86}},\ \bibinfo {pages} {103503} (\bibinfo {year} {2012})},\ \Eprint
  {http://arxiv.org/abs/1203.6163} {arXiv:1203.6163 [astro-ph.CO]} \BibitemShut
  {NoStop}%
%%CITATION = ARXIV:1203.6163;%%
\bibitem [{\citenamefont {Terukina}\ \emph {et~al.}(2014)\citenamefont
  {Terukina}, \citenamefont {Lombriser}, \citenamefont {Yamamoto},
  \citenamefont {Bacon}, \citenamefont {Koyama} \emph
  {et~al.}}]{Terukina:2013eqa}%
  \BibitemOpen
  \bibfield  {author} {\bibinfo {author} {\bibfnamefont {A.}~\bibnamefont
  {Terukina}}, \bibinfo {author} {\bibfnamefont {L.}~\bibnamefont {Lombriser}},
  \bibinfo {author} {\bibfnamefont {K.}~\bibnamefont {Yamamoto}}, \bibinfo
  {author} {\bibfnamefont {D.}~\bibnamefont {Bacon}}, \bibinfo {author}
  {\bibfnamefont {K.}~\bibnamefont {Koyama}},  \emph {et~al.},\ }\href
  {\doibase 10.1088/1475-7516/2014/04/013} {\bibfield  {journal} {\bibinfo
  {journal} {JCAP}\ }\textbf {\bibinfo {volume} {1404}},\ \bibinfo {pages}
  {013} (\bibinfo {year} {2014})},\ \Eprint {http://arxiv.org/abs/1312.5083}
  {arXiv:1312.5083 [astro-ph.CO]} \BibitemShut {NoStop}%
%%CITATION = ARXIV:1312.5083;%%
\bibitem [{\citenamefont {Yamamoto}\ \emph {et~al.}(2010)\citenamefont
  {Yamamoto}, \citenamefont {Nakamura}, \citenamefont {Hutsi}, \citenamefont
  {Narikawa},\ and\ \citenamefont {Sato}}]{Yamamoto:2010ie}%
  \BibitemOpen
  \bibfield  {author} {\bibinfo {author} {\bibfnamefont {K.}~\bibnamefont
  {Yamamoto}}, \bibinfo {author} {\bibfnamefont {G.}~\bibnamefont {Nakamura}},
  \bibinfo {author} {\bibfnamefont {G.}~\bibnamefont {Hutsi}}, \bibinfo
  {author} {\bibfnamefont {T.}~\bibnamefont {Narikawa}}, \ and\ \bibinfo
  {author} {\bibfnamefont {T.}~\bibnamefont {Sato}},\ }\href {\doibase
  10.1103/PhysRevD.81.103517} {\bibfield  {journal} {\bibinfo  {journal}
  {Phys.Rev.}\ }\textbf {\bibinfo {volume} {D81}},\ \bibinfo {pages} {103517}
  (\bibinfo {year} {2010})},\ \Eprint {http://arxiv.org/abs/1004.3231}
  {arXiv:1004.3231 [astro-ph.CO]} \BibitemShut {NoStop}%
%%CITATION = ARXIV:1004.3231;%%
\bibitem [{\citenamefont {Jain}\ and\ \citenamefont
  {Zhang}(2008)}]{Jain:2007yk}%
  \BibitemOpen
  \bibfield  {author} {\bibinfo {author} {\bibfnamefont {B.}~\bibnamefont
  {Jain}}\ and\ \bibinfo {author} {\bibfnamefont {P.}~\bibnamefont {Zhang}},\
  }\href {\doibase 10.1103/PhysRevD.78.063503} {\bibfield  {journal} {\bibinfo
  {journal} {Phys.Rev.}\ }\textbf {\bibinfo {volume} {D78}},\ \bibinfo {pages}
  {063503} (\bibinfo {year} {2008})},\ \Eprint {http://arxiv.org/abs/0709.2375}
  {arXiv:0709.2375 [astro-ph]} \BibitemShut {NoStop}%
%%CITATION = ARXIV:0709.2375;%%
\bibitem [{\citenamefont {Zhao}\ \emph
  {et~al.}(2009{\natexlab{a}})\citenamefont {Zhao}, \citenamefont {Pogosian},
  \citenamefont {Silvestri},\ and\ \citenamefont {Zylberberg}}]{Zhao:2008bn}%
  \BibitemOpen
  \bibfield  {author} {\bibinfo {author} {\bibfnamefont {G.-B.}\ \bibnamefont
  {Zhao}}, \bibinfo {author} {\bibfnamefont {L.}~\bibnamefont {Pogosian}},
  \bibinfo {author} {\bibfnamefont {A.}~\bibnamefont {Silvestri}}, \ and\
  \bibinfo {author} {\bibfnamefont {J.}~\bibnamefont {Zylberberg}},\ }\href
  {\doibase 10.1103/PhysRevD.79.083513} {\bibfield  {journal} {\bibinfo
  {journal} {Phys.Rev.}\ }\textbf {\bibinfo {volume} {D79}},\ \bibinfo {pages}
  {083513} (\bibinfo {year} {2009}{\natexlab{a}})},\ \Eprint
  {http://arxiv.org/abs/0809.3791} {arXiv:0809.3791 [astro-ph]} \BibitemShut
  {NoStop}%
%%CITATION = ARXIV:0809.3791;%%
\bibitem [{\citenamefont {Zhao}\ \emph
  {et~al.}(2009{\natexlab{b}})\citenamefont {Zhao}, \citenamefont {Pogosian},
  \citenamefont {Silvestri},\ and\ \citenamefont {Zylberberg}}]{Zhao:2009fn}%
  \BibitemOpen
  \bibfield  {author} {\bibinfo {author} {\bibfnamefont {G.-B.}\ \bibnamefont
  {Zhao}}, \bibinfo {author} {\bibfnamefont {L.}~\bibnamefont {Pogosian}},
  \bibinfo {author} {\bibfnamefont {A.}~\bibnamefont {Silvestri}}, \ and\
  \bibinfo {author} {\bibfnamefont {J.}~\bibnamefont {Zylberberg}},\ }\href
  {\doibase 10.1103/PhysRevLett.103.241301} {\bibfield  {journal} {\bibinfo
  {journal} {Phys.Rev.Lett.}\ }\textbf {\bibinfo {volume} {103}},\ \bibinfo
  {pages} {241301} (\bibinfo {year} {2009}{\natexlab{b}})},\ \Eprint
  {http://arxiv.org/abs/0905.1326} {arXiv:0905.1326 [astro-ph.CO]} \BibitemShut
  {NoStop}%
%%CITATION = ARXIV:0905.1326;%%
\bibitem [{\citenamefont {Asaba}\ \emph
  {et~al.}(2013{\natexlab{a}})\citenamefont {Asaba}, \citenamefont {Hikage},
  \citenamefont {Koyama}, \citenamefont {Zhao}, \citenamefont {Hojjati} \emph
  {et~al.}}]{Asaba:2013xql}%
  \BibitemOpen
  \bibfield  {author} {\bibinfo {author} {\bibfnamefont {S.}~\bibnamefont
  {Asaba}}, \bibinfo {author} {\bibfnamefont {C.}~\bibnamefont {Hikage}},
  \bibinfo {author} {\bibfnamefont {K.}~\bibnamefont {Koyama}}, \bibinfo
  {author} {\bibfnamefont {G.-B.}\ \bibnamefont {Zhao}}, \bibinfo {author}
  {\bibfnamefont {A.}~\bibnamefont {Hojjati}},  \emph {et~al.},\ }\href
  {\doibase 10.1088/1475-7516/2013/08/029} {\bibfield  {journal} {\bibinfo
  {journal} {JCAP}\ }\textbf {\bibinfo {volume} {1308}},\ \bibinfo {pages}
  {029} (\bibinfo {year} {2013}{\natexlab{a}})},\ \Eprint
  {http://arxiv.org/abs/1306.2546} {arXiv:1306.2546 [astro-ph.CO]} \BibitemShut
  {NoStop}%
%%CITATION = ARXIV:1306.2546;%%
\bibitem [{\citenamefont {Kaiser}(1987)}]{Kaiser:1987qv}%
  \BibitemOpen
  \bibfield  {author} {\bibinfo {author} {\bibfnamefont {N.}~\bibnamefont
  {Kaiser}},\ }\href@noop {} {\bibfield  {journal} {\bibinfo  {journal} {Mon.
  Not. Roy. Astron. Soc.}\ }\textbf {\bibinfo {volume} {227}},\ \bibinfo
  {pages} {1} (\bibinfo {year} {1987})}\BibitemShut {NoStop}%
%%CITATION = MNRAA,227,1;%%
\bibitem [{\citenamefont {Linder}(2008)}]{Linder:2007nu}%
  \BibitemOpen
  \bibfield  {author} {\bibinfo {author} {\bibfnamefont {E.~V.}\ \bibnamefont
  {Linder}},\ }\href {\doibase 10.1016/j.astropartphys.2008.03.002} {\bibfield
  {journal} {\bibinfo  {journal} {Astropart. Phys.}\ }\textbf {\bibinfo
  {volume} {29}},\ \bibinfo {pages} {336} (\bibinfo {year} {2008})},\ \Eprint
  {http://arxiv.org/abs/0709.1113} {arXiv:0709.1113 [astro-ph]} \BibitemShut
  {NoStop}%
%%CITATION = ARXIV:0709.1113;%%
\bibitem [{\citenamefont {Guzzo}\ \emph {et~al.}(2008)\citenamefont {Guzzo}
  \emph {et~al.}}]{Guzzo:2008ac}%
  \BibitemOpen
  \bibfield  {author} {\bibinfo {author} {\bibfnamefont {L.}~\bibnamefont
  {Guzzo}} \emph {et~al.},\ }\href {\doibase 10.1038/nature06555} {\bibfield
  {journal} {\bibinfo  {journal} {Nature}\ }\textbf {\bibinfo {volume} {451}},\
  \bibinfo {pages} {541} (\bibinfo {year} {2008})},\ \Eprint
  {http://arxiv.org/abs/0802.1944} {arXiv:0802.1944 [astro-ph]} \BibitemShut
  {NoStop}%
%%CITATION = ARXIV:0802.1944;%%
\bibitem [{\citenamefont {Yamamoto}\ \emph {et~al.}(2008)\citenamefont
  {Yamamoto}, \citenamefont {Sato},\ and\ \citenamefont
  {Huetsi}}]{Yamamoto:2008gr}%
  \BibitemOpen
  \bibfield  {author} {\bibinfo {author} {\bibfnamefont {K.}~\bibnamefont
  {Yamamoto}}, \bibinfo {author} {\bibfnamefont {T.}~\bibnamefont {Sato}}, \
  and\ \bibinfo {author} {\bibfnamefont {G.}~\bibnamefont {Huetsi}},\ }\href
  {\doibase 10.1143/PTP.120.609} {\bibfield  {journal} {\bibinfo  {journal}
  {Prog. Theor. Phys.}\ }\textbf {\bibinfo {volume} {120}},\ \bibinfo {pages}
  {609} (\bibinfo {year} {2008})},\ \Eprint {http://arxiv.org/abs/0805.4789}
  {arXiv:0805.4789 [astro-ph]} \BibitemShut {NoStop}%
%%CITATION = ARXIV:0805.4789;%%
\bibitem [{\citenamefont {Song}\ and\ \citenamefont
  {Percival}(2009)}]{Song:2008qt}%
  \BibitemOpen
  \bibfield  {author} {\bibinfo {author} {\bibfnamefont {Y.-S.}\ \bibnamefont
  {Song}}\ and\ \bibinfo {author} {\bibfnamefont {W.~J.}\ \bibnamefont
  {Percival}},\ }\href {\doibase 10.1088/1475-7516/2009/10/004} {\bibfield
  {journal} {\bibinfo  {journal} {JCAP}\ }\textbf {\bibinfo {volume} {0910}},\
  \bibinfo {pages} {004} (\bibinfo {year} {2009})},\ \Eprint
  {http://arxiv.org/abs/0807.0810} {arXiv:0807.0810 [astro-ph]} \BibitemShut
  {NoStop}%
%%CITATION = ARXIV:0807.0810;%%
\bibitem [{\citenamefont {Song}\ and\ \citenamefont
  {Kayo}(2010)}]{Song:2010bk}%
  \BibitemOpen
  \bibfield  {author} {\bibinfo {author} {\bibfnamefont {Y.-S.}\ \bibnamefont
  {Song}}\ and\ \bibinfo {author} {\bibfnamefont {I.}~\bibnamefont {Kayo}},\
  }\href {\doibase 10.1111/j.1365-2966.2010.16955.x} {\bibfield  {journal}
  {\bibinfo  {journal} {Mon. Not. Roy. Astron. Soc.}\ }\textbf {\bibinfo
  {volume} {407}},\ \bibinfo {pages} {1123} (\bibinfo {year} {2010})},\ \Eprint
  {http://arxiv.org/abs/1003.2420} {arXiv:1003.2420 [astro-ph.CO]} \BibitemShut
  {NoStop}%
%%CITATION = ARXIV:1003.2420;%%
\bibitem [{\citenamefont {Guzik}\ \emph {et~al.}(2010)\citenamefont {Guzik},
  \citenamefont {Jain},\ and\ \citenamefont {Takada}}]{Guzik:2009cm}%
  \BibitemOpen
  \bibfield  {author} {\bibinfo {author} {\bibfnamefont {J.}~\bibnamefont
  {Guzik}}, \bibinfo {author} {\bibfnamefont {B.}~\bibnamefont {Jain}}, \ and\
  \bibinfo {author} {\bibfnamefont {M.}~\bibnamefont {Takada}},\ }\href
  {\doibase 10.1103/PhysRevD.81.023503} {\bibfield  {journal} {\bibinfo
  {journal} {Phys. Rev.}\ }\textbf {\bibinfo {volume} {D81}},\ \bibinfo {pages}
  {023503} (\bibinfo {year} {2010})},\ \Eprint {http://arxiv.org/abs/0906.2221}
  {arXiv:0906.2221 [astro-ph.CO]} \BibitemShut {NoStop}%
%%CITATION = ARXIV:0906.2221;%%
\bibitem [{\citenamefont {Song}\ \emph {et~al.}(2011)\citenamefont {Song},
  \citenamefont {Zhao}, \citenamefont {Bacon}, \citenamefont {Koyama},
  \citenamefont {Nichol},\ and\ \citenamefont {Pogosian}}]{Song:2010fg}%
  \BibitemOpen
  \bibfield  {author} {\bibinfo {author} {\bibfnamefont {Y.-S.}\ \bibnamefont
  {Song}}, \bibinfo {author} {\bibfnamefont {G.-B.}\ \bibnamefont {Zhao}},
  \bibinfo {author} {\bibfnamefont {D.}~\bibnamefont {Bacon}}, \bibinfo
  {author} {\bibfnamefont {K.}~\bibnamefont {Koyama}}, \bibinfo {author}
  {\bibfnamefont {R.~C.}\ \bibnamefont {Nichol}}, \ and\ \bibinfo {author}
  {\bibfnamefont {L.}~\bibnamefont {Pogosian}},\ }\href {\doibase
  10.1103/PhysRevD.84.083523} {\bibfield  {journal} {\bibinfo  {journal} {Phys.
  Rev.}\ }\textbf {\bibinfo {volume} {D84}},\ \bibinfo {pages} {083523}
  (\bibinfo {year} {2011})},\ \Eprint {http://arxiv.org/abs/1011.2106}
  {arXiv:1011.2106 [astro-ph.CO]} \BibitemShut {NoStop}%
%%CITATION = ARXIV:1011.2106;%%
\bibitem [{\citenamefont {Asaba}\ \emph
  {et~al.}(2013{\natexlab{b}})\citenamefont {Asaba}, \citenamefont {Hikage},
  \citenamefont {Koyama}, \citenamefont {Zhao}, \citenamefont {Hojjati},\ and\
  \citenamefont {Pogosian}}]{Asaba:2013mxj}%
  \BibitemOpen
  \bibfield  {author} {\bibinfo {author} {\bibfnamefont {S.}~\bibnamefont
  {Asaba}}, \bibinfo {author} {\bibfnamefont {C.}~\bibnamefont {Hikage}},
  \bibinfo {author} {\bibfnamefont {K.}~\bibnamefont {Koyama}}, \bibinfo
  {author} {\bibfnamefont {G.-B.}\ \bibnamefont {Zhao}}, \bibinfo {author}
  {\bibfnamefont {A.}~\bibnamefont {Hojjati}}, \ and\ \bibinfo {author}
  {\bibfnamefont {L.}~\bibnamefont {Pogosian}},\ }\href {\doibase
  10.1088/1475-7516/2013/08/029} {\bibfield  {journal} {\bibinfo  {journal}
  {JCAP}\ }\textbf {\bibinfo {volume} {1308}},\ \bibinfo {pages} {029}
  (\bibinfo {year} {2013}{\natexlab{b}})},\ \Eprint
  {http://arxiv.org/abs/1306.2546} {arXiv:1306.2546 [astro-ph.CO]} \BibitemShut
  {NoStop}%
%%CITATION = ARXIV:1306.2546;%%
\bibitem [{\citenamefont {Hellwing}\ \emph {et~al.}(2014)\citenamefont
  {Hellwing}, \citenamefont {Barreira}, \citenamefont {Frenk}, \citenamefont
  {Li},\ and\ \citenamefont {Cole}}]{Hellwing:2014nma}%
  \BibitemOpen
  \bibfield  {author} {\bibinfo {author} {\bibfnamefont {W.~A.}\ \bibnamefont
  {Hellwing}}, \bibinfo {author} {\bibfnamefont {A.}~\bibnamefont {Barreira}},
  \bibinfo {author} {\bibfnamefont {C.~S.}\ \bibnamefont {Frenk}}, \bibinfo
  {author} {\bibfnamefont {B.}~\bibnamefont {Li}}, \ and\ \bibinfo {author}
  {\bibfnamefont {S.}~\bibnamefont {Cole}},\ }\href {\doibase
  10.1103/PhysRevLett.112.221102} {\bibfield  {journal} {\bibinfo  {journal}
  {Phys. Rev. Lett.}\ }\textbf {\bibinfo {volume} {112}},\ \bibinfo {pages}
  {221102} (\bibinfo {year} {2014})},\ \Eprint {http://arxiv.org/abs/1401.0706}
  {arXiv:1401.0706 [astro-ph.CO]} \BibitemShut {NoStop}%
%%CITATION = ARXIV:1401.0706;%%
\bibitem [{\citenamefont {Peebles}(1980)}]{Peebles1980}%
  \BibitemOpen
  \bibfield  {author} {\bibinfo {author} {\bibfnamefont {P.~J.~E.}\
  \bibnamefont {Peebles}},\ }\href@noop {} {\emph {\bibinfo {title} {{The
  large-scale structure of the universe}}}}\ (\bibinfo  {publisher} {Research
  supported by the National Science Foundation.~Princeton, N.J., Princeton
  University Press, 1980.~435 p.},\ \bibinfo {year} {1980})\BibitemShut
  {NoStop}%
\bibitem [{\citenamefont {Oka}\ \emph {et~al.}(2014)\citenamefont {Oka},
  \citenamefont {Saito}, \citenamefont {Nishimichi}, \citenamefont {Taruya},\
  and\ \citenamefont {Yamamoto}}]{Oka:2013cba}%
  \BibitemOpen
  \bibfield  {author} {\bibinfo {author} {\bibfnamefont {A.}~\bibnamefont
  {Oka}}, \bibinfo {author} {\bibfnamefont {S.}~\bibnamefont {Saito}}, \bibinfo
  {author} {\bibfnamefont {T.}~\bibnamefont {Nishimichi}}, \bibinfo {author}
  {\bibfnamefont {A.}~\bibnamefont {Taruya}}, \ and\ \bibinfo {author}
  {\bibfnamefont {K.}~\bibnamefont {Yamamoto}},\ }\href {\doibase
  10.1093/mnras/stu111} {\bibfield  {journal} {\bibinfo  {journal} {Mon. Not.
  Roy. Astron. Soc.}\ }\textbf {\bibinfo {volume} {439}},\ \bibinfo {pages}
  {2515} (\bibinfo {year} {2014})},\ \Eprint {http://arxiv.org/abs/1310.2820}
  {arXiv:1310.2820 [astro-ph.CO]} \BibitemShut {NoStop}%
%%CITATION = ARXIV:1310.2820;%%
\bibitem [{\citenamefont {Samushia}\ \emph {et~al.}(2014)\citenamefont
  {Samushia} \emph {et~al.}}]{Samushia:2013yga}%
  \BibitemOpen
  \bibfield  {author} {\bibinfo {author} {\bibfnamefont {L.}~\bibnamefont
  {Samushia}} \emph {et~al.},\ }\href {\doibase 10.1093/mnras/stu197}
  {\bibfield  {journal} {\bibinfo  {journal} {Mon. Not. Roy. Astron. Soc.}\
  }\textbf {\bibinfo {volume} {439}},\ \bibinfo {pages} {3504} (\bibinfo {year}
  {2014})},\ \Eprint {http://arxiv.org/abs/1312.4899} {arXiv:1312.4899
  [astro-ph.CO]} \BibitemShut {NoStop}%
%%CITATION = ARXIV:1312.4899;%%
\bibitem [{\citenamefont {Sanchez}\ \emph {et~al.}(2013)\citenamefont {Sanchez}
  \emph {et~al.}}]{Sanchez:2013uxa}%
  \BibitemOpen
  \bibfield  {author} {\bibinfo {author} {\bibfnamefont {A.~G.}\ \bibnamefont
  {Sanchez}} \emph {et~al.},\ }\href {\doibase 10.1093/mnras/stt799} {\bibfield
   {journal} {\bibinfo  {journal} {Mon. Not. Roy. Astron. Soc.}\ }\textbf
  {\bibinfo {volume} {433}},\ \bibinfo {pages} {1202} (\bibinfo {year}
  {2013})},\ \Eprint {http://arxiv.org/abs/1303.4396} {arXiv:1303.4396
  [astro-ph.CO]} \BibitemShut {NoStop}%
%%CITATION = ARXIV:1303.4396;%%
\bibitem [{\citenamefont {Sanchez}\ \emph {et~al.}(2014)\citenamefont {Sanchez}
  \emph {et~al.}}]{Sanchez:2013tga}%
  \BibitemOpen
  \bibfield  {author} {\bibinfo {author} {\bibfnamefont {A.~G.}\ \bibnamefont
  {Sanchez}} \emph {et~al.},\ }\href {\doibase 10.1093/mnras/stu342} {\bibfield
   {journal} {\bibinfo  {journal} {Mon. Not. Roy. Astron. Soc.}\ }\textbf
  {\bibinfo {volume} {440}},\ \bibinfo {pages} {2692} (\bibinfo {year}
  {2014})},\ \Eprint {http://arxiv.org/abs/1312.4854} {arXiv:1312.4854
  [astro-ph.CO]} \BibitemShut {NoStop}%
%%CITATION = ARXIV:1312.4854;%%
\bibitem [{\citenamefont {Gil-Mar'n}\ \emph {et~al.}(2016)\citenamefont
  {Gil-Mar'n} \emph {et~al.}}]{Gil-Marin:2015sqa}%
  \BibitemOpen
  \bibfield  {author} {\bibinfo {author} {\bibfnamefont {H.}~\bibnamefont
  {Gil-Mar'n}} \emph {et~al.},\ }\href {\doibase 10.1093/mnras/stw1096}
  {\bibfield  {journal} {\bibinfo  {journal} {Mon. Not. Roy. Astron. Soc.}\
  }\textbf {\bibinfo {volume} {460}},\ \bibinfo {pages} {4188} (\bibinfo {year}
  {2016})},\ \Eprint {http://arxiv.org/abs/1509.06386} {arXiv:1509.06386
  [astro-ph.CO]} \BibitemShut {NoStop}%
%%CITATION = ARXIV:1509.06386;%%
\bibitem [{\citenamefont {Sanchez}\ \emph {et~al.}(2016)\citenamefont {Sanchez}
  \emph {et~al.}}]{Sanchez:2016sas}%
  \BibitemOpen
  \bibfield  {author} {\bibinfo {author} {\bibfnamefont {A.~G.}\ \bibnamefont
  {Sanchez}} \emph {et~al.} (\bibinfo {collaboration} {BOSS}),\ }\href
  {\doibase 10.1093/mnras/stw2443} {\bibfield  {journal} {\bibinfo  {journal}
  {Mon. Not. Roy. Astron. Soc.}\ } (\bibinfo {year} {2016}),\
  10.1093/mnras/stw2443},\ \Eprint {http://arxiv.org/abs/1607.03147}
  {arXiv:1607.03147 [astro-ph.CO]} \BibitemShut {NoStop}%
%%CITATION = ARXIV:1607.03147;%%
\bibitem [{\citenamefont {de~la Torre}\ \emph {et~al.}(2013)\citenamefont
  {de~la Torre} \emph {et~al.}}]{delaTorre:2013rpa}%
  \BibitemOpen
  \bibfield  {author} {\bibinfo {author} {\bibfnamefont {S.}~\bibnamefont
  {de~la Torre}} \emph {et~al.},\ }\href {\doibase 10.1051/0004-6361/201321463}
  {\bibfield  {journal} {\bibinfo  {journal} {Astron. Astrophys.}\ }\textbf
  {\bibinfo {volume} {557}},\ \bibinfo {pages} {A54} (\bibinfo {year}
  {2013})},\ \Eprint {http://arxiv.org/abs/1303.2622} {arXiv:1303.2622
  [astro-ph.CO]} \BibitemShut {NoStop}%
%%CITATION = ARXIV:1303.2622;%%
\bibitem [{\citenamefont {Taruya}\ \emph {et~al.}(2010)\citenamefont {Taruya},
  \citenamefont {Nishimichi},\ and\ \citenamefont {Saito}}]{Taruya:2010mx}%
  \BibitemOpen
  \bibfield  {author} {\bibinfo {author} {\bibfnamefont {A.}~\bibnamefont
  {Taruya}}, \bibinfo {author} {\bibfnamefont {T.}~\bibnamefont {Nishimichi}},
  \ and\ \bibinfo {author} {\bibfnamefont {S.}~\bibnamefont {Saito}},\ }\href
  {\doibase 10.1103/PhysRevD.82.063522} {\bibfield  {journal} {\bibinfo
  {journal} {Phys.Rev.}\ }\textbf {\bibinfo {volume} {D82}},\ \bibinfo {pages}
  {063522} (\bibinfo {year} {2010})},\ \Eprint {http://arxiv.org/abs/1006.0699}
  {arXiv:1006.0699 [astro-ph.CO]} \BibitemShut {NoStop}%
%%CITATION = ARXIV:1006.0699;%%
\bibitem [{\citenamefont {Beutler}\ \emph {et~al.}(2016)\citenamefont {Beutler}
  \emph {et~al.}}]{Beutler:2016arn}%
  \BibitemOpen
  \bibfield  {author} {\bibinfo {author} {\bibfnamefont {F.}~\bibnamefont
  {Beutler}} \emph {et~al.} (\bibinfo {collaboration} {BOSS}),\ }\href@noop {}
  {\bibfield  {journal} {\bibinfo  {journal} {Submitted to: Mon. Not. Roy.
  Astron. Soc.}\ } (\bibinfo {year} {2016})},\ \Eprint
  {http://arxiv.org/abs/1607.03150} {arXiv:1607.03150 [astro-ph.CO]}
  \BibitemShut {NoStop}%
%%CITATION = ARXIV:1607.03150;%%
\bibitem [{\citenamefont {Taruya}\ \emph {et~al.}(2013)\citenamefont {Taruya},
  \citenamefont {Nishimichi},\ and\ \citenamefont
  {Bernardeau}}]{Taruya:2013my}%
  \BibitemOpen
  \bibfield  {author} {\bibinfo {author} {\bibfnamefont {A.}~\bibnamefont
  {Taruya}}, \bibinfo {author} {\bibfnamefont {T.}~\bibnamefont {Nishimichi}},
  \ and\ \bibinfo {author} {\bibfnamefont {F.}~\bibnamefont {Bernardeau}},\
  }\href {\doibase 10.1103/PhysRevD.87.083509} {\bibfield  {journal} {\bibinfo
  {journal} {Phys. Rev.}\ }\textbf {\bibinfo {volume} {D87}},\ \bibinfo {pages}
  {083509} (\bibinfo {year} {2013})},\ \Eprint {http://arxiv.org/abs/1301.3624}
  {arXiv:1301.3624 [astro-ph.CO]} \BibitemShut {NoStop}%
%%CITATION = ARXIV:1301.3624;%%
\bibitem [{\citenamefont {Barreira}\ \emph {et~al.}(2016)\citenamefont
  {Barreira}, \citenamefont {Sanchez},\ and\ \citenamefont
  {Schmidt}}]{Barreira:2016mg}%
  \BibitemOpen
  \bibfield  {author} {\bibinfo {author} {\bibfnamefont {A.}~\bibnamefont
  {Barreira}}, \bibinfo {author} {\bibfnamefont {A.~G.}\ \bibnamefont
  {Sanchez}}, \ and\ \bibinfo {author} {\bibfnamefont {F.}~\bibnamefont
  {Schmidt}},\ }\href@noop {} {\  (\bibinfo {year} {2016})},\ \Eprint
  {http://arxiv.org/abs/1605.03965} {arXiv:1605.03965 [astro-ph.CO]}
  \BibitemShut {NoStop}%
%%CITATION = ARXIV:1605.03965;%%
\bibitem [{\citenamefont {Bose}\ and\ \citenamefont
  {Koyama}(2016)}]{Bose:2016qun}%
  \BibitemOpen
  \bibfield  {author} {\bibinfo {author} {\bibfnamefont {B.}~\bibnamefont
  {Bose}}\ and\ \bibinfo {author} {\bibfnamefont {K.}~\bibnamefont {Koyama}},\
  }\href {\doibase 10.1088/1475-7516/2016/08/032} {\bibfield  {journal}
  {\bibinfo  {journal} {JCAP}\ }\textbf {\bibinfo {volume} {1608}},\ \bibinfo
  {pages} {032} (\bibinfo {year} {2016})},\ \Eprint
  {http://arxiv.org/abs/1606.02520} {arXiv:1606.02520 [astro-ph.CO]}
  \BibitemShut {NoStop}%
%%CITATION = ARXIV:1606.02520;%%
\bibitem [{\citenamefont {Bernardeau}\ \emph {et~al.}(2002)\citenamefont
  {Bernardeau}, \citenamefont {Colombi}, \citenamefont {Gaztanaga},\ and\
  \citenamefont {Scoccimarro}}]{Bernardeau:2001qr}%
  \BibitemOpen
  \bibfield  {author} {\bibinfo {author} {\bibfnamefont {F.}~\bibnamefont
  {Bernardeau}}, \bibinfo {author} {\bibfnamefont {S.}~\bibnamefont {Colombi}},
  \bibinfo {author} {\bibfnamefont {E.}~\bibnamefont {Gaztanaga}}, \ and\
  \bibinfo {author} {\bibfnamefont {R.}~\bibnamefont {Scoccimarro}},\ }\href
  {\doibase 10.1016/S0370-1573(02)00135-7} {\bibfield  {journal} {\bibinfo
  {journal} {Phys. Rept.}\ }\textbf {\bibinfo {volume} {367}},\ \bibinfo
  {pages} {1} (\bibinfo {year} {2002})},\ \Eprint
  {http://arxiv.org/abs/astro-ph/0112551} {arXiv:astro-ph/0112551 [astro-ph]}
  \BibitemShut {NoStop}%
%%CITATION = ASTRO-PH/0112551;%%
\bibitem [{\citenamefont {Carlson}\ \emph {et~al.}(2009)\citenamefont
  {Carlson}, \citenamefont {White},\ and\ \citenamefont
  {Padmanabhan}}]{Carlson:2009it}%
  \BibitemOpen
  \bibfield  {author} {\bibinfo {author} {\bibfnamefont {J.}~\bibnamefont
  {Carlson}}, \bibinfo {author} {\bibfnamefont {M.}~\bibnamefont {White}}, \
  and\ \bibinfo {author} {\bibfnamefont {N.}~\bibnamefont {Padmanabhan}},\
  }\href {\doibase 10.1103/PhysRevD.80.043531} {\bibfield  {journal} {\bibinfo
  {journal} {Phys. Rev.}\ }\textbf {\bibinfo {volume} {D80}},\ \bibinfo {pages}
  {043531} (\bibinfo {year} {2009})},\ \Eprint {http://arxiv.org/abs/0905.0479}
  {arXiv:0905.0479 [astro-ph.CO]} \BibitemShut {NoStop}%
%%CITATION = ARXIV:0905.0479;%%
\bibitem [{\citenamefont {Winther}\ \emph {et~al.}(2017)\citenamefont
  {Winther}, \citenamefont {Koyama}, \citenamefont {Manera}, \citenamefont
  {Wright},\ and\ \citenamefont {Zhao}}]{Winther:2017jof}%
  \BibitemOpen
  \bibfield  {author} {\bibinfo {author} {\bibfnamefont {H.~A.}\ \bibnamefont
  {Winther}}, \bibinfo {author} {\bibfnamefont {K.}~\bibnamefont {Koyama}},
  \bibinfo {author} {\bibfnamefont {M.}~\bibnamefont {Manera}}, \bibinfo
  {author} {\bibfnamefont {B.~S.}\ \bibnamefont {Wright}}, \ and\ \bibinfo
  {author} {\bibfnamefont {G.-B.}\ \bibnamefont {Zhao}},\ }\href@noop {} {\
  (\bibinfo {year} {2017})},\ \Eprint {http://arxiv.org/abs/1703.00879}
  {arXiv:1703.00879 [astro-ph.CO]} \BibitemShut {NoStop}%
%%CITATION = ARXIV:1703.00879;%%
\bibitem [{\citenamefont {Pichon}\ and\ \citenamefont
  {Bernardeau}(1999)}]{Pichon:1999tk}%
  \BibitemOpen
  \bibfield  {author} {\bibinfo {author} {\bibfnamefont {C.}~\bibnamefont
  {Pichon}}\ and\ \bibinfo {author} {\bibfnamefont {F.}~\bibnamefont
  {Bernardeau}},\ }\href@noop {} {\bibfield  {journal} {\bibinfo  {journal}
  {Astron. Astrophys.}\ }\textbf {\bibinfo {volume} {343}},\ \bibinfo {pages}
  {663} (\bibinfo {year} {1999})},\ \Eprint
  {http://arxiv.org/abs/astro-ph/9902142} {arXiv:astro-ph/9902142 [astro-ph]}
  \BibitemShut {NoStop}%
%%CITATION = ASTRO-PH/9902142;%%
\bibitem [{\citenamefont {Pueblas}\ and\ \citenamefont
  {Scoccimarro}(2009)}]{Pueblas:2008uv}%
  \BibitemOpen
  \bibfield  {author} {\bibinfo {author} {\bibfnamefont {S.}~\bibnamefont
  {Pueblas}}\ and\ \bibinfo {author} {\bibfnamefont {R.}~\bibnamefont
  {Scoccimarro}},\ }\href {\doibase 10.1103/PhysRevD.80.043504} {\bibfield
  {journal} {\bibinfo  {journal} {Phys. Rev.}\ }\textbf {\bibinfo {volume}
  {D80}},\ \bibinfo {pages} {043504} (\bibinfo {year} {2009})},\ \Eprint
  {http://arxiv.org/abs/0809.4606} {arXiv:0809.4606 [astro-ph]} \BibitemShut
  {NoStop}%
%%CITATION = ARXIV:0809.4606;%%
\bibitem [{\citenamefont {Libeskind}\ \emph {et~al.}(2013)\citenamefont
  {Libeskind}, \citenamefont {Hoffman}, \citenamefont {Steinmetz},
  \citenamefont {Gottloeber}, \citenamefont {Knebe},\ and\ \citenamefont
  {Hess}}]{Libeskind:2012ya}%
  \BibitemOpen
  \bibfield  {author} {\bibinfo {author} {\bibfnamefont {N.~I.}\ \bibnamefont
  {Libeskind}}, \bibinfo {author} {\bibfnamefont {Y.}~\bibnamefont {Hoffman}},
  \bibinfo {author} {\bibfnamefont {M.}~\bibnamefont {Steinmetz}}, \bibinfo
  {author} {\bibfnamefont {S.}~\bibnamefont {Gottloeber}}, \bibinfo {author}
  {\bibfnamefont {A.}~\bibnamefont {Knebe}}, \ and\ \bibinfo {author}
  {\bibfnamefont {S.}~\bibnamefont {Hess}},\ }\href {\doibase
  10.1088/2041-8205/766/2/L15} {\bibfield  {journal} {\bibinfo  {journal}
  {Astrophys. J.}\ }\textbf {\bibinfo {volume} {766}},\ \bibinfo {pages} {L15}
  (\bibinfo {year} {2013})},\ \Eprint {http://arxiv.org/abs/1212.1454}
  {arXiv:1212.1454 [astro-ph.CO]} \BibitemShut {NoStop}%
%%CITATION = ARXIV:1212.1454;%%
\bibitem [{\citenamefont {Koyama}\ \emph {et~al.}(2009)\citenamefont {Koyama},
  \citenamefont {Taruya},\ and\ \citenamefont {Hiramatsu}}]{Koyama:2009me}%
  \BibitemOpen
  \bibfield  {author} {\bibinfo {author} {\bibfnamefont {K.}~\bibnamefont
  {Koyama}}, \bibinfo {author} {\bibfnamefont {A.}~\bibnamefont {Taruya}}, \
  and\ \bibinfo {author} {\bibfnamefont {T.}~\bibnamefont {Hiramatsu}},\ }\href
  {\doibase 10.1103/PhysRevD.79.123512} {\bibfield  {journal} {\bibinfo
  {journal} {Phys.Rev.}\ }\textbf {\bibinfo {volume} {D79}},\ \bibinfo {pages}
  {123512} (\bibinfo {year} {2009})},\ \Eprint {http://arxiv.org/abs/0902.0618}
  {arXiv:0902.0618 [astro-ph.CO]} \BibitemShut {NoStop}%
%%CITATION = ARXIV:0902.0618;%%
\bibitem [{\citenamefont {Taruya}(2016)}]{Taruya:2016jdt}%
  \BibitemOpen
  \bibfield  {author} {\bibinfo {author} {\bibfnamefont {A.}~\bibnamefont
  {Taruya}},\ }\href {\doibase 10.1103/PhysRevD.94.023504} {\bibfield
  {journal} {\bibinfo  {journal} {Phys. Rev.}\ }\textbf {\bibinfo {volume}
  {D94}},\ \bibinfo {pages} {023504} (\bibinfo {year} {2016})},\ \Eprint
  {http://arxiv.org/abs/1606.02168} {arXiv:1606.02168 [astro-ph.CO]}
  \BibitemShut {NoStop}%
%%CITATION = ARXIV:1606.02168;%%
\bibitem [{\citenamefont {Jeong}\ and\ \citenamefont
  {Komatsu}(2006)}]{Jeong:2006xd}%
  \BibitemOpen
  \bibfield  {author} {\bibinfo {author} {\bibfnamefont {D.}~\bibnamefont
  {Jeong}}\ and\ \bibinfo {author} {\bibfnamefont {E.}~\bibnamefont
  {Komatsu}},\ }\href {\doibase 10.1086/507781} {\bibfield  {journal} {\bibinfo
   {journal} {Astrophys. J.}\ }\textbf {\bibinfo {volume} {651}},\ \bibinfo
  {pages} {619} (\bibinfo {year} {2006})},\ \Eprint
  {http://arxiv.org/abs/astro-ph/0604075} {arXiv:astro-ph/0604075 [astro-ph]}
  \BibitemShut {NoStop}%
%%CITATION = ASTRO-PH/0604075;%%
\bibitem [{\citenamefont {Taruya}\ and\ \citenamefont
  {Hiramatsu}(2008)}]{Taruya:2007xy}%
  \BibitemOpen
  \bibfield  {author} {\bibinfo {author} {\bibfnamefont {A.}~\bibnamefont
  {Taruya}}\ and\ \bibinfo {author} {\bibfnamefont {T.}~\bibnamefont
  {Hiramatsu}},\ }\href {\doibase 10.1086/526515} {\bibfield  {journal}
  {\bibinfo  {journal} {Astrophys. J.}\ }\textbf {\bibinfo {volume} {674}},\
  \bibinfo {pages} {617} (\bibinfo {year} {2008})},\ \Eprint
  {http://arxiv.org/abs/0708.1367} {arXiv:0708.1367 [astro-ph]} \BibitemShut
  {NoStop}%
%%CITATION = ARXIV:0708.1367;%%
\bibitem [{\citenamefont {Matarrese}\ and\ \citenamefont
  {Pietroni}(2007)}]{Matarrese:2007wc}%
  \BibitemOpen
  \bibfield  {author} {\bibinfo {author} {\bibfnamefont {S.}~\bibnamefont
  {Matarrese}}\ and\ \bibinfo {author} {\bibfnamefont {M.}~\bibnamefont
  {Pietroni}},\ }\href {\doibase 10.1088/1475-7516/2007/06/026} {\bibfield
  {journal} {\bibinfo  {journal} {JCAP}\ }\textbf {\bibinfo {volume} {0706}},\
  \bibinfo {pages} {026} (\bibinfo {year} {2007})},\ \Eprint
  {http://arxiv.org/abs/astro-ph/0703563} {arXiv:astro-ph/0703563 [astro-ph]}
  \BibitemShut {NoStop}%
%%CITATION = ASTRO-PH/0703563;%%
\bibitem [{\citenamefont {Valageas}(2007)}]{Valageas:2006bi}%
  \BibitemOpen
  \bibfield  {author} {\bibinfo {author} {\bibfnamefont {P.}~\bibnamefont
  {Valageas}},\ }\href {\doibase 10.1051/0004-6361:20066832} {\bibfield
  {journal} {\bibinfo  {journal} {Astron. Astrophys.}\ }\textbf {\bibinfo
  {volume} {465}},\ \bibinfo {pages} {725} (\bibinfo {year} {2007})},\ \Eprint
  {http://arxiv.org/abs/astro-ph/0611849} {arXiv:astro-ph/0611849 [astro-ph]}
  \BibitemShut {NoStop}%
%%CITATION = ASTRO-PH/0611849;%%
\bibitem [{\citenamefont {Matsubara}(2008)}]{Matsubara:2007wj}%
  \BibitemOpen
  \bibfield  {author} {\bibinfo {author} {\bibfnamefont {T.}~\bibnamefont
  {Matsubara}},\ }\href {\doibase 10.1103/PhysRevD.77.063530} {\bibfield
  {journal} {\bibinfo  {journal} {Phys. Rev.}\ }\textbf {\bibinfo {volume}
  {D77}},\ \bibinfo {pages} {063530} (\bibinfo {year} {2008})},\ \Eprint
  {http://arxiv.org/abs/0711.2521} {arXiv:0711.2521 [astro-ph]} \BibitemShut
  {NoStop}%
%%CITATION = ARXIV:0711.2521;%%
\bibitem [{\citenamefont {Pietroni}(2008)}]{Pietroni:2008jx}%
  \BibitemOpen
  \bibfield  {author} {\bibinfo {author} {\bibfnamefont {M.}~\bibnamefont
  {Pietroni}},\ }\href {\doibase 10.1088/1475-7516/2008/10/036} {\bibfield
  {journal} {\bibinfo  {journal} {JCAP}\ }\textbf {\bibinfo {volume} {0810}},\
  \bibinfo {pages} {036} (\bibinfo {year} {2008})},\ \Eprint
  {http://arxiv.org/abs/0806.0971} {arXiv:0806.0971 [astro-ph]} \BibitemShut
  {NoStop}%
%%CITATION = ARXIV:0806.0971;%%
\bibitem [{\citenamefont {Anselmi}\ and\ \citenamefont
  {Pietroni}(2012)}]{Anselmi:2012cn}%
  \BibitemOpen
  \bibfield  {author} {\bibinfo {author} {\bibfnamefont {S.}~\bibnamefont
  {Anselmi}}\ and\ \bibinfo {author} {\bibfnamefont {M.}~\bibnamefont
  {Pietroni}},\ }\href {\doibase 10.1088/1475-7516/2012/12/013} {\bibfield
  {journal} {\bibinfo  {journal} {JCAP}\ }\textbf {\bibinfo {volume} {1212}},\
  \bibinfo {pages} {013} (\bibinfo {year} {2012})},\ \Eprint
  {http://arxiv.org/abs/1205.2235} {arXiv:1205.2235 [astro-ph.CO]} \BibitemShut
  {NoStop}%
%%CITATION = ARXIV:1205.2235;%%
\bibitem [{\citenamefont {Blas}\ \emph {et~al.}(2016)\citenamefont {Blas},
  \citenamefont {Garny}, \citenamefont {Ivanov},\ and\ \citenamefont
  {Sibiryakov}}]{Blas:2015qsi}%
  \BibitemOpen
  \bibfield  {author} {\bibinfo {author} {\bibfnamefont {D.}~\bibnamefont
  {Blas}}, \bibinfo {author} {\bibfnamefont {M.}~\bibnamefont {Garny}},
  \bibinfo {author} {\bibfnamefont {M.~M.}\ \bibnamefont {Ivanov}}, \ and\
  \bibinfo {author} {\bibfnamefont {S.}~\bibnamefont {Sibiryakov}},\ }\href
  {\doibase 10.1088/1475-7516/2016/07/052} {\bibfield  {journal} {\bibinfo
  {journal} {JCAP}\ }\textbf {\bibinfo {volume} {1607}},\ \bibinfo {pages}
  {052} (\bibinfo {year} {2016})},\ \Eprint {http://arxiv.org/abs/1512.05807}
  {arXiv:1512.05807 [astro-ph.CO]} \BibitemShut {NoStop}%
%%CITATION = ARXIV:1512.05807;%%
\bibitem [{\citenamefont {Baumann}\ \emph {et~al.}(2012)\citenamefont
  {Baumann}, \citenamefont {Nicolis}, \citenamefont {Senatore},\ and\
  \citenamefont {Zaldarriaga}}]{Baumann:2010tm}%
  \BibitemOpen
  \bibfield  {author} {\bibinfo {author} {\bibfnamefont {D.}~\bibnamefont
  {Baumann}}, \bibinfo {author} {\bibfnamefont {A.}~\bibnamefont {Nicolis}},
  \bibinfo {author} {\bibfnamefont {L.}~\bibnamefont {Senatore}}, \ and\
  \bibinfo {author} {\bibfnamefont {M.}~\bibnamefont {Zaldarriaga}},\ }\href
  {\doibase 10.1088/1475-7516/2012/07/051} {\bibfield  {journal} {\bibinfo
  {journal} {JCAP}\ }\textbf {\bibinfo {volume} {1207}},\ \bibinfo {pages}
  {051} (\bibinfo {year} {2012})},\ \Eprint {http://arxiv.org/abs/1004.2488}
  {arXiv:1004.2488 [astro-ph.CO]} \BibitemShut {NoStop}%
%%CITATION = ARXIV:1004.2488;%%
\bibitem [{\citenamefont {Carrasco}\ \emph {et~al.}(2012)\citenamefont
  {Carrasco}, \citenamefont {Hertzberg},\ and\ \citenamefont
  {Senatore}}]{Carrasco:2012cv}%
  \BibitemOpen
  \bibfield  {author} {\bibinfo {author} {\bibfnamefont {J.~J.~M.}\
  \bibnamefont {Carrasco}}, \bibinfo {author} {\bibfnamefont {M.~P.}\
  \bibnamefont {Hertzberg}}, \ and\ \bibinfo {author} {\bibfnamefont
  {L.}~\bibnamefont {Senatore}},\ }\href {\doibase 10.1007/JHEP09(2012)082}
  {\bibfield  {journal} {\bibinfo  {journal} {JHEP}\ }\textbf {\bibinfo
  {volume} {09}},\ \bibinfo {pages} {082} (\bibinfo {year} {2012})},\ \Eprint
  {http://arxiv.org/abs/1206.2926} {arXiv:1206.2926 [astro-ph.CO]} \BibitemShut
  {NoStop}%
%%CITATION = ARXIV:1206.2926;%%
\bibitem [{\citenamefont {Scoccimarro}(2004)}]{Scoccimarro:2004tg}%
  \BibitemOpen
  \bibfield  {author} {\bibinfo {author} {\bibfnamefont {R.}~\bibnamefont
  {Scoccimarro}},\ }\href {\doibase 10.1103/PhysRevD.70.083007} {\bibfield
  {journal} {\bibinfo  {journal} {Phys.Rev.}\ }\textbf {\bibinfo {volume}
  {D70}},\ \bibinfo {pages} {083007} (\bibinfo {year} {2004})},\ \Eprint
  {http://arxiv.org/abs/astro-ph/0407214} {arXiv:astro-ph/0407214 [astro-ph]}
  \BibitemShut {NoStop}%
%%CITATION = ASTRO-PH/0407214;%%
\bibitem [{\citenamefont {Percival}\ and\ \citenamefont
  {White}(2009)}]{Percival:2008sh}%
  \BibitemOpen
  \bibfield  {author} {\bibinfo {author} {\bibfnamefont {W.~J.}\ \bibnamefont
  {Percival}}\ and\ \bibinfo {author} {\bibfnamefont {M.}~\bibnamefont
  {White}},\ }\href {\doibase 10.1111/j.1365-2966.2008.14211.x} {\bibfield
  {journal} {\bibinfo  {journal} {Mon. Not. Roy. Astron. Soc.}\ }\textbf
  {\bibinfo {volume} {393}},\ \bibinfo {pages} {297} (\bibinfo {year}
  {2009})},\ \Eprint {http://arxiv.org/abs/0808.0003} {arXiv:0808.0003
  [astro-ph]} \BibitemShut {NoStop}%
%%CITATION = ARXIV:0808.0003;%%
\bibitem [{\citenamefont {Cole}\ \emph {et~al.}(1995)\citenamefont {Cole},
  \citenamefont {Fisher},\ and\ \citenamefont {Weinberg}}]{Cole:1994wf}%
  \BibitemOpen
  \bibfield  {author} {\bibinfo {author} {\bibfnamefont {S.}~\bibnamefont
  {Cole}}, \bibinfo {author} {\bibfnamefont {K.~B.}\ \bibnamefont {Fisher}}, \
  and\ \bibinfo {author} {\bibfnamefont {D.~H.}\ \bibnamefont {Weinberg}},\
  }\href {\doibase 10.1093/mnras/275.2.515} {\bibfield  {journal} {\bibinfo
  {journal} {Mon. Not. Roy. Astron. Soc.}\ }\textbf {\bibinfo {volume} {275}},\
  \bibinfo {pages} {515} (\bibinfo {year} {1995})},\ \Eprint
  {http://arxiv.org/abs/astro-ph/9412062} {arXiv:astro-ph/9412062 [astro-ph]}
  \BibitemShut {NoStop}%
%%CITATION = ASTRO-PH/9412062;%%
\bibitem [{\citenamefont {Peacock}\ and\ \citenamefont
  {Dodds}(1994)}]{Peacock:1993xg}%
  \BibitemOpen
  \bibfield  {author} {\bibinfo {author} {\bibfnamefont {J.~A.}\ \bibnamefont
  {Peacock}}\ and\ \bibinfo {author} {\bibfnamefont {S.~J.}\ \bibnamefont
  {Dodds}},\ }\href {\doibase 10.1093/mnras/267.4.1020} {\bibfield  {journal}
  {\bibinfo  {journal} {Mon. Not. Roy. Astron. Soc.}\ }\textbf {\bibinfo
  {volume} {267}},\ \bibinfo {pages} {1020} (\bibinfo {year} {1994})},\ \Eprint
  {http://arxiv.org/abs/astro-ph/9311057} {arXiv:astro-ph/9311057 [astro-ph]}
  \BibitemShut {NoStop}%
%%CITATION = ASTRO-PH/9311057;%%
\bibitem [{\citenamefont {Park}\ \emph {et~al.}(1994)\citenamefont {Park},
  \citenamefont {Vogeley}, \citenamefont {Geller},\ and\ \citenamefont
  {Huchra}}]{Park:1994fa}%
  \BibitemOpen
  \bibfield  {author} {\bibinfo {author} {\bibfnamefont {C.}~\bibnamefont
  {Park}}, \bibinfo {author} {\bibfnamefont {M.~S.}\ \bibnamefont {Vogeley}},
  \bibinfo {author} {\bibfnamefont {M.~J.}\ \bibnamefont {Geller}}, \ and\
  \bibinfo {author} {\bibfnamefont {J.~P.}\ \bibnamefont {Huchra}},\ }\href
  {\doibase 10.1086/174508} {\bibfield  {journal} {\bibinfo  {journal}
  {Astrophys. J.}\ }\textbf {\bibinfo {volume} {431}},\ \bibinfo {pages} {569}
  (\bibinfo {year} {1994})}\BibitemShut {NoStop}%
%%CITATION = ASJOA,431,569;%%
\bibitem [{\citenamefont {Ballinger}\ \emph {et~al.}(1996)\citenamefont
  {Ballinger}, \citenamefont {Peacock},\ and\ \citenamefont
  {Heavens}}]{Ballinger:1996cd}%
  \BibitemOpen
  \bibfield  {author} {\bibinfo {author} {\bibfnamefont {W.~E.}\ \bibnamefont
  {Ballinger}}, \bibinfo {author} {\bibfnamefont {J.~A.}\ \bibnamefont
  {Peacock}}, \ and\ \bibinfo {author} {\bibfnamefont {A.~F.}\ \bibnamefont
  {Heavens}},\ }\href {\doibase 10.1093/mnras/282.3.877} {\bibfield  {journal}
  {\bibinfo  {journal} {Mon. Not. Roy. Astron. Soc.}\ }\textbf {\bibinfo
  {volume} {282}},\ \bibinfo {pages} {877} (\bibinfo {year} {1996})},\ \Eprint
  {http://arxiv.org/abs/astro-ph/9605017} {arXiv:astro-ph/9605017 [astro-ph]}
  \BibitemShut {NoStop}%
%%CITATION = ASTRO-PH/9605017;%%
\bibitem [{\citenamefont {Magira}\ \emph {et~al.}(2000)\citenamefont {Magira},
  \citenamefont {Jing},\ and\ \citenamefont {Suto}}]{Magira:1999bn}%
  \BibitemOpen
  \bibfield  {author} {\bibinfo {author} {\bibfnamefont {H.}~\bibnamefont
  {Magira}}, \bibinfo {author} {\bibfnamefont {Y.~P.}\ \bibnamefont {Jing}}, \
  and\ \bibinfo {author} {\bibfnamefont {Y.}~\bibnamefont {Suto}},\ }\href
  {\doibase 10.1086/308170} {\bibfield  {journal} {\bibinfo  {journal}
  {Astrophys. J.}\ }\textbf {\bibinfo {volume} {528}},\ \bibinfo {pages} {30}
  (\bibinfo {year} {2000})},\ \Eprint {http://arxiv.org/abs/astro-ph/9907438}
  {arXiv:astro-ph/9907438 [astro-ph]} \BibitemShut {NoStop}%
%%CITATION = ASTRO-PH/9907438;%%
\bibitem [{\citenamefont {Nishimichi}\ and\ \citenamefont
  {Taruya}(2011)}]{Nishimichi:2011jm}%
  \BibitemOpen
  \bibfield  {author} {\bibinfo {author} {\bibfnamefont {T.}~\bibnamefont
  {Nishimichi}}\ and\ \bibinfo {author} {\bibfnamefont {A.}~\bibnamefont
  {Taruya}},\ }\href {\doibase 10.1103/PhysRevD.84.043526} {\bibfield
  {journal} {\bibinfo  {journal} {Phys. Rev.}\ }\textbf {\bibinfo {volume}
  {D84}},\ \bibinfo {pages} {043526} (\bibinfo {year} {2011})},\ \Eprint
  {http://arxiv.org/abs/1106.4562} {arXiv:1106.4562 [astro-ph.CO]} \BibitemShut
  {NoStop}%
%%CITATION = ARXIV:1106.4562;%%
\bibitem [{\citenamefont {Ishikawa}\ \emph {et~al.}(2014)\citenamefont
  {Ishikawa}, \citenamefont {Totani}, \citenamefont {Nishimichi}, \citenamefont
  {Takahashi}, \citenamefont {Yoshida},\ and\ \citenamefont
  {Tonegawa}}]{Ishikawa:2013aea}%
  \BibitemOpen
  \bibfield  {author} {\bibinfo {author} {\bibfnamefont {T.}~\bibnamefont
  {Ishikawa}}, \bibinfo {author} {\bibfnamefont {T.}~\bibnamefont {Totani}},
  \bibinfo {author} {\bibfnamefont {T.}~\bibnamefont {Nishimichi}}, \bibinfo
  {author} {\bibfnamefont {R.}~\bibnamefont {Takahashi}}, \bibinfo {author}
  {\bibfnamefont {N.}~\bibnamefont {Yoshida}}, \ and\ \bibinfo {author}
  {\bibfnamefont {M.}~\bibnamefont {Tonegawa}},\ }\href {\doibase
  10.1093/mnras/stu1382} {\bibfield  {journal} {\bibinfo  {journal} {Mon. Not.
  Roy. Astron. Soc.}\ }\textbf {\bibinfo {volume} {443}},\ \bibinfo {pages}
  {3359} (\bibinfo {year} {2014})},\ \Eprint {http://arxiv.org/abs/1308.6087}
  {arXiv:1308.6087 [astro-ph.CO]} \BibitemShut {NoStop}%
%%CITATION = ARXIV:1308.6087;%%
\bibitem [{\citenamefont {Blake}\ \emph {et~al.}(2011)\citenamefont {Blake}
  \emph {et~al.}}]{Blake:2011rj}%
  \BibitemOpen
  \bibfield  {author} {\bibinfo {author} {\bibfnamefont {C.}~\bibnamefont
  {Blake}} \emph {et~al.},\ }\href {\doibase 10.1111/j.1365-2966.2011.18903.x}
  {\bibfield  {journal} {\bibinfo  {journal} {Mon. Not. Roy. Astron. Soc.}\
  }\textbf {\bibinfo {volume} {415}},\ \bibinfo {pages} {2876} (\bibinfo {year}
  {2011})},\ \Eprint {http://arxiv.org/abs/1104.2948} {arXiv:1104.2948
  [astro-ph.CO]} \BibitemShut {NoStop}%
%%CITATION = ARXIV:1104.2948;%%
\bibitem [{\citenamefont {Pezzotta}\ \emph {et~al.}(2016)\citenamefont
  {Pezzotta} \emph {et~al.}}]{Pezzotta:2016gbo}%
  \BibitemOpen
  \bibfield  {author} {\bibinfo {author} {\bibfnamefont {A.}~\bibnamefont
  {Pezzotta}} \emph {et~al.},\ }\href@noop {} {\  (\bibinfo {year} {2016})},\
  \Eprint {http://arxiv.org/abs/1612.05645} {arXiv:1612.05645 [astro-ph.CO]}
  \BibitemShut {NoStop}%
%%CITATION = ARXIV:1612.05645;%%
\bibitem [{\citenamefont {Peacock}(1992)}]{Peacock1992}%
  \BibitemOpen
  \bibfield  {author} {\bibinfo {author} {\bibfnamefont {J.}~\bibnamefont
  {Peacock}},\ }\href@noop {} {\bibfield  {journal} {\bibinfo  {journal} {Mon.
  Not. Roy. Astron. Soc.}\ }\textbf {\bibinfo {volume} {258}},\ \bibinfo
  {pages} {581} (\bibinfo {year} {1992})}\BibitemShut {NoStop}%
%%CITATION = MNRAA,227,1;%%
\bibitem [{\citenamefont {Taruya}\ \emph {et~al.}(2014)\citenamefont {Taruya},
  \citenamefont {Koyama}, \citenamefont {Hiramatsu},\ and\ \citenamefont
  {Oka}}]{Taruya:2013quf}%
  \BibitemOpen
  \bibfield  {author} {\bibinfo {author} {\bibfnamefont {A.}~\bibnamefont
  {Taruya}}, \bibinfo {author} {\bibfnamefont {K.}~\bibnamefont {Koyama}},
  \bibinfo {author} {\bibfnamefont {T.}~\bibnamefont {Hiramatsu}}, \ and\
  \bibinfo {author} {\bibfnamefont {A.}~\bibnamefont {Oka}},\ }\href {\doibase
  10.1103/PhysRevD.89.043509} {\bibfield  {journal} {\bibinfo  {journal}
  {Phys.Rev.}\ }\textbf {\bibinfo {volume} {D89}},\ \bibinfo {pages} {043509}
  (\bibinfo {year} {2014})},\ \Eprint {http://arxiv.org/abs/1309.6783}
  {arXiv:1309.6783 [astro-ph.CO]} \BibitemShut {NoStop}%
%%CITATION = ARXIV:1309.6783;%%
\bibitem [{\citenamefont {Taruya}\ \emph {et~al.}(2009)\citenamefont {Taruya},
  \citenamefont {Nishimichi}, \citenamefont {Saito},\ and\ \citenamefont
  {Hiramatsu}}]{Taruya:2009ir}%
  \BibitemOpen
  \bibfield  {author} {\bibinfo {author} {\bibfnamefont {A.}~\bibnamefont
  {Taruya}}, \bibinfo {author} {\bibfnamefont {T.}~\bibnamefont {Nishimichi}},
  \bibinfo {author} {\bibfnamefont {S.}~\bibnamefont {Saito}}, \ and\ \bibinfo
  {author} {\bibfnamefont {T.}~\bibnamefont {Hiramatsu}},\ }\href {\doibase
  10.1103/PhysRevD.80.123503} {\bibfield  {journal} {\bibinfo  {journal} {Phys.
  Rev.}\ }\textbf {\bibinfo {volume} {D80}},\ \bibinfo {pages} {123503}
  (\bibinfo {year} {2009})},\ \Eprint {http://arxiv.org/abs/0906.0507}
  {arXiv:0906.0507 [astro-ph.CO]} \BibitemShut {NoStop}%
%%CITATION = ARXIV:0906.0507;%%
\bibitem [{\citenamefont {Okamura}\ \emph {et~al.}(2011)\citenamefont
  {Okamura}, \citenamefont {Taruya},\ and\ \citenamefont
  {Matsubara}}]{Okamura:2011nu}%
  \BibitemOpen
  \bibfield  {author} {\bibinfo {author} {\bibfnamefont {T.}~\bibnamefont
  {Okamura}}, \bibinfo {author} {\bibfnamefont {A.}~\bibnamefont {Taruya}}, \
  and\ \bibinfo {author} {\bibfnamefont {T.}~\bibnamefont {Matsubara}},\ }\href
  {\doibase 10.1088/1475-7516/2011/08/012} {\bibfield  {journal} {\bibinfo
  {journal} {JCAP}\ }\textbf {\bibinfo {volume} {1108}},\ \bibinfo {pages}
  {012} (\bibinfo {year} {2011})},\ \Eprint {http://arxiv.org/abs/1105.1491}
  {arXiv:1105.1491 [astro-ph.CO]} \BibitemShut {NoStop}%
%%CITATION = ARXIV:1105.1491;%%
\bibitem [{\citenamefont {Crocce}\ and\ \citenamefont
  {Scoccimarro}(2008)}]{Crocce:2007dt}%
  \BibitemOpen
  \bibfield  {author} {\bibinfo {author} {\bibfnamefont {M.}~\bibnamefont
  {Crocce}}\ and\ \bibinfo {author} {\bibfnamefont {R.}~\bibnamefont
  {Scoccimarro}},\ }\href {\doibase 10.1103/PhysRevD.77.023533} {\bibfield
  {journal} {\bibinfo  {journal} {Phys. Rev.}\ }\textbf {\bibinfo {volume}
  {D77}},\ \bibinfo {pages} {023533} (\bibinfo {year} {2008})},\ \Eprint
  {http://arxiv.org/abs/0704.2783} {arXiv:0704.2783 [astro-ph]} \BibitemShut
  {NoStop}%
%%CITATION = ARXIV:0704.2783;%%
\bibitem [{\citenamefont {Crocce}\ \emph {et~al.}(2012)\citenamefont {Crocce},
  \citenamefont {Scoccimarro},\ and\ \citenamefont
  {Bernardeau}}]{Crocce:2012fa}%
  \BibitemOpen
  \bibfield  {author} {\bibinfo {author} {\bibfnamefont {M.}~\bibnamefont
  {Crocce}}, \bibinfo {author} {\bibfnamefont {R.}~\bibnamefont {Scoccimarro}},
  \ and\ \bibinfo {author} {\bibfnamefont {F.}~\bibnamefont {Bernardeau}},\
  }\href {\doibase 10.1111/j.1365-2966.2012.22127.x} {\bibfield  {journal}
  {\bibinfo  {journal} {Mon. Not. Roy. Astron. Soc.}\ }\textbf {\bibinfo
  {volume} {427}},\ \bibinfo {pages} {2537} (\bibinfo {year} {2012})},\ \Eprint
  {http://arxiv.org/abs/1207.1465} {arXiv:1207.1465 [astro-ph.CO]} \BibitemShut
  {NoStop}%
%%CITATION = ARXIV:1207.1465;%%
\bibitem [{\citenamefont {Taruya}\ \emph {et~al.}(2012)\citenamefont {Taruya},
  \citenamefont {Bernardeau}, \citenamefont {Nishimichi},\ and\ \citenamefont
  {Codis}}]{Taruya:2012ut}%
  \BibitemOpen
  \bibfield  {author} {\bibinfo {author} {\bibfnamefont {A.}~\bibnamefont
  {Taruya}}, \bibinfo {author} {\bibfnamefont {F.}~\bibnamefont {Bernardeau}},
  \bibinfo {author} {\bibfnamefont {T.}~\bibnamefont {Nishimichi}}, \ and\
  \bibinfo {author} {\bibfnamefont {S.}~\bibnamefont {Codis}},\ }\href
  {\doibase 10.1103/PhysRevD.86.103528} {\bibfield  {journal} {\bibinfo
  {journal} {Phys. Rev.}\ }\textbf {\bibinfo {volume} {D86}},\ \bibinfo {pages}
  {103528} (\bibinfo {year} {2012})},\ \Eprint {http://arxiv.org/abs/1208.1191}
  {arXiv:1208.1191 [astro-ph.CO]} \BibitemShut {NoStop}%
%%CITATION = ARXIV:1208.1191;%%
\bibitem [{\citenamefont {Takahashi}\ \emph {et~al.}(2009)\citenamefont
  {Takahashi}, \citenamefont {Yoshida}, \citenamefont {Takada}, \citenamefont
  {Matsubara}, \citenamefont {Sugiyama}, \citenamefont {Kayo}, \citenamefont
  {Nishizawa}, \citenamefont {Nishimichi}, \citenamefont {Saito},\ and\
  \citenamefont {Taruya}}]{Takahashi:2009bq}%
  \BibitemOpen
  \bibfield  {author} {\bibinfo {author} {\bibfnamefont {R.}~\bibnamefont
  {Takahashi}}, \bibinfo {author} {\bibfnamefont {N.}~\bibnamefont {Yoshida}},
  \bibinfo {author} {\bibfnamefont {M.}~\bibnamefont {Takada}}, \bibinfo
  {author} {\bibfnamefont {T.}~\bibnamefont {Matsubara}}, \bibinfo {author}
  {\bibfnamefont {N.}~\bibnamefont {Sugiyama}}, \bibinfo {author}
  {\bibfnamefont {I.}~\bibnamefont {Kayo}}, \bibinfo {author} {\bibfnamefont
  {A.~J.}\ \bibnamefont {Nishizawa}}, \bibinfo {author} {\bibfnamefont
  {T.}~\bibnamefont {Nishimichi}}, \bibinfo {author} {\bibfnamefont
  {S.}~\bibnamefont {Saito}}, \ and\ \bibinfo {author} {\bibfnamefont
  {A.}~\bibnamefont {Taruya}},\ }\href {\doibase 10.1088/0004-637X/700/1/479}
  {\bibfield  {journal} {\bibinfo  {journal} {Astrophys. J.}\ }\textbf
  {\bibinfo {volume} {700}},\ \bibinfo {pages} {479} (\bibinfo {year}
  {2009})},\ \Eprint {http://arxiv.org/abs/0902.0371} {arXiv:0902.0371
  [astro-ph.CO]} \BibitemShut {NoStop}%
%%CITATION = ARXIV:0902.0371;%%
\bibitem [{\citenamefont {Chodorowski}\ and\ \citenamefont
  {Ciecielag}(2002)}]{Chodorowski:2001id}%
  \BibitemOpen
  \bibfield  {author} {\bibinfo {author} {\bibfnamefont {M.}~\bibnamefont
  {Chodorowski}}\ and\ \bibinfo {author} {\bibfnamefont {P.}~\bibnamefont
  {Ciecielag}},\ }\href {\doibase 10.1046/j.1365-8711.2002.05161.x} {\bibfield
  {journal} {\bibinfo  {journal} {Mon. Not. Roy. Astron. Soc.}\ }\textbf
  {\bibinfo {volume} {331}},\ \bibinfo {pages} {133} (\bibinfo {year}
  {2002})},\ \Eprint {http://arxiv.org/abs/astro-ph/0109291}
  {arXiv:astro-ph/0109291 [astro-ph]} \BibitemShut {NoStop}%
%%CITATION = ASTRO-PH/0109291;%%
\bibitem [{\citenamefont {Aghamousa}\ \emph {et~al.}(2016)\citenamefont
  {Aghamousa} \emph {et~al.}}]{Aghamousa:2016zmz}%
  \BibitemOpen
  \bibfield  {author} {\bibinfo {author} {\bibfnamefont {A.}~\bibnamefont
  {Aghamousa}} \emph {et~al.} (\bibinfo {collaboration} {DESI}),\ }\href@noop
  {} {\  (\bibinfo {year} {2016})},\ \Eprint {http://arxiv.org/abs/1611.00036}
  {arXiv:1611.00036 [astro-ph.IM]} \BibitemShut {NoStop}%
%%CITATION = ARXIV:1611.00036;%%
\bibitem [{\citenamefont {Laureijs}\ \emph {et~al.}(2011)\citenamefont
  {Laureijs} \emph {et~al.}}]{Laureijs:2011gra}%
  \BibitemOpen
  \bibfield  {author} {\bibinfo {author} {\bibfnamefont {R.}~\bibnamefont
  {Laureijs}} \emph {et~al.} (\bibinfo {collaboration} {EUCLID}),\ }\href@noop
  {} {\  (\bibinfo {year} {2011})},\ \Eprint {http://arxiv.org/abs/1110.3193}
  {arXiv:1110.3193 [astro-ph.CO]} \BibitemShut {NoStop}%
%%CITATION = ARXIV:1110.3193;%%
\bibitem [{\citenamefont {Ross}\ \emph {et~al.}(2015)\citenamefont {Ross},
  \citenamefont {Samushia}, \citenamefont {Howlett}, \citenamefont {Percival},
  \citenamefont {Burden},\ and\ \citenamefont {Manera}}]{Ross:2014qpa}%
  \BibitemOpen
  \bibfield  {author} {\bibinfo {author} {\bibfnamefont {A.~J.}\ \bibnamefont
  {Ross}}, \bibinfo {author} {\bibfnamefont {L.}~\bibnamefont {Samushia}},
  \bibinfo {author} {\bibfnamefont {C.}~\bibnamefont {Howlett}}, \bibinfo
  {author} {\bibfnamefont {W.~J.}\ \bibnamefont {Percival}}, \bibinfo {author}
  {\bibfnamefont {A.}~\bibnamefont {Burden}}, \ and\ \bibinfo {author}
  {\bibfnamefont {M.}~\bibnamefont {Manera}},\ }\href {\doibase
  10.1093/mnras/stv154} {\bibfield  {journal} {\bibinfo  {journal} {Mon. Not.
  Roy. Astron. Soc.}\ }\textbf {\bibinfo {volume} {449}},\ \bibinfo {pages}
  {835} (\bibinfo {year} {2015})},\ \Eprint {http://arxiv.org/abs/1409.3242}
  {arXiv:1409.3242 [astro-ph.CO]} \BibitemShut {NoStop}%
%%CITATION = ARXIV:1409.3242;%%
\bibitem [{\citenamefont {Zhao}(2014)}]{Zhao:2013dza}%
  \BibitemOpen
  \bibfield  {author} {\bibinfo {author} {\bibfnamefont {G.-B.}\ \bibnamefont
  {Zhao}},\ }\href {\doibase 10.1088/0067-0049/211/2/23} {\bibfield  {journal}
  {\bibinfo  {journal} {Astrophys. J. Suppl.}\ }\textbf {\bibinfo {volume}
  {211}},\ \bibinfo {pages} {23} (\bibinfo {year} {2014})},\ \Eprint
  {http://arxiv.org/abs/1312.1291} {arXiv:1312.1291 [astro-ph.CO]} \BibitemShut
  {NoStop}%
%%CITATION = ARXIV:1312.1291;%%
\bibitem [{\citenamefont {Howlett}\ \emph {et~al.}(2015)\citenamefont
  {Howlett}, \citenamefont {Manera},\ and\ \citenamefont
  {Percival}}]{Howlett:2015hfa}%
  \BibitemOpen
  \bibfield  {author} {\bibinfo {author} {\bibfnamefont {C.}~\bibnamefont
  {Howlett}}, \bibinfo {author} {\bibfnamefont {M.}~\bibnamefont {Manera}}, \
  and\ \bibinfo {author} {\bibfnamefont {W.~J.}\ \bibnamefont {Percival}},\
  }\href {\doibase 10.1016/j.ascom.2015.07.003} {\bibfield  {journal} {\bibinfo
   {journal} {Astron. Comput.}\ }\textbf {\bibinfo {volume} {12}},\ \bibinfo
  {pages} {109} (\bibinfo {year} {2015})},\ \Eprint
  {http://arxiv.org/abs/1506.03737} {arXiv:1506.03737 [astro-ph.CO]}
  \BibitemShut {NoStop}%
%%CITATION = ARXIV:1506.03737;%%
\bibitem [{\citenamefont {Li}\ \emph {et~al.}(2012)\citenamefont {Li},
  \citenamefont {Zhao}, \citenamefont {Teyssier},\ and\ \citenamefont
  {Koyama}}]{Li:2011vk}%
  \BibitemOpen
  \bibfield  {author} {\bibinfo {author} {\bibfnamefont {B.}~\bibnamefont
  {Li}}, \bibinfo {author} {\bibfnamefont {G.-B.}\ \bibnamefont {Zhao}},
  \bibinfo {author} {\bibfnamefont {R.}~\bibnamefont {Teyssier}}, \ and\
  \bibinfo {author} {\bibfnamefont {K.}~\bibnamefont {Koyama}},\ }\href
  {\doibase 10.1088/1475-7516/2012/01/051} {\bibfield  {journal} {\bibinfo
  {journal} {JCAP}\ }\textbf {\bibinfo {volume} {1201}},\ \bibinfo {pages}
  {051} (\bibinfo {year} {2012})},\ \Eprint {http://arxiv.org/abs/1110.1379}
  {arXiv:1110.1379 [astro-ph.CO]} \BibitemShut {NoStop}%
%%CITATION = ARXIV:1110.1379;%%
\bibitem [{\citenamefont {Hinshaw}\ \emph {et~al.}(2013)\citenamefont {Hinshaw}
  \emph {et~al.}}]{Hinshaw:2012aka}%
  \BibitemOpen
  \bibfield  {author} {\bibinfo {author} {\bibfnamefont {G.}~\bibnamefont
  {Hinshaw}} \emph {et~al.} (\bibinfo {collaboration} {WMAP}),\ }\href
  {\doibase 10.1088/0067-0049/208/2/19} {\bibfield  {journal} {\bibinfo
  {journal} {Astrophys. J. Suppl.}\ }\textbf {\bibinfo {volume} {208}},\
  \bibinfo {pages} {19} (\bibinfo {year} {2013})},\ \Eprint
  {http://arxiv.org/abs/1212.5226} {arXiv:1212.5226 [astro-ph.CO]} \BibitemShut
  {NoStop}%
%%CITATION = ARXIV:1212.5226;%%
\bibitem [{\citenamefont {{Cautun}}\ and\ \citenamefont {{van de
  Weygaert}}(2011)}]{cv2011}%
  \BibitemOpen
  \bibfield  {author} {\bibinfo {author} {\bibfnamefont {M.~C.}\ \bibnamefont
  {{Cautun}}}\ and\ \bibinfo {author} {\bibfnamefont {R.}~\bibnamefont {{van de
  Weygaert}}},\ }\href@noop {} {\bibfield  {journal} {\bibinfo  {journal}
  {ArXiv e-prints}\ } (\bibinfo {year} {2011})},\ \Eprint
  {http://arxiv.org/abs/1105.0370} {arXiv:1105.0370 [astro-ph.IM]} \BibitemShut
  {NoStop}%
\bibitem [{\citenamefont {{Schaap}}\ and\ \citenamefont {{van de
  Weygaert}}(2000)}]{sv2000}%
  \BibitemOpen
  \bibfield  {author} {\bibinfo {author} {\bibfnamefont {W.~E.}\ \bibnamefont
  {{Schaap}}}\ and\ \bibinfo {author} {\bibfnamefont {R.}~\bibnamefont {{van de
  Weygaert}}},\ }\href@noop {} {\bibfield  {journal} {\bibinfo  {journal}
  {ArXiv e-prints}\ }\textbf {\bibinfo {volume} {363}},\ \bibinfo {pages} {L29}
  (\bibinfo {year} {2000})},\ \Eprint
  {http://arxiv.org/abs/arXiv:astro-ph/0011007} {arXiv:astro-ph/0011007}
  \BibitemShut {NoStop}%
\bibitem [{\citenamefont {{van de Weygaert}}\ and\ \citenamefont
  {{Schaap}}(2009)}]{vs2009}%
  \BibitemOpen
  \bibfield  {author} {\bibinfo {author} {\bibfnamefont {R.}~\bibnamefont {{van
  de Weygaert}}}\ and\ \bibinfo {author} {\bibfnamefont {W.}~\bibnamefont
  {{Schaap}}},\ }in\ \href {\doibase 10.1007/978-3-540-44767-2_11} {\emph
  {\bibinfo {booktitle} {Data Analysis in Cosmology}}},\ \bibinfo {series}
  {Lecture Notes in Physics, Berlin Springer Verlag}, Vol.\ \bibinfo {volume}
  {665},\ \bibinfo {editor} {edited by\ \bibinfo {editor} {\bibfnamefont
  {V.~J.}\ \bibnamefont {{Mart{\'{\i}}nez}}}, \bibinfo {editor} {\bibfnamefont
  {E.}~\bibnamefont {{Saar}}}, \bibinfo {editor} {\bibfnamefont
  {E.}~\bibnamefont {{Mart{\'{\i}}nez-Gonz{\'a}lez}}}, \ and\ \bibinfo {editor}
  {\bibfnamefont {M.-J.}\ \bibnamefont {{Pons-Border{\'{\i}}a}}}}\ (\bibinfo
  {year} {2009})\ pp.\ \bibinfo {pages} {291--413}\BibitemShut {NoStop}%
\bibitem [{\citenamefont {Li}\ \emph {et~al.}(2013)\citenamefont {Li},
  \citenamefont {Hellwing}, \citenamefont {Koyama}, \citenamefont {Zhao},
  \citenamefont {Jennings} \emph {et~al.}}]{Li:2012by}%
  \BibitemOpen
  \bibfield  {author} {\bibinfo {author} {\bibfnamefont {B.}~\bibnamefont
  {Li}}, \bibinfo {author} {\bibfnamefont {W.~A.}\ \bibnamefont {Hellwing}},
  \bibinfo {author} {\bibfnamefont {K.}~\bibnamefont {Koyama}}, \bibinfo
  {author} {\bibfnamefont {G.-B.}\ \bibnamefont {Zhao}}, \bibinfo {author}
  {\bibfnamefont {E.}~\bibnamefont {Jennings}},  \emph {et~al.},\ }\href
  {\doibase 10.1093/mnras/sts072} {\bibfield  {journal} {\bibinfo  {journal}
  {Mon.Not.Roy.Astron.Soc.}\ }\textbf {\bibinfo {volume} {428}},\ \bibinfo
  {pages} {743} (\bibinfo {year} {2013})},\ \Eprint
  {http://arxiv.org/abs/1206.4317} {arXiv:1206.4317 [astro-ph.CO]} \BibitemShut
  {NoStop}%
%%CITATION = ARXIV:1206.4317;%%
\bibitem [{\citenamefont {{Hellwing}}\ \emph {et~al.}(2013)\citenamefont
  {{Hellwing}}, \citenamefont {{Li}}, \citenamefont {{Frenk}},\ and\
  \citenamefont {{Cole}}}]{Hellwing2013}%
  \BibitemOpen
  \bibfield  {author} {\bibinfo {author} {\bibfnamefont {W.~A.}\ \bibnamefont
  {{Hellwing}}}, \bibinfo {author} {\bibfnamefont {B.}~\bibnamefont {{Li}}},
  \bibinfo {author} {\bibfnamefont {C.~S.}\ \bibnamefont {{Frenk}}}, \ and\
  \bibinfo {author} {\bibfnamefont {S.}~\bibnamefont {{Cole}}},\ }\href
  {\doibase 10.1093/mnras/stt1430} {\bibfield  {journal} {\bibinfo  {journal}
  {"Mon. Not. Roy. Astron. Soc."}\ }\textbf {\bibinfo {volume} {435}},\
  \bibinfo {pages} {2806} (\bibinfo {year} {2013})},\ \Eprint
  {http://arxiv.org/abs/1305.7486} {arXiv:1305.7486} \BibitemShut {NoStop}%
\bibitem [{\citenamefont {{Hellwing}}\ \emph {et~al.}(2016)\citenamefont
  {{Hellwing}}, \citenamefont {{Schaller}}, \citenamefont {{Frenk}},
  \citenamefont {{Theuns}}, \citenamefont {{Schaye}}, \citenamefont {{Bower}},\
  and\ \citenamefont {{Crain}}}]{Hellwing2016}%
  \BibitemOpen
  \bibfield  {author} {\bibinfo {author} {\bibfnamefont {W.~A.}\ \bibnamefont
  {{Hellwing}}}, \bibinfo {author} {\bibfnamefont {M.}~\bibnamefont
  {{Schaller}}}, \bibinfo {author} {\bibfnamefont {C.~S.}\ \bibnamefont
  {{Frenk}}}, \bibinfo {author} {\bibfnamefont {T.}~\bibnamefont {{Theuns}}},
  \bibinfo {author} {\bibfnamefont {J.}~\bibnamefont {{Schaye}}}, \bibinfo
  {author} {\bibfnamefont {R.~G.}\ \bibnamefont {{Bower}}}, \ and\ \bibinfo
  {author} {\bibfnamefont {R.~A.}\ \bibnamefont {{Crain}}},\ }\href {\doibase
  10.1093/mnrasl/slw081} {\bibfield  {journal} {\bibinfo  {journal} {"Mon. Not.
  Roy. Astron. Soc."}\ }\textbf {\bibinfo {volume} {461}},\ \bibinfo {pages}
  {L11} (\bibinfo {year} {2016})},\ \Eprint {http://arxiv.org/abs/1603.03328}
  {arXiv:1603.03328} \BibitemShut {NoStop}%
\bibitem [{\citenamefont {Schaye}\ \emph {et~al.}(2015)\citenamefont {Schaye}
  \emph {et~al.}}]{Schaye:2014tpa}%
  \BibitemOpen
  \bibfield  {author} {\bibinfo {author} {\bibfnamefont {J.}~\bibnamefont
  {Schaye}} \emph {et~al.},\ }\href {\doibase 10.1093/mnras/stu2058} {\bibfield
   {journal} {\bibinfo  {journal} {Mon. Not. Roy. Astron. Soc.}\ }\textbf
  {\bibinfo {volume} {446}},\ \bibinfo {pages} {521} (\bibinfo {year}
  {2015})},\ \Eprint {http://arxiv.org/abs/1407.7040} {arXiv:1407.7040
  [astro-ph.GA]} \BibitemShut {NoStop}%
%%CITATION = ARXIV:1407.7040;%%
\bibitem [{\citenamefont {Winther}\ \emph {et~al.}(2015)\citenamefont {Winther}
  \emph {et~al.}}]{Winther:2015wla}%
  \BibitemOpen
  \bibfield  {author} {\bibinfo {author} {\bibfnamefont {H.~A.}\ \bibnamefont
  {Winther}} \emph {et~al.},\ }\href {\doibase 10.1093/mnras/stv2253}
  {\bibfield  {journal} {\bibinfo  {journal} {Mon. Not. Roy. Astron. Soc.}\
  }\textbf {\bibinfo {volume} {454}},\ \bibinfo {pages} {4208} (\bibinfo {year}
  {2015})},\ \Eprint {http://arxiv.org/abs/1506.06384} {arXiv:1506.06384
  [astro-ph.CO]} \BibitemShut {NoStop}%
%%CITATION = ARXIV:1506.06384;%%
\bibitem [{\citenamefont {Falck}\ \emph {et~al.}(2015)\citenamefont {Falck},
  \citenamefont {Koyama},\ and\ \citenamefont {Zhao}}]{Falck:2015rsa}%
  \BibitemOpen
  \bibfield  {author} {\bibinfo {author} {\bibfnamefont {B.}~\bibnamefont
  {Falck}}, \bibinfo {author} {\bibfnamefont {K.}~\bibnamefont {Koyama}}, \
  and\ \bibinfo {author} {\bibfnamefont {G.-B.}\ \bibnamefont {Zhao}},\ }\href
  {\doibase 10.1088/1475-7516/2015/07/049} {\bibfield  {journal} {\bibinfo
  {journal} {JCAP}\ }\textbf {\bibinfo {volume} {1507}},\ \bibinfo {pages}
  {049} (\bibinfo {year} {2015})},\ \Eprint {http://arxiv.org/abs/1503.06673}
  {arXiv:1503.06673 [astro-ph.CO]} \BibitemShut {NoStop}%
%%CITATION = ARXIV:1503.06673;%%
\bibitem [{\citenamefont {Lewis}\ and\ \citenamefont
  {Bridle}(2002)}]{Lewis:2002ah}%
  \BibitemOpen
  \bibfield  {author} {\bibinfo {author} {\bibfnamefont {A.}~\bibnamefont
  {Lewis}}\ and\ \bibinfo {author} {\bibfnamefont {S.}~\bibnamefont {Bridle}},\
  }\href {\doibase 10.1103/PhysRevD.66.103511} {\bibfield  {journal} {\bibinfo
  {journal} {Phys. Rev.}\ }\textbf {\bibinfo {volume} {D66}},\ \bibinfo {pages}
  {103511} (\bibinfo {year} {2002})},\ \Eprint
  {http://arxiv.org/abs/astro-ph/0205436} {arXiv:astro-ph/0205436 [astro-ph]}
  \BibitemShut {NoStop}%
%%CITATION = ASTRO-PH/0205436;%%
\end{thebibliography}%
\end{document}